\documentclass[12pt]{article}
\usepackage{multirow}
\usepackage{geometry}
\usepackage{bbold}
\usepackage{amsmath,amssymb,amsfonts}
\usepackage{graphicx}
\usepackage[footnotesize]{caption2}
\usepackage{mathrsfs}
\usepackage{bbm}
\usepackage{braket}
\usepackage{cancel}
\usepackage{color}
\usepackage{enumitem}

\definecolor{Orange}{cmyk}{0,0.61,0.87,0}
\definecolor{JungleGreen}{cmyk}{0.99,0,0.52,0}
\definecolor{OliveGreen}{cmyk}{0.64,0,0.95,0.40}
\definecolor{Brown}{cmyk}{0,0.81,1,0.60}
\definecolor{RoyalBlue}{cmyk}{0.71,0.53,0,0.12}

\usepackage[colorlinks=true, linkcolor=OliveGreen, citecolor=RoyalBlue,
urlcolor=RoyalBlue]{hyperref}

\allowdisplaybreaks

\definecolor{darkblue}{rgb}{0,  0,  .5}
\definecolor{lightgray}{gray}{0.5}

\newcommand{\phPF}[2]{{#1} \,  {\mathrm{e}\ ^{  i \:  #2}}}
\newcommand{\phNF}[2]{{#1} \,  {\mathrm{e}\ ^{ - i \: #2}}}

\newcommand{\Mg}{M_{\textnormal{GUT}}}
\newcommand{\Mi}{M_{\rm in}}
\newcommand{\Mw}{M_{\textnormal{EW}}}

\newcommand{\V}{V_{\textnormal{CKM}}}
\newcommand{\VG}{V_{\textnormal{GCKM}}}

\newcommand{\eq}[1]{Eq.~\eqref{#1}}
\newcommand{\Figref}[1]{Fig~\ref{#1}}

\newcommand{\e}{\mathrm{e}}

\newcommand{\BBsm}{\text{BR}(B_s\rightarrow\mu^+\mu^-)}

\newcommand{\Bmueg}{\text{BR}(\mu\rightarrow e\gamma)}
\newcommand{\muegD}{\mu\rightarrow e\gamma}

\newcommand{\BtauegE}{\text{BR}_{\rm{EXP}}(\mu\rightarrow e\gamma)}

\newcommand{\amuegR}{a_{\mu e \gamma R}}

\usepackage{cleveref}

\crefformat{section}{\S#2#1#3} 
\crefformat{subsection}{\S#2#1#3}
\crefformat{subsubsection}{\S#2#1#3}

\newcommand{\TPDK}{\tau\left(p\rightarrow K^+ \bar{\nu}\right)}

\newcommand{\Dt}{(4\pi)^2\Lambda\frac{d}{d\Lambda}}

\newcommand{\nn}{\nonumber}

\def\beq{\begin{equation}}
\def\eeq{\end{equation}}
\def\bea{\begin{eqnarray} }
\def\eea{ \end{eqnarray} }

\def\mgut{M_{\rm GUT}}

\def\afiv{A_\mathbf{\overline{5}}}
\def\aten{A_\mathbf{10}}
\DeclareMathOperator{\Tr}{Tr}

\def\alpfiv{\alpha_\mathbf{\overline{5}}}
\def\alpten{\alpha_\mathbf{10}}
\def\alplam{\alpha_{\lambda}}
\def\alplamp{\alpha_{\lambda'}}
\def\betH{\beta_H}
\def\betSg{\beta_\Sigma}

\begin{document}

\begin{flushright}
{\tt KCL-PH-TH/2020-61}, {\tt CERN-TH-2020-172}  \\
{\tt UMN-TH-4002/20, FTPI-MINN-20/33} \\
{\tt KIAS-P20064 }\\
\end{flushright}

\vspace{0.7cm}
\begin{center}
{\Large\bf 
Low-Energy  Probes of  No-Scale  SU(5) Super-GUTs\\
}
\end{center}

\vspace{0.5cm}
\begin{center}{\large
{\bf John~Ellis}$^{1}$,
{\bf Jason~L.~Evans}$^2$,
{\bf Natsumi~Nagata}$^3$,  \\
{\bf Keith A. Olive}$^4$ and {\bf L. Velasco-Sevilla}$^{5}$\\
}
\end{center}

\begin{center}
{\em $^1$Theoretical Particle Physics and Cosmology Group, Department of
  Physics, King's~College~London, London WC2R 2LS, United Kingdom;\\
Theoretical Physics Department, CERN, CH-1211 Geneva 23,
  Switzerland};\\
  {\em National Institute of Chemical Physics \& Biophysics, R{\" a}vala 10, 10143 Tallinn, Estonia}\\[0.2cm]
  {\em $^2$T. D. Lee Institute, Shanghai Jiao Tong University, Shanghai 200240, China}\\[0.2cm]
  {\em $^3$Department of Physics, University of Tokyo, 
Tokyo 113--0033,
Japan}\\[0.2cm]
  {\em $^4$William I. Fine Theoretical Physics Institute, School of Physics and Astronomy,\\
University of Minnesota, Minneapolis, MN 55455, USA}\\[0.2cm]
{\em $^5$ Department of Physics and Technology, University of Bergen,\\
PO Box 7803, 5020 Bergen, Norway; \\
Korea Institute for Advanced Study, Seoul 02455, Korea}\\
\vspace{0.5cm}
{\bf Abstract}\\
\end{center}

We explore the possible values of the $\mu \to e \gamma$ branching ratio, $\Bmueg$, and the electron
dipole moment (eEDM), $d_e$, in no-scale SU(5) super-GUT models with the boundary conditions that soft
supersymmetry-breaking matter scalar masses vanish at some high input
scale, $\Mi$, above the GUT scale, $\Mg$. We take into account the constraints from the cosmological cold dark
matter density, $\Omega_{CDM} h^2$, the Higgs mass, $M_h$, and the experimental lower limit on the lifetime for 
$p \to K^+ \bar \nu$, the dominant proton decay mode in these super-GUT models. Reconciling this limit with
$\Omega_{CDM} h^2$ and $M_h$ requires the Higgs field responsible for the charge-2/3 quark masses
to be twisted, and possibly also that responsible for the charge-1/3 and charged-lepton masses, with model-dependent soft supersymmetry-breaking masses. 
We consider six possible models for the super-GUT initial conditions, and two possible choices 
for quark flavor mixing, contrasting their predictions for proton decay with versions of the models 
in which mixing effects are neglected. We find that $\TPDK$ may be 
accessible to the upcoming Hyper-Kamiokande experiment, whereas
all the models predict $\Bmueg$ and $d_e$ below 
the current and prospective future experimental sensitivities
or both flavor choices, when the dark matter density, Higgs mass
and current proton decay constraints are taken into account. 
However, there are limited regions with one of the 
flavor choices in two of the models where $\mu \to e$
conversion on a heavy nucleus may be observable in the future.
{\it Our results indicate that there is no supersymmetric
flavor problem in the class of no-scale models we consider.}

\newpage

\section{Introduction}

Supersymmetry remains an attractive prospective extension of the Standard Model (SM), despite its non-appearance during Runs 1 and 2 of the LHC \cite{ATLAS20,CMS20}. Indeed, the discovery of a 125-GeV Higgs boson at the LHC \cite{lhch} has supplemented the traditional arguments for supersymmetry, which include the naturalness of the electroweak scale \cite{Maiani:1979cx}, the unification of the fundamental interactions \cite{Ellis:1990zq} and the existence of a cold dark matter candidate (if R-parity is conserved) \cite{ehnos}. The minimal supersymmetric extension of the SM (MSSM) predicted the existence of a Higgs boson with mass $M_h \lesssim 130$ GeV \cite{mh}, and is a prime example of new physics capable of stabilizing the electroweak vacuum for $M_h \sim 125$ GeV \cite{Ellis:2000ig}. 
Furthermore, global fits in the framework of simple supersymmetric models suggest that the couplings of the lightest supersymmetric Higgs boson should be very similar to those of the Higgs boson in the SM, as is indicated by the ATLAS and CMS experiments \cite{ATLASmu,CMSmu}. When the supersymmetric particle masses are large, which is the case we consider, the Higgs couplings resemble even more 
closely the couplings predicted by the SM. 
 
 However, the continuing absence of supersymmetry at the LHC~\cite{ATLAS20,CMS20} reinforces the need to seek complementary indications of supersymmetry outside colliders. It is in this context that we address the questions of proton decay,
 contributions to the electron dipole moment and $\mu$ flavor violation observables in the SU(5) models based on no-scale supergravity that were introduced in  \cite{Ellis:2017djk}. There, Higgs fields were assigned to twisted chiral supermultiplets with a suitable choices of modular weights in order to obtain the correct mass of the observed Higgs boson and the cold dark matter density, while avoiding proton decay in violation of the current limits.  
 
 Contrary to what happens in the Standard Model, where flavor and CP violation are controlled by the Cabibbo-Kobayashi-Maskawa (CKM) matrix, there is no established mechanism for flavor and CP violation in supersymmetry, the so-called {\it supersymmetric flavor problem}. Experiments show that many low-energy predictions of CKM mixing must be reproduced in any extension of the SM, which is therefore an important constraint on any supersymmetric model that is studied.  
 
 In a previous study of super-GUT no-scale models in~\cite{Ellis:2016qra} we adopted a pragmatic approach to this challenge, using particular  Ans{\"a}tze for Yukawa couplings to study flavor violation constraints in a scenario with
 maximal sfermion flavor violation at the input scale $\Mi > \Mg$. Here we revisit flavor violation and proton decay,
 considering alternative options for the flavor mixing  associated with different embeddings of the MSSM fields in GUT multiplets.~\footnote{For reviews of supersymmetry, GUTs and flavor mixing, see~\cite{Reviews}. 
 }
 
 In the SM, Yukawa couplings in the up- and down-quark sectors are described by a couple of $3\times3$ complex matrices whose diagonalizations each require two unitary matrices, one acting on left-handed quarks and the other on right-handed quarks. The two left-handed matrices, one in the up-quark sector and the other in the down-quark sector, combine to form the CKM matrix, whereas the right-handed matrices remain unobservable. In supersymmetry, however, the right-handed matrices propagate into the soft-breaking terms and hence become constrained by flavor observables. These observables clearly indicate that off-diagonal elements of the right-handed sfermion mixing matrices should be tiny~\footnote{See~\cite{Altmannshofer:2009ne} for a comprehensive review and an analysis of the particular case of $\BBsm$.}. Any model of supersymmetric flavor must specify how to reproduce the CKM matrix via the two down- and up-quark left-handed matrices that diagonalize the Yukawa couplings. One choice is to associate the CKM matrix with the up-quark Yukawa matrix, for which electroweak (EW) precision observables play an important role in constraining how this is propagated into the supersymmetric sector, as was studied for the CMSSM in \cite{Gomez:2015ila}. Another is to associate the CKM matrix with the down-quark sector, as we considered in~\cite{Ellis:2017djk}. In this case the constraints from flavor observables are more stringent than those from EW observables, particularly for the low $\tan\beta$ values that we use. 
 
 We study in this paper six different no-scale super-GUT SU(5) models, some with both electroweak
 Higgs representations in twisted chiral supermultiplets, and some with only one twisted Higgs
 supermultiplet. The soft supersymmetry-breaking masses of the MSSM matter sfermions vanish at the input scale $\Mi$
 in all the models, but they have different boundary conditions for other supersymmetry-breaking parameters.
 Four of the models have $\Mi = 10^{16.5}$~GeV, whereas the other two have
 $\Mi = 10^{18}$~GeV, in which case there are larger renormalization-group running effects above the GUT scale,
 $\Mg$. For each model, we study predictions for proton decay, $\mu \to e \gamma$
 and the electron EDM, using two possible choices for the flavor embeddings of the
 quarks and leptons into SU(5) multiplets that illustrate the ambiguity discussed in the previous
 paragraph. We find that proton decay rates are relatively insensitive to the treatment of flavor
 mixing, whereas $\mu \to e \gamma$ and the electron EDM are more sensitive. In general, the
 predictions for these flavor observables are below the present experimental limits when the
 cosmological dark matter density and the proton lifetime are taken into account, though there are limited regions with one of the 
flavor choices in two of the models where $\mu \to e$
conversion on a heavy nucleus may be observable in the future.
 {\it These no-scale super-GUT models have no supersymmetric flavor problem,}
 as also argued in~\cite{Ellis:2016qra}.

This paper is organized as follows. In Section~\ref{sec:sectionModelFramework} we introduce the class of no-scale SU(5)
super-GUT models we study, including the specification of different choices for the embedding of MSSM fields in 
GUT multiplets and the corresponding Ans{\"a}tze for matter Yukawa coupling matrices, 
the no-scale boundary conditions on soft supersymmetry breaking at $\Mi$, and our treatment of the
renormalization-group running down to the electroweak scale. 
Then in Section~\ref{sec:section_ExpConst} we discuss how proton decay, $\mu \to e$ flavor-violating observables  
and the electron EDM arise in these models, and review the available experimental information.
In Section~\ref{sec:section_Analysis} we introduce the specific no-scale models we study, and analyze their predictions for
these observables. We then present our conclusions in Section~\ref{sec:section_Conclusions}.\\

\section{Model Framework \label{sec:sectionModelFramework}}
\subsection{Embedding the MSSM in SU(5) \label{subsec:MSSMEmbedding}}

In the minimal supersymmetric SU(5) GUT model, the three generations of
matter superfields are embedded into
three pairs of $\overline{\bf 5}$ and ${\bf 10}$ representations.
There are also two chiral electroweak Higgs superfields
$H_u$ and $H_d$, whose vacuum expectation values (vevs) break the electroweak SU(2)$\times$U(1) gauge group down 
spontaneously to U(1)$_{\rm EM}$. They are embedded in ${\bf 5}$ and $\overline{\bf 5}$
representations, $H$ and $\overline{H}$, which also contain ${\bf 3}$ and $\overline{\bf 3}$ 
colored Higgs superfields $H_C$ and $\overline{H}_C$, respectively. 
The SU(5) GUT gauge group is broken spontaneously down to the Standard Model (SM) gauge group
by the vev of a ${\bf 24}$ chiral superfield, $\Sigma \equiv \sqrt{2} \, \Sigma^A \, T^A$, 
where $T^A$ ($A=1, \dots, 24$) are the generators of SU(5) with ${\rm Tr}(T^A T^B) =
\delta_{AB}/2$. 
The vev of the adjoint is given by $\langle \Sigma \rangle =V\cdot {\rm diag}(2,2,2,-3,-3)$, 
with $V = 4 \mu_\Sigma/\lambda^\prime$.
We follow the notation of \cite{Ellis:2010jb, Ellis:2016qra,Ellis:2016tjc,Ellis:2019fwf} for the SU(5) superpotential
parameters:
\begin{align}
 W_5 &=  \mu_\Sigma {\rm Tr}\Sigma^2 + \frac{1}{6} \lambda^\prime {\rm
 Tr} \Sigma^3 + \mu_H \overline{H} H + \lambda \overline{H} \Sigma H
\nonumber \\
&+ \left(h_{\bf 10}\right)_{ij} 
 {\bf 10}_i {\bf 10}_j H +
 \left(h_{\overline{\bf 5}}\right)_{ij} {\bf 10}_i \overline{\bf 5}_{j}
 \overline{H} ~,
\label{W5}
\end{align}
where we have suppressed all SU(5) indices.

Once SU(5) is broken, the GUT gauge bosons acquire masses $M_X = 5 g_5 V$,
where $g_5$ is the SU(5) gauge coupling. Doublet-triplet separation within the $H$ and $\overline{H}$ representations
can be achieved by a fine-tuning condition: $\mu_H -3\lambda V \ll V$, in which case the color-triplet Higgs
states have masses $M_{H_C} = 5\lambda V$.
We note also that the masses of the
color and weak adjoint components of $\Sigma$ are equal to $M_\Sigma =
5\lambda^\prime V/2$, while the singlet component of $\Sigma$ acquires a
mass $M_{\Sigma_{24}} = \lambda^\prime V/2$. 

Our notation for the Yukawa couplings of MSSM fields is specified by the following low-energy superpotential:
\bea
\label{wMSSM}
W_Y&=&h_E^{ij}\epsilon_{\alpha \beta} H_d^{\alpha} L_i^{\beta} E_j^c  
+ h_{D}^{ij} \epsilon_{\alpha \beta} H_d^{\alpha} Q_i^{\beta} D_j^c  
- h_{U}^{ij} \epsilon_{\alpha \beta} H_u^{\alpha} Q_i^{\beta} U_j^c \, .
\eea
Note that we use a ``Left-Right" (LR) notation for Yukawa couplings, which means that the first index of the Yukawa couplings corresponds to the SU(2) doublets, and the second index to the SU(2) singlets.

In order to match the GUT theory (\ref{W5}) to the MSSM (\ref{wMSSM}), 
in particular for the proton decay operators we discuss below,
we decompose the second row of the SU(5) superpotential (\ref{W5}) into MSSM component fields, yielding the Yukawa couplings of the MSSM fields in terms of the SU(5) field
couplings, as follows: 
\bea
\sqrt{2}~ \mathbf{10}_i
(h_{\overline{\bf 5}})_{ij}
\bar{\mathbf{5}}_j H&=&-E^c_i (h_{\overline{\bf 5}})_{ij} L_j \overline{H} -Q_i (h_{\overline{\bf 5}})_{ij} L_j \overline{H^C}-U^c_i (h_{\overline {\bf{5}}})_{ij} D^c_j \overline{H^C}
-Q_i (h_{\overline{\bf 5}})_{ij} D^c_j \overline{H} \, , \nonumber\\
\frac{1}{4}~\mathbf{10}_i (h_{\bf 10})_{ij}\mathbf{10}_j H &=& Q_i (h_{\bf 10})_{ij} U^c_j H + \frac{1}{2}Q_i (h_{\bf 10})_{ij} Q_j H^C - U^c_i (h_{10})_{ij}  E^c_j H^C \, ,
\label{eq:sup_rel_PD}
\eea
where the superscripts $C$ on Higgs multiplets indicate 
their color triplet components.

We recall that the embedding of the MSSM fields into the SU(5) model is ambiguous, and various Ans{\"a}tze are possible. 
In particular, the following SU(5) Yukawa couplings were chosen in \cite{Ellis:2019fwf}~\footnote{Throughout this work,  
$\hat h$ denotes a diagonalized Yukawa matrix.}
\bea
\label{eq:h10h5or}
(h_{10})_{ij}=\ \hat h_{10 i} \delta_{ij} \ \mathrm{e}^{i \phi_i},\quad
(h_{\bar 5})_ {ij}=\left(\VG^{*} \hat h_{{\bar 5} } V_R^T\right)_{ij} \, ,
\eea
where $\VG$ is the CKM matrix at the GUT scale.
Transforming the fields $E_i^c \rightarrow (\VG \ E^c)_i $ and $U_i^c \rightarrow  {\rm{e}}^{-\phi_i} U_i^c$,
we choose the embedding 
\bea
\label{eq:orembedding}
10_i=\left\{Q_i, \mathrm{e}^{-i\phi_i} U^c_i, (\VG E^c)_i \right\}, \quad 
\bar 5_i=\left\{D^c_i, L_i  \right\}\, ,
\eea
where the phase factors $\phi_i$ satisfy the condition 
\begin{equation}
    \sum_{i=1}^3 \phi_i=0 ~,
\end{equation}
so that only two of them are independent \footnote{Note that these phases contribute only to the running of the off-diagonal elements of the soft mass terms, which are very small, and we neglect this effect here.}. 

It is well known that the masses of the leptons and down-type quarks of the first two generations
are not consistent with unification at the GUT scale,~\footnote{The differences could be
accommodated by postulating dimension-5 terms in the SU(5) superpotential~\cite{Ellis:1979fg}.} whereas
those of the third generation are  in reasonable agreement with Yukawa unification. 
We determine the SU(5) Yukawa couplings by using the following matching conditions for the 
MSSM couplings after renormalization group (RG) running them from the electroweak scale up to the GUT scale:
\bea
h_{10,i}&=&\frac{1}{4} h_{U,i}(M_{\mathrm{GUT}}),\quad i=1,2,3, \nonumber\\
 h_{\overline{5},(i,j)} &=&\sqrt{2}  h_{D (i,j)}, \quad i,j =1,2,3 \quad \text{(except~for} \; (i,j) = (3,3)), \nonumber\\
 h_{\overline{5},{(3,3)}}&=&\frac{1}{\sqrt{2}}\left[ h_{D_{(3,3)}}(M_{\mathrm{GUT}}) + h_{E_{(3,3)}}(M_{\mathrm{GUT}})    \right].
\label{eq:h10h5A2}
\eea
Thus, the Yukawa couplings of the charge-2/3 quarks are matched directly to the GUT-scale couplings 
of the {\bf 10} representations, up to a
numerical factor, as are those of the first two generations of quarks in the $\mathbf{\bar{5}}$
representations.~\footnote{ This choice is conservative, in the sense that it leads to a longer 
proton lifetime than if $m_\mu$ and $m_e$ were used instead of $m_s$ and $m_d$ for matching the Yukawa couplings
of the first two generations of $\mathbf{\bar{5}}$ fermions. See Section~4.8 of~\cite{Ellis:2019fwf} for
a more detailed discussion.}  Recalling that the third-generation Yukawa couplings for $b$ and $\tau$ are similar, we match an average of these Yukawa couplings to that of the third generation of $\mathbf{\bar{5}}$ fermions.

Using as input the values for the Yukawa couplings at the EW scale discussed further below, 
we use Eq.~(\ref{eq:h10h5A2}) to determine the SU(5) Yukawa couplings at the GUT scale, which we then run up to $\Mi$.
Note that we also run the Yukawa couplings of the
first two generations of charged leptons up to the GUT scale. These are not used as a basis for further running to $\Mi$, 
but are subsequently run back down to the EW scale.

There are ambiguities in the description of flavor mixing in the
supersymmetric GUT model. Various options were considered in~\cite{Ellis:2016qra},
including the contrasting cases $V_R= \mathbf{1}$ and $V_R=V_{\rm GCKM}$.
If we choose $V_R= \mathbf{1}$ in (\ref{eq:h10h5or}),  we obtain from Eq.~(\ref{eq:h10h5A2}) 
and the embedding (\ref{eq:orembedding}) the following
relations between the MSSM couplings and the diagonal GUT-scale couplings (after running down from $\Mi$ to $\mgut$):
\bea
h_U &=& 4 \hat h_{10} \qquad {\rm except} \quad (i,j) = (3,3) \, , \nonumber \\
h_D &=&\VG^{*} \hat h_{\bar 5}/\sqrt{2} \qquad {\rm except} \quad (i,j) = (3,3) \, ,
\label{eq:hMSSMembA}
\eea
Because of the lack of Yukawa coupling unification, we do not relate $h_E$ and ${h_D}_{(3,3)}$ to $h_{\bar 5}$ at the GUT scale. We also do not relate  ${h_U}_{(3,3)}$ to $h_{10}$
at the GUT scale, in order to converge more efficiently to the observed top quark mass. 
For $h_E$, ${h_D}_{(3,3)}$ and ${h_U}_{(3,3)}$, the previous values at $\mgut$ are used for running back down to the EW scale, as will become clear when we discuss the RGE boundary conditions below. 

This is one of three choices for the treatment of flavor
that we consider in this paper: 

$\bullet$ We call choice {\bf A} the embedding (\ref{eq:orembedding}) combined with
$V_R = {\bf 1}$ in (\ref{eq:h10h5or}). This is the Ansatz
 A2 
 considered in \cite{Ellis:2016qra}.\\

We consider also the embedding
(after shifting only $U_i^c \rightarrow  {\rm{e}}^{-\phi_i} U_i^c$),
\bea
\label{eq:orembedding2}
10_i=\left\{Q_i, \mathrm{e}^{-i\phi_i} U^c_i, E^c_i \right\}, \quad 
{\bar 5}_i=\left\{D^c_i, L_i  \right\}.
\eea
Choosing again $V_R= {\bf 1}$, we obtain once again Eq.~(\ref{eq:hMSSMembA}) for matching when running down from $\mgut$ to the EW scale.

$\bullet$ We call this choice of embedding {\bf B}, noting that it is equivalent to Ansatz A3 of  \cite{Ellis:2016qra}.~\footnote{If we take $V_R=V_{GCKM}$
with this embedding,  we obtain Ansatz A4 of \cite{Ellis:2016qra}.
This choice turns out to be problematic for the observables we discuss below,
and is not considered further here.} \\

At this point A and B are identical. 
There would be no difference if we had Yukawa unification, since $h_E$ in case ({\bf B}) would be $h_E=h_D^T=h_5^T/\sqrt{2}$ as opposed to $h_E = \hat h_{\bar{5}}/\sqrt{2}$, i.e., equal to the diagonal SU(5) coupling as in case {\bf A}. However, since we do not match $h_E$ from the 5-plet, we can only``mimic" this condition at the EW scale and, as we see below, the boundary conditions for {\bf A} and {\bf B} differ at the EW scale. 
We emphasize that in the case of perfect unification the choices {\bf A} and {\bf B} would make identical predictions for all observables. {\bf A} and {\bf B} would not be distinct cases but rather different ways of formulating the same model for specifying the lepton sector in terms of the  {\bf 5}-plet of SU(5) and possibly additional operators. The motivation to consider cases {\bf A} and {\bf B} here is to explore the sensitivity to the precise way the couplings in the charged-lepton sector alter flavor observables.\\

$\bullet$ We also compare our results for $\TPDK$ with these flavor choices to models that ignore the flavor structure by limiting the RG running to diagonal matrix elements. We label this choice {\bf NF}.\\

The Yukawa couplings of the MSSM fields entering the dimension-six operators mediating proton decay can be defined from the Yukawa couplings of the SU(5) theory, \eq{eq:sup_rel_PD},  as follows:
\bea
h^{U^c_k E^c_l} = (4 \hat h_{10})_{kk} \  (V_{\rm{10}})_{kl}
\, , \quad \quad \quad &\quad &  
h^{U^c_k D^c_l}= e^{-i \phi_k} (\V)^*_{ks} \left( \frac{(\hat{h}_5)_{ss}}{\sqrt{2}} \right) (V_R)^T_{sl} \, , \nonumber \\
h^{Q_k L_l} =     (\V)^*_{ks} \left(\frac{(\hat{h}_5)_{ss}}{\sqrt{2}}\right)  (V_R)^T_{sl} \, , && \frac{1}{2}h^{Q_k Q_l} = e^{i \phi_k} (2\hat{h}_{10})_k \delta_{kl} \, , 
\label{eq:Yuk5_initialcond}
\eea
where $(V_{\rm{10}})_{kl}$=$\VG$ for {\bf A} and $\mathbf{1}$ for {\bf B}, while  
$V_R={\bf 1}$ for both of the choices {\bf A} and {\bf B}.

\subsection{Soft Supersymmetry Breaking \label{sec:susyX}}

We write the soft supersymmetry-breaking terms in the Lagrangian in the
SU(5) GUT symmetry limit as 
\bea 
\mathcal{L}_{\rm{soft}}
&=&
 -(m^2_{\bar{5}})_{ij} \overline{\bf 5}^*_i\overline{\bf 5}_j   - (m^2_{10})_{ij}  {\bf 10}^\dagger_i {\bf 10}_j - m_{H}^2 \left| H \right|^2  - m_{\bar{H}}^2 \left| \overline{H} \right|^2 -
\frac{1}{2} M_5 \ \hat\lambda^A \hat \lambda^A \nonumber\\
 &- & \left[ m^2_{\Sigma} \ \mathrm{Tr}\left( \Sigma^\dagger\Sigma \right) +    b_\Sigma \ \Tr {\Sigma}^2 + \frac{1}{6} a' \ \Tr {\Sigma}^3 + b_H \ \overline{H} H 
 + a \ \overline{H} {\Sigma} H \right] \nonumber \\
 &-& \left[ \ a_{10}\   {\bf 10} \ {\bf 10} \ H
 +  \ a_{\bar 5} \ {\bf 10} \ \overline{\bf 5} \ \overline{H} + \mathrm{h.c.} \right] \, ,
 \label{eq:lagr}
\eea
where the $\hat \lambda^A$ are the SU(5) gaugino fields.
For convenience, we make no distinction in notation between chiral superfields and their scalar components.

In super-GUT models \cite{super-GUT,emo,Ellis:2010jb,
Ellis:2016tjc,Ellis:2017djk,Ellis:2019fwf}, the soft supersymmetry-breaking mass parameters are taken to be universal at some input scale, $\Mi$, that is greater than the GUT scale, $M_{\rm GUT}$. 
The RG running of the couplings and masses then takes place in two stages. 
We run the 2-loop MSSM beta functions for Yukawa couplings,
trilinear terms, soft masses-squared, $m_{H_d}$,  $m_{H_u}$, $B$, and $\mu$ between the electroweak scale, $M_{\rm EW}$, and $\Mg$, including three generations of fermions and sfermions, the SU(3)$\times$SU(2)$\times$U(1) gauge bosons and gauginos, and the SU(2)-doublet Higgs bosons and Higgsinos. Then, between  $\Mg$ and $\Mi$ the SU(5) GUT parameters are run also with three generations of fermions and sfermions, SU(5) gauge bosons and gauginos, Higgses and Higgsinos. For the sake of clarity we now specify all the boundary conditions we impose at $\Mi$ and $\Mg$.

Our boundary conditions at $\Mi$ are derived from
no-scale supergravity \cite{no-scale1,no-scale2,LN}. 
We assume a K\"ahler potential of the form
\beq
K \; = \; - 3 \ln \left(T + {\bar T} - \frac{1}{3} \sum_i |\phi_i|^2\right) + \sum_a \frac{|\varphi_a|^2}{(T + {\bar T})^{n_a}} \, ,
\eeq
where $T$ is a volume modulus, the $\phi_i$ are untwisted matter fields and include the
SU(5) matter multiplets. The $\varphi_a$ are
twisted fields, which include $H$ and/or ${\bar H}$, and the $n_a$ are the modular weights of the twisted fields.  
We also allow for modular weights in the superpotential, writing
\beq\label{w_phi}
W = (T+c)^{\beta}W_2(\phi_i,\varphi_a) + (T+c)^{\alpha}W_3(\phi_i,
\varphi_a)  +  \mu_\Lambda \, ,
\eeq
where $c$ is an arbitrary constant, and $W_{2,3}$ denote bilinear and trilinear terms with modular weights $\beta, \alpha$
that are in general non-zero and can differ for each superpotential term.
When $\langle \phi,\varphi \rangle=0$,
the effective potential for $T$ is completely flat at the tree level, with an undetermined vev, and
the gravitino mass
\beq
m_{3/2} = \frac{\mu_\Lambda}{(T+\bar{T})^{3/2}} 
\label{gravmass}
\eeq
is undetermined, varying with the value of this volume modulus~\footnote{
The parameter $\mu_\Lambda$ does not play any other role in our construction, and its precise value is unimportant for our analysis.}.
We assume here that some Planck-scale dynamics fixes $T = \bar{T} = c$, and assume the representative
value $c = 1/2$ in the following.~\footnote{Our results are insensitive to this choice, as its only phenomenological impact is on the parameterization of the bilinear and trilinear soft supersymmetry-breaking parameters $A_F$ and $B_S$ in (\ref{eq:boundarycond_Mi}).} 
Finally, we assume a universal gauge kinetic function 
$f_{ab} = \delta_{ab}$, so that at $\Mi$ there is a universal gaugino mass, $m_{1/2}$.

We work with the no-scale framework introduced in \cite{Ellis:2016tjc}, 
where $m_0=0$, but allow for the possibility that the Higgs 5-plets are twisted, 
in which case either one or both of their soft masses may be non-zero. It was shown in~\cite{Ellis:2017djk}
that in models in which matter and both Higgs supermultiplets are untwisted, the minimal SU(5) super-GUT model considered here is unable to provide simultaneously a
dark matter relic density and Higgs mass in agreement with experimental values, and at the same time provide a sufficiently long proton lifetime. It was concluded in \cite{Ellis:2017djk} that either one or both of the Higgs multiplets must be twisted. 
The bilinear and trilinear soft supersymmetry-breaking terms $b_\Sigma, b_H, a^\prime, a, a_{10}, a_5$ may also be non-zero. 
Each gets a contribution from the modular weight in Eq. (\ref{w_phi}) and an additional contribution
that depends on the specific superpotential term and whether the 5-plets are twisted or not.  Our boundary conditions at $\Mi$ are therefore:
\bea
\label{eq:boundarycond_Mi}
&& M_5=m_{1/2} \, ,\nonumber\\
&&(m^2_{\mathbf{10}})_{ij}=(m^2_{\overline{\mathbf{5}}})_{ij} =  m_\Sigma^2=0 \, , \nonumber\\
&&m_2^2 \equiv m_H^2 
=p\  m_{3/2}^2,\quad  m_1^2 \equiv m_{\bar{H}}^2= q\ m_{3/2}^2 \, , \nonumber\\
&&  (A_{F})_{ij}=(r_F-\alpha_F)m_{3/2}\delta_{ij}  \quad
{\small{(F=\mathbf{10}, \overline{\mathbf{5}}})} \, ,
 \nonumber\\
 &&  A_{F}=(r_F-\alpha_F)m_{3/2}  \quad
{\small{(F= \lambda, \lambda^\prime})} \, ,
 \nonumber\\
&&B_S= (p_S-\beta_S) m_{3/2} \quad (S=H,\Sigma) \, , \nonumber\\
&& (a_{\mathbf{10}})_{ij}=(A_{\mathbf{10}})_{ii} (h_{\mathbf{10}})_{ij} \, ,\nonumber\\
&& (a_{\overline{\mathbf{5}}})_{ij}=(A_{\overline{\mathbf{5}}})_{ii} (h_{\overline{\mathbf{5}}})_{ij} \, .
\eea
The parameters $p,q = (0,1)$ depend whether $(H,{\bar H})$ is untwisted (0) or twisted (1). 
The parameters $r_F = p, q, p+q, 0$, for $F=\mathbf{10}, \overline{\mathbf{5}}, \lambda, \lambda^\prime$, 
and $p_S = p+q, 0$ for $S=H,\Sigma$.
The different modular weights, $\alpha$, $\beta$, chosen for the different models are specified in 
 Section \ref{Models}. We take all the $n_a = 0$.
Other quantities run up to $\Mi$, such as the SU(5) Yukawa couplings, are not reset at $\Mi$.

\subsection{Renormalization-Group Running of Parameters \label{sec:parruns}}

Having specified the theoretical boundary conditions at $\Mi$, we now discuss the renormaliz- ation-group (RG) 
running of the model parameters. This involves matching parameters at 
$\Mg$, since the fundamental degrees of freedom and hence the RG equations
differ above and below this scale, and the phenomenological inputs for the gauge and Yukawa couplings
are measured at the electroweak scale.
The RG equations are run up and down between the electroweak scale and $\Mi$ iteratively until
a convergent solution is found.
We use the following matching and boundary conditions.
 
\paragraph{Matching boundary conditions at $\mathbf{\Mg}$:}
There are two sets of boundary conditions at $\mgut$, one corresponding to RG running from the EW scale to $\Mi$, and the other when running back down.

We first specify the matching conditions for the gauge couplings when running up from the EW scale.
At one-loop level in the $\overline{\rm DR}$ renormalization
scheme \cite{Siegel:1979wq}, we have 
\begin{align}
 \frac{1}{g_1^2(Q)}&=\frac{1}{g_5^2(Q)}+\frac{1}{8\pi^2}\biggl[
\frac{2}{5}
\ln \frac{Q}{M_{H_C}}-10\ln\frac{Q}{M_X}
\biggr] - \frac{8c_5 V}{M_P} 
~, \label{eq:matchg1} \\
 \frac{1}{g_2^2(Q)}&=\frac{1}{g_5^2(Q)}+\frac{1}{8\pi^2}\biggl[
2\ln \frac{Q}{M_\Sigma}-6\ln\frac{Q}{M_X}
\biggr] - \frac{24 c_5 V}{M_P} 
~, \label{eq:matchg2} \\
 \frac{1}{g_3^2(Q)}&=\frac{1}{g_5^2(Q)}+\frac{1}{8\pi^2}\biggl[
\ln \frac{Q}{M_{H_C}}+3\ln \frac{Q}{M_\Sigma}-4\ln\frac{Q}{M_X}
\biggr]+\frac{16 c_5 V}{M_P} \label{eq:matchg3} ~,
\end{align}
where $g_1$, $g_2$, and $g_3$, are the U(1), SU(2), and SU(3) gauge
couplings, respectively, and $Q$ is a renormalization scale taken in our analysis to be
the unification scale: $Q = M_{\rm GUT}$. 

The last terms in
Eqs. (\ref{eq:matchg1}) - (\ref{eq:matchg3}) represent a possible contribution from the dimension-five
operator 
\begin{equation}
 W_{\rm eff}^{\Delta g} = \frac{c_5}{M_P} {\rm Tr}\left[
\Sigma {\cal W} {\cal W}
\right] ~,
\label{eq:SigmaWW}
\end{equation}
where ${\cal W}\equiv  {\cal W}^A T^A$ denotes the
superfields corresponding to the field strengths of the SU(5) gauge vector bosons
${\cal V} \equiv {\cal V}^A T^A$. Since $V/M_P \simeq 10^{-2}$, these terms
can be comparable to the one-loop threshold corrections, and their 
possible presence should
be taken into account when discussing gauge-coupling unification~\cite{Tobe:2003yj}. Including the $c_5$
coupling is essential for our purposes, as it allows us to choose 
independently the 
Higgs couplings $\lambda$ and $\lambda^\prime$, which we specify at the GUT scale. 

Eqs.~(\ref{eq:matchg1} - \ref{eq:matchg3}) can be combined to give
\beq
\frac{1}{g_5^2} = -\frac{1}{g_1^2} +\frac{1}{g_2^2}+\frac{1}{g_3^2}- \frac{1}{8\pi^2} \left( \frac35 \ln \frac{Q}{M_{H_C}} + 5 \ln \frac{Q}{M_\Sigma} \right) \, .
\eeq
The masses, $M_{H_C}$ and $M_\Sigma$, have implicit dependences on the gauge couplings, including $g_5$,
making it impossible to write an analytic expression for the matching of the three low-energy gauge couplings, $g_i$, to $g_5$. Nevertheless, we can solve
for $g_5$ iteratively.

The matching conditions for the Yukawa couplings were given in \eq{eq:h10h5A2}.
 As noted there, we take the average of ${h_E}_ {3,3}$ and ${h_D}_{3,3}$ for the third-generation charged-lepton and charge-1/3 quark Yukawa couplings, which are close to the unification expected in SU(5). We adopt a similar approach for the trilinear terms and the soft squared masses. For the embedding {\bf{A}}, when matching from $\Mg$ to $\Mi$ we take for the trilinear couplings
\bea
\label{eq:av5A}
a_{\bar{\mathbf{5}}}= \left(a_D + \VG^* a^{\rm{T}}_E\right)/\sqrt{2} \, ,
\eea
\bea
\label{eq:av10A}
a_{\mathbf{10}}= a_U /4 \, ,
\eea
and for the soft squared masses
\bea
\label{eq:m25A}
m^2_{\bar{\mathbf{5}}}= \left(m^2_L +m^2_D \right)/2 \, ,
\eea
\bea
m^2_{\mathbf{10}}= \left(m^2_Q +m^2_U + \VG m^2_E \VG^\dagger \right)/3 \, .
\label{eq:m10}
\eea
For the embedding {\bf{B}}, when matching from $\Mg$ to $\Mi$ we take the same matching conditions for $a_{\mathbf{10}}$ and  $m^2_{\overline{\mathbf{5}}}$ as for the embedding {\bf{A}}, see Eqs.~(\ref{eq:av10A},\ref{eq:m25A}), respectively, with
\bea
a_{\overline{\mathbf{5}}}= \left(a_D + a^{\rm{T}}_E\right)/\sqrt{2}
\label{eq:a5B}
\eea
and
\bea
\label{eq:avm210B}
m^2_{\mathbf{10}}= \left(m^2_Q +m^2_U + m^2_E  \right)/3 \, .
\eea
We note that by taking these averages we are effectively generating two inequivalent models at the GUT scale, which in turn produce different values for observable quantities. If one was not required to use the averages in  \eq{eq:h10h5A2} and Eqs.~ (\ref{eq:av5A}, \ref{eq:m10}, \ref{eq:a5B}, \ref{eq:avm210B}), perfect unification would allow us simply to formulate the SU(5) theory with the quark-sector couplings, and all quark-sector differences between the two models would vanish. 

The remaining matching conditions for masses at $\Mg$ when running up from the EW scale are:
\begin{equation}
\label{eq:bcatMGUT1}
M_5 = g_5^2 \biggl(-\frac{M_1}{g_1^2} + \frac{M_2}{g_2^2} + \frac{M_3}{g_3^2}\biggr) \, , \quad
m^2_{\bar{H}} = m^2_{H_d} \, , \quad m^2_{H} = m^2_{H_u} \, , 
\end{equation}
The matching of the gaugino masses to $M_5$ 
when running up to 
$\mgut$ is chosen to be consistent with
the matching of the gaugino masses to $M_5$ when running down from $\mgut$ to the EW scale as discussed below. Finally, when running from $\Mg$ to $\Mi$, $m^2_\Sigma$ is set equal to its value from the previous iterative run down from $\Mi$ where it was initially set to 0 as in Eq.~(\ref{eq:boundarycond_Mi}).

At $\Mi$, the soft mass terms are reset according to \eq{eq:boundarycond_Mi} and the theory is run down to $\Mg$, where the matching conditions for the soft squared-mass terms and Yukawa couplings are
 \begin{align}
\label{eq:bcatMGUT}
\begin{array}{ll}
m^2_{D}=m^2_{L}=m^2_{\bar{5}} \, , & \\
m^2_{Q}=m^2_{U}=m^2_{10} \, , &  \\
m^2_{E}=\VG^\dagger m^2_{10} \VG \quad (\rm choice \; \bf{A}) \, ,  
   &  m^2_{E}= m^2_{10} \quad (\rm choice \; \bf{B}) \, , \\
m^2_{H_d}=m^2_{\bar{H}}, \quad m^2_{H_u}=m^2_{H} \, . &
\end{array}
\end{align}
For the trilinear terms, we use
\begin{align}
 & a_U= {4} a_{10}\, , \nonumber \\
& a_D= a_5 \sqrt{2}\, ,  \nonumber \\
& a_E= a_5^{\mathrm{T}} \VG /\sqrt{2}\quad (\text{choice  {\bf A}})\, ,  \qquad a_E= a_5^{\mathrm{T}} /\sqrt{2}\quad (\text{choice  {\bf B}})  \label{eq:trilinearpropaE} \, ,
\end{align}
where $a_U$, $a_D$ and $a_E$ correspond to the MSSM up-type quarks, down-type quarks and lepton trilinear couplings, respectively, and we recall that we assume $V_R=\mathbf{1}$ for both the choices {\bf A}  and {\bf B}, with the embeddings given in \eq{eq:orembedding} and \eq{eq:orembedding2}, respectively. 
The Yukawa matching conditions were given in (\ref{eq:hMSSMembA}),
and the soft terms in \eq{eq:lagr} must be embedded in the same way,
once the MSSM is embedded in SU(5). Hence the trilinear couplings in Eq.~(\ref{eq:trilinearpropaE}) are rotated in the same ways as the Yukawa couplings in \eq{eq:h10h5A2}, while all the soft squared-mass terms remain invariant with the exception of $m_E^2$ in choice {\bf A} as seen in Eq.~\eqref{eq:bcatMGUT}.

From linear combinations of the matching conditions for the gauge couplings in Eqs.~(\ref{eq:matchg1} - \ref{eq:matchg3}) we obtain~\cite{Ellis:2016tjc, Hisano:1992jj, Hisano:1992mh, Hisano:2013cqa}: 
\begin{align}
 \frac{3}{g_2^2(Q)} - \frac{2}{g_3^2(Q)} -\frac{1}{g_1^2(Q)}
&=-\frac{3}{10\pi^2} \ln \left(\frac{Q}{M_{H_C}}\right)
-\frac{96c_5 V}{M_P}
~,\label{eq:matchmhc} \\[3pt]
 \frac{5}{g_1^2(Q)} -\frac{3}{g_2^2(Q)} -\frac{2}{g_3^2(Q)}
&= -\frac{3}{2\pi^2}\ln\left(\frac{Q^3}{M_X^2 M_\Sigma}\right) ~,
\label{eq:matchmgut}
\\[3pt]
 \frac{5}{g_1^2(Q)} +\frac{3}{g_2^2(Q)} -\frac{2}{g_3^2(Q)}&= -\frac{15}{2\pi^2} \ln\left(\frac{Q}{M_X}\right) + \frac{6}{g_5^2(Q)} -\frac{144c_5 V}{M_P} ~.\label{eq:matchg5}
\end{align}
Eqs.~(\ref{eq:matchmhc}--\ref{eq:matchg5}) provide three conditions on the masses $M_{H_C}$, $M_\Sigma$ and $M_X$, which
can related to the GUT Higgs vev $V$ through the couplings $\lambda$, $\lambda^\prime$, and $g_5$ respectively. 
As a result, if $c_5 =0$ only one of the two GUT couplings $\lambda$ or $\lambda^\prime$ 
can be chosen as a free parameter.
If, however, $c_5 \ne 0$, $\lambda$ and $\lambda^\prime$ can be chosen independently with the following condition on the 
dimension-five coupling:
\begin{equation}
c_5 = \frac{M_P}{8 V} \left[ \frac{1}{6 g_3^2 (M_{\rm GUT})}- \frac{1}{6 g_1^2 (M_{\rm GUT})}- \frac{1}{40\pi^2}\ln \biggl(\frac{M_{\rm GUT}}{M_{H_C}}\biggr) \right] ~,
\end{equation}
which can be obtained from Eq.~\eqref{eq:matchmhc} by setting $g_1 (M_{\rm GUT}) = g_2 (M_{\rm GUT})$. 
It is important to note that allowing $c_5 \ne 0$ enables us to increase the colored Higgs mass, thereby increasing the proton lifetime \cite{Ellis:2017djk,Ellis:2019fwf}.

The matching conditions for
the gaugino masses \cite{Tobe:2003yj, Hisano:1993zu,Evans:2019oyw,Ellis:2019fwf} are
\begin{align}
 M_1 &= \frac{g_1^2}{g_5^2} M_5
-\frac{g_1^2}{16\pi^2}\left[10 M_5 -10(A_{\lambda^\prime} -B_\Sigma)
 +\frac{2}{5}B_H\right]
-\frac{4c_5 g_1^2V(A_{\lambda^\prime} -B_\Sigma)}{M_P} ~,
\label{eq:m1match}
\\[3pt]
M_2 &= \frac{g_2^2}{g_5^2} M_5
-\frac{g_2^2}{16\pi^2}\left[6 M_5 -6A_{\lambda^\prime} +4B_\Sigma
 \right]
-\frac{12c_5 g_2^2V(A_{\lambda^\prime} -B_\Sigma)}{M_P} ~,
\label{eq:m2match}
\\[3pt]
M_3 &= \frac{g_3^2}{g_5^2} M_5
-\frac{g_3^2}{16\pi^2}\left[4 M_5 -4A_{\lambda^\prime} +B_\Sigma
- B_H \right]
+\frac{8c_5 g_3^2V(A_{\lambda^\prime} -B_\Sigma)}{M_P}
~.
\label{eq:m3match}
\end{align}
Finally, we must match the MSSM $\mu$ and $B$-terms
to their SU(5) counterparts \cite{Borzumati:2009hu}
\begin{align}
 \mu &= \mu_H - 3 \lambda V\left[
1+ \frac{A_{\lambda^\prime} -B_\Sigma}{2 \mu_\Sigma}
\right] ~,
\label{eq:matchingmu}
 \\[3pt]
 B &= B_H + \frac{3\lambda V \Delta}{\mu}
+ \frac{6 \lambda}{\lambda^\prime \mu} \left[
(A_{\lambda^\prime} -B_\Sigma) (2 B_\Sigma -A_{\lambda^\prime}
 +\Delta) -m_\Sigma^2
\right]~,
\label{eq:matchingb}
\end{align}
with
\begin{equation}
 \Delta \equiv A_{\lambda^\prime} - B_\Sigma - A_\lambda +B_H ~.
\label{eq:deltadef}
\end{equation}
As noted earlier, in the minimal SU(5) GUT model studied here we must tune $|\mu_H -3\lambda V|$
to be ${\cal O}(M_{\rm SUSY})$. The parameters
$\mu$ and $B$
can be determined at the electroweak scale
by the minimization of the Higgs potential as in the CMSSM. These are then run up to the scale where Eqs.~\eqref{eq:matchingmu}
and \eqref{eq:matchingb} are applied. 
However, the GUT $A$- and $B$-terms are specified at the
input scale by Eq.~(\ref{eq:boundarycond_Mi})
and, in general, the condition (\ref{eq:matchingb}) will not be satisfied.

This mismatch can be rectified by adding a Giudice-Masiero (GM) term to the K\"ahler potential \cite{GM}:
\beq\label{GM1}
\Delta K =  c_H (T+c)^{\gamma_H} H \bar{H} + c_\Sigma (T+c)^{\gamma_\Sigma} \Sigma^2 + {\rm h.c.} \, ,
\eeq
where we have allowed for the possibility of additional
modular weights, $\gamma_H$ and $\gamma_\Sigma$.
This term induces shifts in both the $\mu$-terms and $B$-terms \cite{egno4,Ellis:2017djk,Ellis:2019fwf}:
\begin{eqnarray}
\Delta \mu_{H} = c_H m_{3/2} \, , & \Delta \mu_{\Sigma} = c_\Sigma m_{3/2} \, , \\
\Delta B_H \mu_H =   (p + q -\gamma_H) c_H m_{3/2}^2 \, , &
\Delta B_\Sigma \mu_\Sigma = -\gamma_\Sigma c_\Sigma m_{3/2}^2 \, .
\end{eqnarray}
As a result, there is a shift in $\Delta$ given by 
\beq
\delta \Delta = \left(\gamma_\Sigma \frac{c_\Sigma}{\mu_\Sigma} +  (p + q -\gamma_H) \frac{c_H}{\mu_H}  \right) m_{3/2}^2 \, .
\eeq
Then any mismatch in (\ref{eq:matchingb}) can be corrected by 
\beq
\frac{3 \lambda V  \delta \Delta}{\mu}  = \left( (p + q -\gamma_H) c_H + \frac{12 \lambda}{\lambda^\prime} 
\gamma_\Sigma c_\Sigma \right)  \frac{m_{3/2}^2}{\mu}  \, ,
\label{fix}
\eeq
where we have used $\mu_\Sigma = \lambda^\prime V / 4$
and $\mu_H = 3 \lambda V$. If $\lambda \gg \lambda^\prime$, we can ignore, $c_H$, and use (\ref{fix}) to determine $c_\Sigma$ (for a given value of $\gamma_\Sigma$).

\paragraph{Boundary conditions at ${\mathbf \Mw}$:}

Although the soft supersymmetry-breaking parameters are input at the high scale, $\Mi$, some of the
phenomenological inputs are set by
boundary conditions at the electroweak scale, $\Mw$, namely the ratio of electroweak Higgs vevs, $\tan\beta, m_{f}$ and $V_{\rm CKM}$.
The Higgs vevs are in principle determined by the minimization of the Higgs potential at the weak scale. However, it is common in constrained models to fix these by using the experimental value of $M_Z$ and $\tan \beta$, 
and solve for $\mu$ and the pseudoscalar Higgs mass, or equivalently the MSSM $B$-term. 
In very constrained models such as 
the no-scale models considered here, $B$ is fixed by the high-scale boundary conditions and as a consequence, either
$\tan \beta$ is an output rather than an input \cite{vcmssm}, or a GM term is
used to fix the matching conditions for the $B$-terms. 
We adopt the latter approach here, and treat $\tan \beta$
as a weak-scale input. 

We also use the experimental values of the masses of the six quarks and the three charged leptons, $m_f$. The matching of Yukawa couplings is done in terms of the CKM matrix elements, using experimental input for the CKM matrix at $\Mw$. In general $h_D$ and $h_E$ can be written as follows
\bea
\label{eq:YukDiagMEW}
h_D=\V^* \hat h_D(\Mw) U^{T D}_R \, ,\quad
h_E=U^{E*}_L \hat h_E(\Mw) U^{T E}_R \, ,
\eea
where $\V$ ($=U^D_L$)~\footnote{In the way we define the Yukawa couplings, these enter the SM interaction Lagrangian as $\mathcal{L}_D=\overline{Q}_L h_D^* D_R + \overline{D}_R h_D^T Q_L$.} is the CKM matrix at the EW scale,
$\hat h_D(\Mw) = diag(y_d,y_s,y_b)$, and $\hat h_E(\Mw) = diag(y_e,y_\mu,y_\tau)$ are the diagonalized mass matrices containing the mass eigenvalues for the $D$-type quarks and charged leptons, respectively. The $U$ matrices aid with the diagonalization of these matrices. When running up to the $\Mg$ scale they should match Eqs. (\ref{eq:hMSSMembA}) at $\Mg$ for the choices {\bf A} and {\bf B}, respectively. Hence, in both cases we start with $U^D_R=\mathbf{1}$ and $U^D_L=\V$, while $U^E_R=U^E_L=\mathbf{1}$ for {\bf A} and $U^E_R=\V^*$ and $U^E_L=\mathbf{1}$ for {\bf B}.

At $\Mg$ the RG evolution determines the evolution of $\V$ into $\VG$, while $U_R^D$, $U^E_L$ and $U^E_R$ are no longer diagonal. However, since we match the SU(5) fields to the MSSM fields at $\Mg$ with \eq{eq:h10h5A2}, once the RG
program has converged, $U^R_D$ is in practice equal to $\mathbf{1}$. We match the Yukawa couplings for the first two generations
of charged leptons at $\Mg$,
so that they converge rapidly to 
satisfy $U^E_R=U^E_L=\mathbf{1}$ for {\bf A} and $U^E_R=\V^*$ and $U^E_L=\mathbf{1}$ for {\bf B} at the EW scale. Any remaining non-diagonality can be absorbed into the embedding of the MSSM fields into SU(5), and does not alter the Yukawa couplings relevant for proton decay.
Finally, all the fermion masses are converted appropriately to the  ${\overline{\rm{DR}}}$ scheme and then matched to the supersymmetric theory at $M_Z$.

\section{Experimental Constraints \label{sec:section_ExpConst}}
\subsection{Proton Decay\label{sec:ProtonDecay}}

The most important constraint on the supersymmetric SU(5) GUT model from searches for proton decay comes from
the decay mode $p\rightarrow K^+ \bar{\nu}$, for which the current experimental limit  is  \cite{Miura:2016krn}
\bea
\tau\left(p\rightarrow K^+ \bar{\nu}\right) > 6.6 \times 10^{33} \quad {\rm{yrs}} \, .
\label{eq:lft_PDKnu}
\eea
In this paper we will refer to this limit as the {\emph{proton life-time limit}} if not otherwise specified.
In the future, the Hyper-Kamiokande (HK) experiment is expected to be sensitive to $\TPDK \sim 5 \times 10^{34}$~yrs~\cite{HK},
an improvement by almost an order of magnitude. Since generic amplitudes for dimension-5 proton decay are
inversely proportional to sparticle masses (see below), the HK reach for proton decay will provide sensitivity to
supersymmetric model parameters $\sim 3$ times larger than the current constraints from $\TPDK$.

\subsubsection*{Dimension-5 Proton Decay Operators}

In \cite{Goto:1998qg} a complete analysis of proton decay operators in supersymmetric SU(5) theories was given, including
in particular the explicit forms of the Wilson Coefficients (WCs) $C_{5L}$ and $C_{5R}$ entering into the 
dimension-five Lagrangian generated by integrating out the colored Higgs multiplets \cite{Sakai:1981pk}:
\beq
{\mathcal{L}}^{\rm{eff}}_5= C_{5L}^{ijkl} \ O^{5L}(Q_i, Q_j,Q_k, L_{l}) +  C_{5R}^{ijkl} \ O^{5R}(\bar{u}_{i}, \bar{e}_j, \bar{u}_{k} ,\bar{d}_{l} ) + {\rm h.c.} \, ,
\eeq
where
$i, j, k$  and $\ell$ are flavor indices, and
\begin{eqnarray}
O^{5L}(Q_i, Q_j,Q_k, L_{\ell}) & \equiv & \int d^2\theta \frac{1}{2} \epsilon_{abc} (Q^a_i\cdot Q^b_j) (Q^c_k \cdot L_{\ell}) \, , \nonumber \\ 
O^{5R}(\bar{u}_{i}, \bar{e}_j, \bar{u}_{k} ,\bar{d}_{l} ) & \equiv & \int d^2\theta \epsilon^{abc} (\bar{u}_{ia} \bar{e}_j \bar{u}_{kb} \bar{d}_{l c}) \, ,
\end{eqnarray}
where $a, b, c$ are color indices. Normalizing these operators at the GUT scale, $\Mg$, and matching the Yukawa matrices using Eq.  (\ref{eq:h10h5A2}), we find
\bea
C_{5L}^{ijkl}(\Mg) &=&
\frac{1}{M_{H_C}}h^{Q_i Q_j} h^{Q_k L_l} \, ,\nonumber \\ 
C_{5R}^{ijkl}(\Mg) &=&
\frac{1}{M_{H_C}}h^{U_i E_j} h^{U_k D_l} \, .
\label{eq:C5L_R_matching}
\eea
The Yukawa matrices appearing in \eq{eq:C5L_R_matching} are different for the different embeddings, as seen in Eq.~(\ref{eq:Yuk5_initialcond}). This is because each of the terms in the superpotential \eq{eq:sup_rel_PD} 
that are relevant for proton decay depend on $V_R$ and the choices of 
the $h_{10}$ and $h_5$ Yukawa matrices in Eqs.~(\ref{eq:h10h5or}).

The leading-order RG evolutions of the $C_{5L}^{ijkl}$ and $C_{5R}^{ijkl}$ between $\Mg$ and the supersymmetry breaking scale are given by~\cite{Goto:1998qg}
\bea
\overline{\beta} (C_{5L}^{ijkl})\equiv  \Dt C_{5L}^{ijkl} &=& 
  \left(
    -8g_3^2 -6g_2^2 -\frac{2}{5}g_1^2
  \right)C_{5L}^{ijkl}
  + C_{5L}^{mjkl}
     \left( h_D h_D^\dagger + h_U h_U^\dagger \right)^{i}_{~m}
\nonumber\\&&
 + C_{5L}^{imkl}
     \left( h_D h_D^\dagger + h_U h_U^\dagger \right)^{j}_{~m} ~
        + C_{5L}^{ijml}
     \left( h_D h_D^\dagger + h_U h_U^\dagger \right)^{k}_{~m}
\nonumber\\&&
     + C_{5L}^{ijkm}
     \left( h_E^\dagger h_E \right)_{m}^{~l} \, , \label{eq:UUdagger} \\
\overline{\beta} (C_{5R}^{ijkl}) \equiv    \Dt C_{5R}^{ijkl} &=& 
  \left(
    -8g_3^2 -\frac{12}{5}g_1^2
  \right)C_{5R}^{ijkl} + C_{5R}^{mjkl}
     \left( 2\, h_U^\dagger h_U \right)_{m}^{~i}
\nonumber\\&&
   +C_{5R}^{imkl}
     \left( 2\, h_E h_E^\dagger \right)^{j}_{~m}
     + C_{5R}^{ijml}
     \left( 2\, h_U^\dagger h_U \right)_{m}^{~k} ~
\nonumber\\&&
   + C_{5R}^{ijkm}
     \left( 2\, h_D^\dagger h_D \right)_{m}^{~l} \, ,
\label{eq:UdaggerU}
\eea 
where $\Lambda$ is the renormalization scale. 
Below the supersymmetry-breaking scale, we use the RGEs given in Ref.~\cite{Ellis:2015rya}.

We write the effective Lagrangian for $p\rightarrow K^+ \bar{\nu}_i$ decay in the following form:
\begin{align}
\label{eq:effLPDEW}
 {\cal L}(p\to K^+\bar{\nu}_i^{})
=&C_{RL}(usd\nu_i)\bigl[\epsilon_{abc}(u_R^as_R^b)(d_L^c\nu_i^{})\bigr]
+C_{RL}(uds\nu_i)\bigl[\epsilon_{abc}(u_R^ad_R^b)(s_L^c\nu_i^{})\bigr]
\nonumber \\
+&C_{LL}(usd\nu_i)\bigl[\epsilon_{abc}(u_L^as_L^b)(d_L^c\nu_i^{})\bigr]
+C_{LL}(uds\nu_i)\bigl[\epsilon_{abc}(u_L^ad_L^b)(s_L^c\nu_i^{})\bigr]
~.
\end{align}
The operators $C_{LL}\left(usd\nu_k   \right)$ and $C_{LL}\left(uds\nu_k   \right)$
are mediated by Wino exchange, and $C_{RL}(usd\nu_\tau)$ and $C_{RL}(uds\nu_\tau)$ 
are mediated by higgsino exchange (see Eqs.~(23) and (27) of \cite{Ellis:2015rya}). At the EW scale, the operators entering into the proton decay amplitudes are
$C_{5L}^{221i}$ and  $C_{5L}^{331i}$, $i=1,2,3$, which contribute to $C_{LL}\left(usd\nu_k   \right)$ and $C_{LL}\left(uds\nu_k   \right)$, and 
 $C_{5R}^{*3311}$ and $C_{5R}^{*3312}$, which contribute to $C_{RL}(usd\nu_\tau)$ and $C_{RL}(uds\nu_\tau)$.

However, due to the off-diagonal nature of the Yukawa matrices, the evolution from 
$\Mg$ down to $\Mw$ induces contributions from some other operators.
Consider as an example $C_{5L}^{3312}$, whose leading-order RG terms are
\bea
\overline{\beta} (C_{5L}^{3312}) &\simeq& C_{5L}^{3312}\left(    -8g_3^2 -6g_2^2 -\frac{2}{3}g_1^2 + 2 (h_Dh_D^\dagger)^3_3+ 2 y^2_t \right)  \nonumber \\
&+&  C_{5L}^{331m} (h^\dagger_E h_E)^2_m + C_{5L}^{33m2} (h_D h^\dagger_D)^1_m \, .
\label{eq:C5L3312}
\eea
The terms in \eq{eq:C5L3312} involving $h_D$ and $h_E$ are not diagonal, and generate contributions to the  $\beta$ functions of the operators mentioned above. In particular
\bea
C_{5L}^{33m2} (h_D h^\dagger_D)^1_m\simeq C_{5L}^{3332} (h_D h^\dagger_D)^1_3 = {\mathcal{O}}\left( C_{5L}^{3312} (h_Dh_D^\dagger)^3_3 \right).
\eea
We are therefore required to run $C_{5L}^{3332}$ between the weak and GUT scales, using the initial condition set by \eq{eq:Yuk5_initialcond}, even though the corresponding operator does not contribute directly to the effective Lagrangian (\ref{eq:effLPDEW}) defined at the EW scale. We note,
on the other hand, that the combinations $h_U^\dagger h_U$ and $h_U h_U^\dagger$ appearing in \eq{eq:UUdagger} for $C_{5L}$ and \eq{eq:UdaggerU} for $C_{5R}$, respectively,  remain diagonal as in the case considered in \cite{Ellis:2015rya} (see Eq.~(22) of that reference).~\footnote{We have omitted contributions that are proportional $y_u$ and $y_c$, given their smallness in comparison to $y_t$, and we have omitted terms proportional to $C_{5L}^{1312}$.}

The dimension-6 operator coefficients $C_{LL}\left(ud_pd_q\nu_k   \right)$  and  $C_{RL}(ud_pd_q\nu_\tau)$, $p,q=1,2$, are related to the dimension-five WCs 
$C_{5L}^{221i}$, $C_{5L}^{331i}$, 
 $C_{5R}^{*3311}$, and $C_{5R}^{*3312}$ (which were obtained by integrating out the colored Higgs multiplets in \eq{eq:C5L_R_matching}) via CKM mixing angle factors and
 loop integrals:
 \begin{align}
 C_{RL}(usd\nu_\tau)&=-V_{td}C^{\widetilde{H}}_{2}(m_Z)~, 
 & &C_{RL}(uds\nu_\tau)=-V_{ts}C^{\widetilde{H}}_{1}(m_Z)~,\nonumber \\
 C_{LL}(usd\nu_k)&=\sum_{j=2,3}V_{j1}V_{j2}
C^{\widetilde{W}}_{jk}(m_Z)~,
 & &C_{LL}(uds\nu_k)=\sum_{j=2,3}V_{j1}V_{j2}
C^{\widetilde{W}}_{jk}(m_Z)~.
\end{align}
where
\begin{align}
 C_i^{\widetilde{H}}
&=\frac{y_ty_\tau}{(4\pi)^2}
F(\mu, m_{\widetilde{t}_R}^2,m_{\tau_R}^2) \, C^{*331i}_{5R}~, \nonumber \\
 C^{\widetilde{W}}_{jk} &=
\frac{\alpha_2}{4\pi}\left[
F(M_2, m_{\widetilde{Q}_1}^2,
 m_{\widetilde{Q}_j}^2) +F(M_2, m_{\widetilde{Q}_j}^2,
 m_{\widetilde{L}_k}^2)\right]C^{jj1k}_{5L} ~. 
\label{Cijk}
\end{align}
Here $m_{\widetilde{t}_R}$, $m_{\widetilde{\tau}_R}$,
$m_{\widetilde{Q}_j}$, and $m_{\widetilde{L}_k}$
are the masses of the right-handed stop, the right-handed stau, left-handed
squarks, and left-handed sleptons, respectively, $\alpha_i \equiv
g_i^2/(4\pi)$, and
 \begin{align}
F(M, m_1^2, m_2^2) &\equiv
\frac{M}{m_1^2-m_2^2}
\biggl[
\frac{m_1^2}{m_1^2-M^2}\ln \biggl(\frac{m_1^2}{M^2}\biggr)
-\frac{m_2^2}{m_2^2-M^2}\ln \biggl(\frac{m_2^2}{M^2}\biggr)
\biggr]~.
\label{eq:funceq}
\end{align}
The loop integrals (\ref{Cijk}) yield the dimension-6 operator coefficients
at the supersymmetry-breaking scale, and they must then 
be run down to the EW scale. The corresponding RGEs
are given in \cite{Ellis:2015rya}, where many other details of the
calculation are provided.

\begin{figure}[!ht]
\centering
\includegraphics[height=6.5cm]{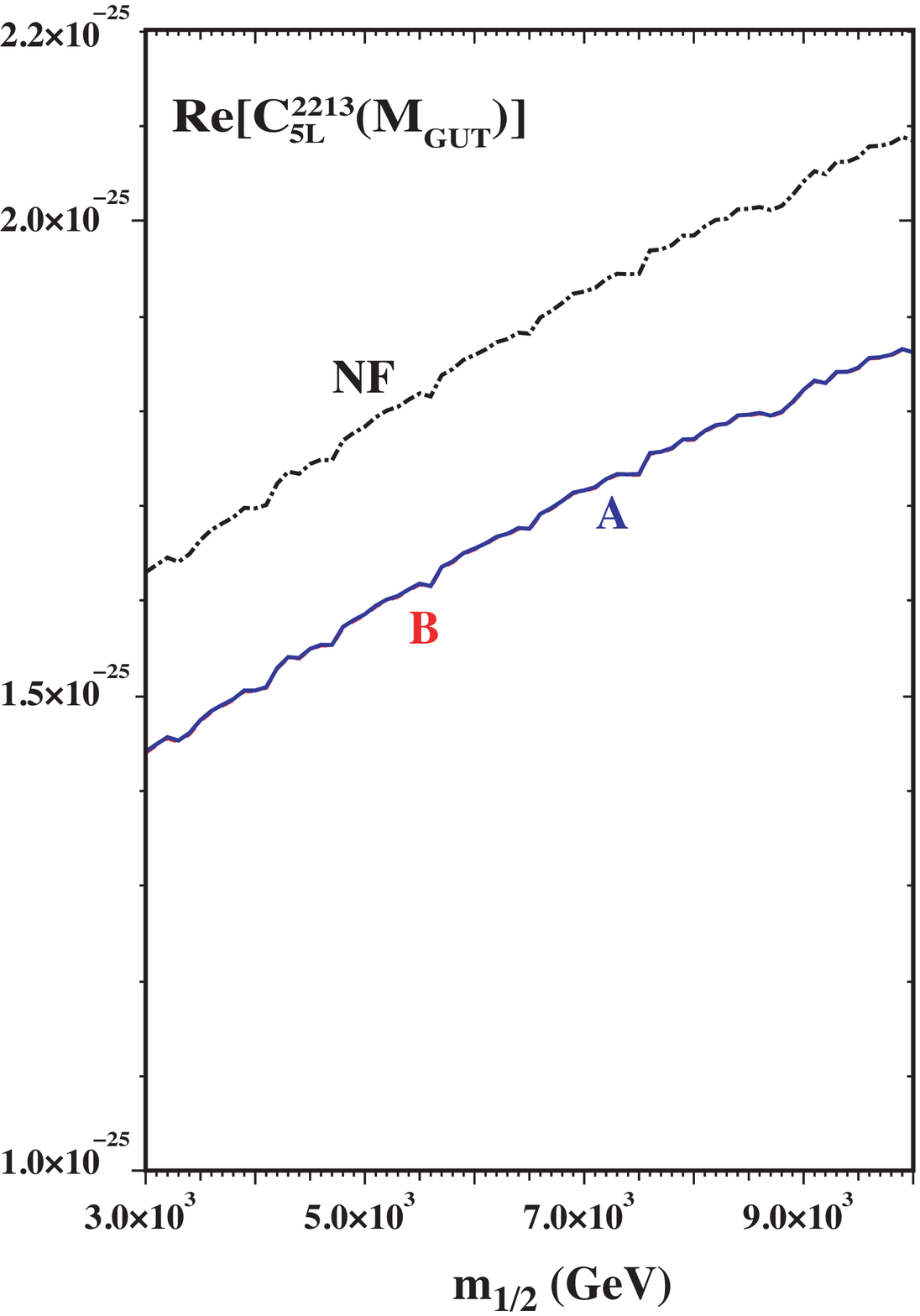}
\hskip 1in
\includegraphics[height=6.5cm]{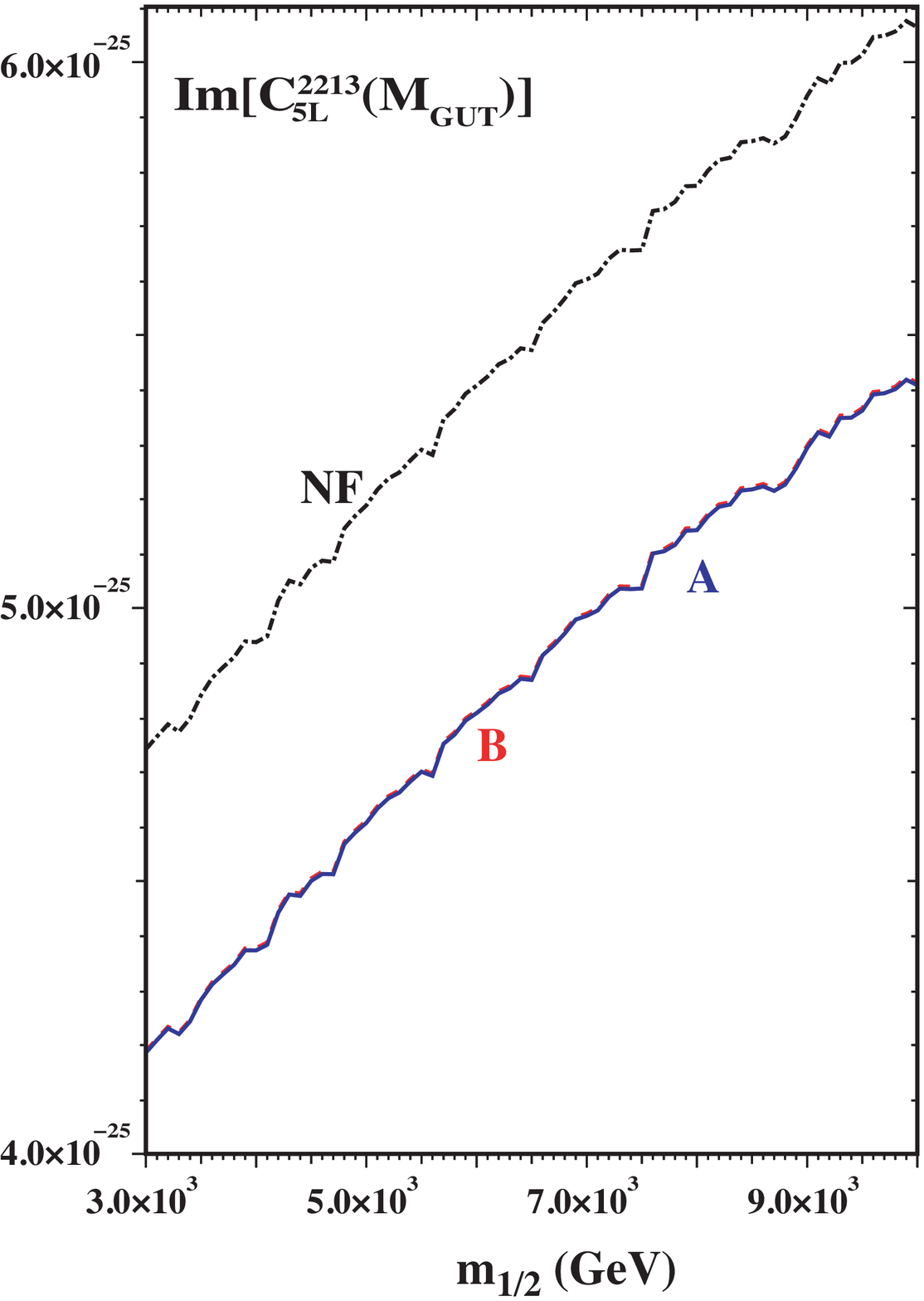}
\caption{\it{Comparison of the values (in units of [GeV]$^{-1}$)
of the real (left panel) and imaginary (right panel) parts of the Wilson coefficient $C_{5L}^{2213}(\Mg)$ as functions of $m_{1/2}$ in Model M1 (defined in Section \ref{Models}), with parameters $\tan\beta=6$, $\Mi = 10^{16.5}$ GeV, and $m_{3/2} = 5$ TeV.  The value for choice {\bf A} is shown as a solid line, that for choice {\bf B} as a dashed line, and that for choice {\bf NF} as a dot-dashed line.
\label{fig:C5L_M1M2} }}
\end{figure}

Finally, as also given in \cite{Ellis:2015rya}, the partial decay width for $p\to K^+ \bar{\nu}_i$ decay is
\begin{equation}
\label{pdkrates}
 \Gamma(p\to K^+\bar{\nu}_i)
=\frac{m_p}{32\pi}\biggl(1-\frac{m_K^2}{m_p^2}\biggr)^2
\vert {\cal A}(p\to K^+\bar{\nu}_i)\vert^2~,
\end{equation}
where 
\begin{align}
 {\cal A}(p\to K^+\bar{\nu}_e)&=
C_{LL}(usd\nu_e)\langle K^+\vert (us)_Ld_L\vert p\rangle
+C_{LL}(uds\nu_e)\langle K^+\vert (ud)_Ls_L\vert p\rangle ~,
\nonumber \\
 {\cal A}(p\to K^+\bar{\nu}_\mu)&=
C_{LL}(usd\nu_\mu)\langle K^+\vert (us)_Ld_L\vert p\rangle
+C_{LL}(uds\nu_\mu)\langle K^+\vert (ud)_Ls_L\vert p\rangle ~,
\nonumber \\
 {\cal A}(p\to K^+\bar{\nu}_\tau)&=
C_{RL}(usd\nu_\tau)\langle K^+\vert (us)_Rd_L\vert p\rangle
+
C_{RL}(uds\nu_\tau)\langle K^+\vert (ud)_Rs_L\vert p\rangle
\nonumber \\
&+
C_{LL}(usd\nu_\tau)\langle K^+\vert (us)_Ld_L\vert p\rangle
+C_{LL}(uds\nu_\tau)\langle K^+\vert (ud)_Ls_L\vert p\rangle
~.
\label{eq:ApKnue}
\end{align}
The proton decay rates (\ref{pdkrates}) depend on the Yukawa coupling matrices 
through the various WCs, and hence on our choice of 
diagonalization scheme. As an illustration of this sensitivity, in Fig.~\ref{fig:C5L_M1M2}
we compare the values of $C_{5L}^{2213}(M_{\rm GUT})$ for the three flavor structures introduced in Section~2.1 as functions of $m_{1/2}$ in Model M1 defined in Section \ref{Models}, with the model parameters $m_{3/2} = 5$ TeV, $\tan\beta=6$, and $\Mi = 10^{16.5}$ GeV.
The solid line is for the choice {\bf A}, the dashed line for the choice {\bf B}, and the dot-dashed line for the ``no-flavor" choice {\bf NF}. Shown separately are the real and imaginary parts of the WC. We see that choices {\bf A} and
{\bf B} yield very similar results, whereas the value of
$C_{5L}^{2213}(M_{\rm GUT})$ is about 10\% larger for choice {\bf NF} (i.e., when 
off-diagonal flavor-violating effects are ignored) mainly because of the treatment
of the Yukawa couplings.
When off-diagonal terms are considered in the Yukawa couplings, off-diagonal terms appear also in the soft masses-squared, trilinear terms, etc., which affect the running of the gauge couplings, with the largest effect being that on $g_3^2$. Note that the off-diagonal terms in $h_d$ affect not only $y_d$ and $y_s$ as shown in Fig. \ref{fig:YukCoups_M1} (see below)
but also the CKM matrix elements at the GUT scale.
At the electroweak scale, 
the difference in the WCs is about the same, (roughly 10\% between choices {\bf A/B} and {\bf NF}) though the magnitudes of the coefficients
are about 3--4 times larger.

The sensitivities to the Yukawa couplings of the charge-1/3 quarks $d$, $s$ and $b$ are also significant. 
We can understand this by considering 
the one-loop $\beta$ function of $h_d$, which is given by 
\bea
\beta^{(1)}_{h_d}=\frac{1}{16\pi^2} h_d\left[{\rm{Tr}}\left[ 3h_d h_d^\dagger + h_eh_e^\dagger  \right] + 3h_d^\dagger h_d + h_u^\dagger h_u+ f(g_1^2,g_2^2, g_3^2)\right],
\eea
where $f(g_1^2,g_2^2, g_3^2)=-\frac{16}{3}g^2_3 -3 g^2_2-\frac{7}{9}g^2_1$. The Yukawa matrix is non-diagonal at $\Mw$. In particular, $h_{d}^{23}$ and $h_{d}^{32}$ are non-zero due to the structure of the Yukawa couplings and the form of the Yukawa matrices in Eq.~(\ref{eq:hMSSMembA}), where $|h_d^{11}|, |h_d^{12}|, |h_d^{21}|, |h_d^{13}|, |h_d^{31}| < |h_d^{22}|, |h_d^{23}|, |h_d^{32}|$ $\ll |h_d^{33}|$. Due to the differences between the $\beta$ functions of the elements of $h_d$, each element evolves differently. In order to determine the change in the evolution with respect to evolving only the diagonal elements, we see from the hierarchy of the elements of the Yukawa couplings that the lightest eigenvalue, corresponding to $y_d$, will be affected mainly by
$|h_d^{11}|, |h_d^{12}|, |h_d^{21}|, |h_d^{13}|, |h_d^{31}|$ and the second eigenvalue, corresponding to $y_s$, by $|h_d^{22}|$, $|h_d^{23}|$ and $|h_d^{32}|$. 

We focus first on $y_s$, for which the relevant $\beta$ functions are
$\beta^{(1)}_{h_{d}^{22}}$, $\beta^{(1)}_{h_{d}^{32}}$ and $\beta^{(1)}_{h_{d}^{23}}$. In the cases of both model choices {\bf A} and {\bf B}, $|h_{d}^{32}|\ll |h_{d}^{23}|,|h_{d}^{22}|$,
whereas $|h_{d}^{32}| =  |h_{d}^{23}| = 0$ for {\bf NF}. 
To a good approximation we have
\bea
\beta^{(1)}_{h_{d}^{22}} &=& \frac{1}{16\pi^2} \ \left\{  h_{d}^{22} \left[  f(g_1^2,g_2^2, g_3^2) + 3 y_b^2 + y_\tau^2 \right] +  \sum_i h_{d}^{2i} (h_u^\dagger h_u)^{i2}  \right\}\nonumber\\
&\approx&\frac{1}{16\pi^2} \ h_{d}^{22} \left[ f(g_1^2,g_2^2, g_3^2)+ y^2_c + 3 y_b^2 + y_\tau^2  \right] \, ,\nonumber\\
\beta^{(1)}_{h_{d}^{23}} &=&\frac{1}{16\pi^2} \ \left\{  h_{d}^{23}  \left[  f(g_1^2,g_2^2, g_3^2) + 3 y_b^2 + y_\tau^2 \right] + \sum_i h_{d}^{2i}(h_u^\dagger h_u)^{i3}   \right\}\nonumber\\
&\approx&\frac{1}{16\pi^2} \ h_{d}^{23} \left[ f(g_1^2,g_2^2, g_3^2)+ y^2_t + 3 y_b^2 + y_\tau^2  \right] \, , \nonumber\\
\beta^{(1)}_{h_{d}^{32}} & \approx &  \frac{1}{16\pi^2} \ h_{d}^{32}  \left[ f(g_1^2,g_2^2, g_3^2)   +   y^2_c 
+  3 y_b^2 + y_\tau^2   \right] \, .
\eea
We see that, due to the term proportional to  $y^2_t$ in $\beta^{(1)}_{h_{d}^{23}}$,  $h_{d}^{23}$ will evolve differently from  $h_{d}^{22}$ and  $h_{d}^{32}$. In particular, when evolving the parameters of the MSSM from $\Mw$ to $\Mg$,  $h_{d}^{23}$ decreases {\emph{less}} than $h_{d}^{22}$ and $h_{d}^{32}$, which in turn produces a higher value of $y^2_s$ at $\Mg$ than when the running of off-diagonal Yukawa couplings is neglected, because no information on the evolution of 
$h_d^{23}$ is considered in that case.

A comparison of the squared Yukawa couplings, $y_d^2$, $y_s^2$, and $y_b^2$ as functions of $m_{1/2}$ for the set of inputs used in Fig.~\ref{fig:C5L_M1M2} is shown in Fig.~\ref{fig:YukCoups_M1}. 
We see that while the differences between {\bf A} and {\bf B} do not manifest themselves in any of the down-quark Yukawa couplings, they do differ from the {\bf NF} choice for the first two generations. 
The flavor-violating contributions
are negligible for the bottom quark because $m_b \gg m_{d,s}$, and the three choices 
considered give results that are nearly identical. The fact that the difference between 
$y_d^2$ and $y_s^2$ is larger for choices {\bf A} and {\bf B} than for the {\bf NF} choice
is a reflection of the larger magnitudes of the off-diagonal Yukawa couplings.

\begin{figure}[!ht]
\centering
\includegraphics[width=5.2cm,height=5.2cm]{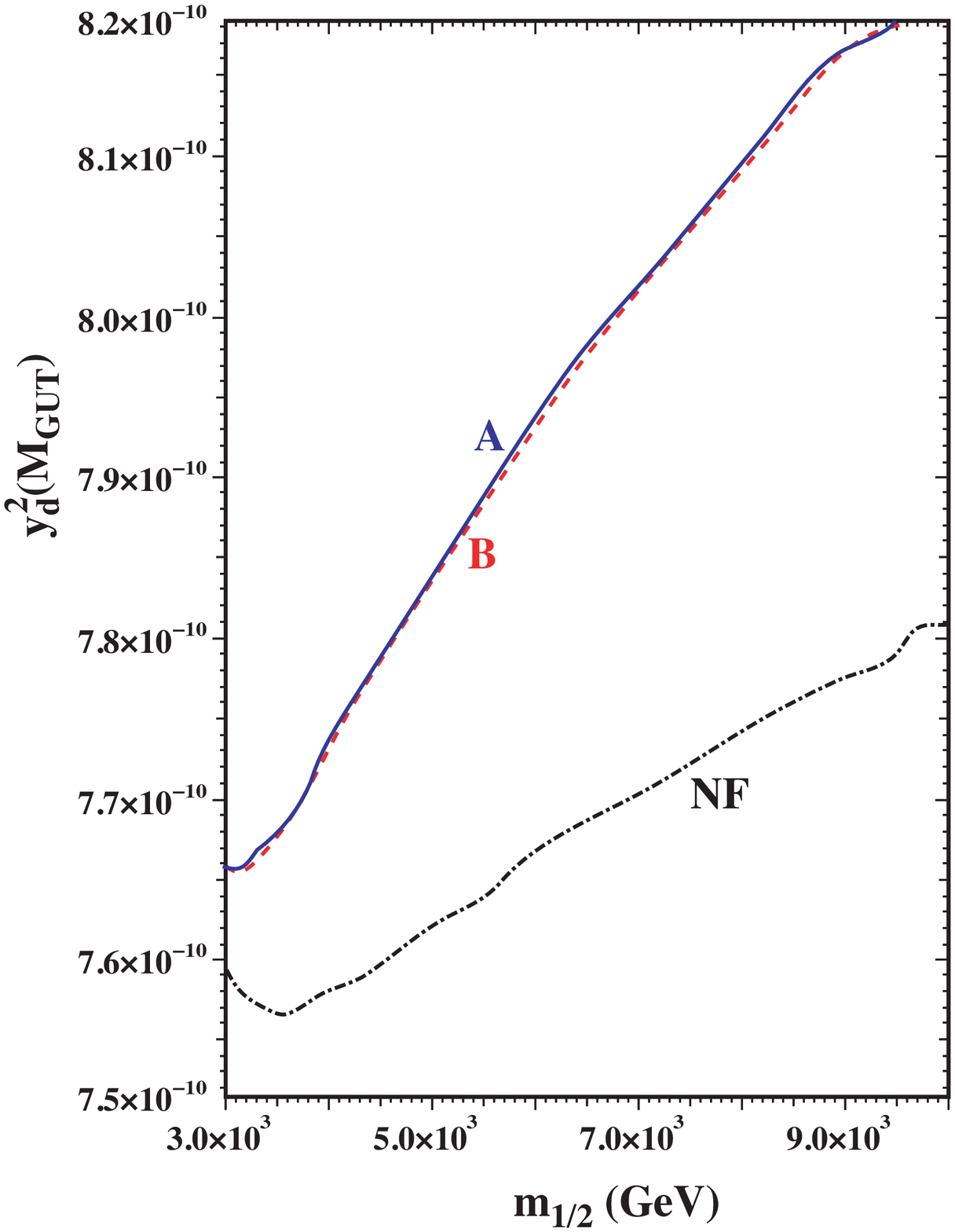} 
\includegraphics[width=5.2cm,height=5.2cm]{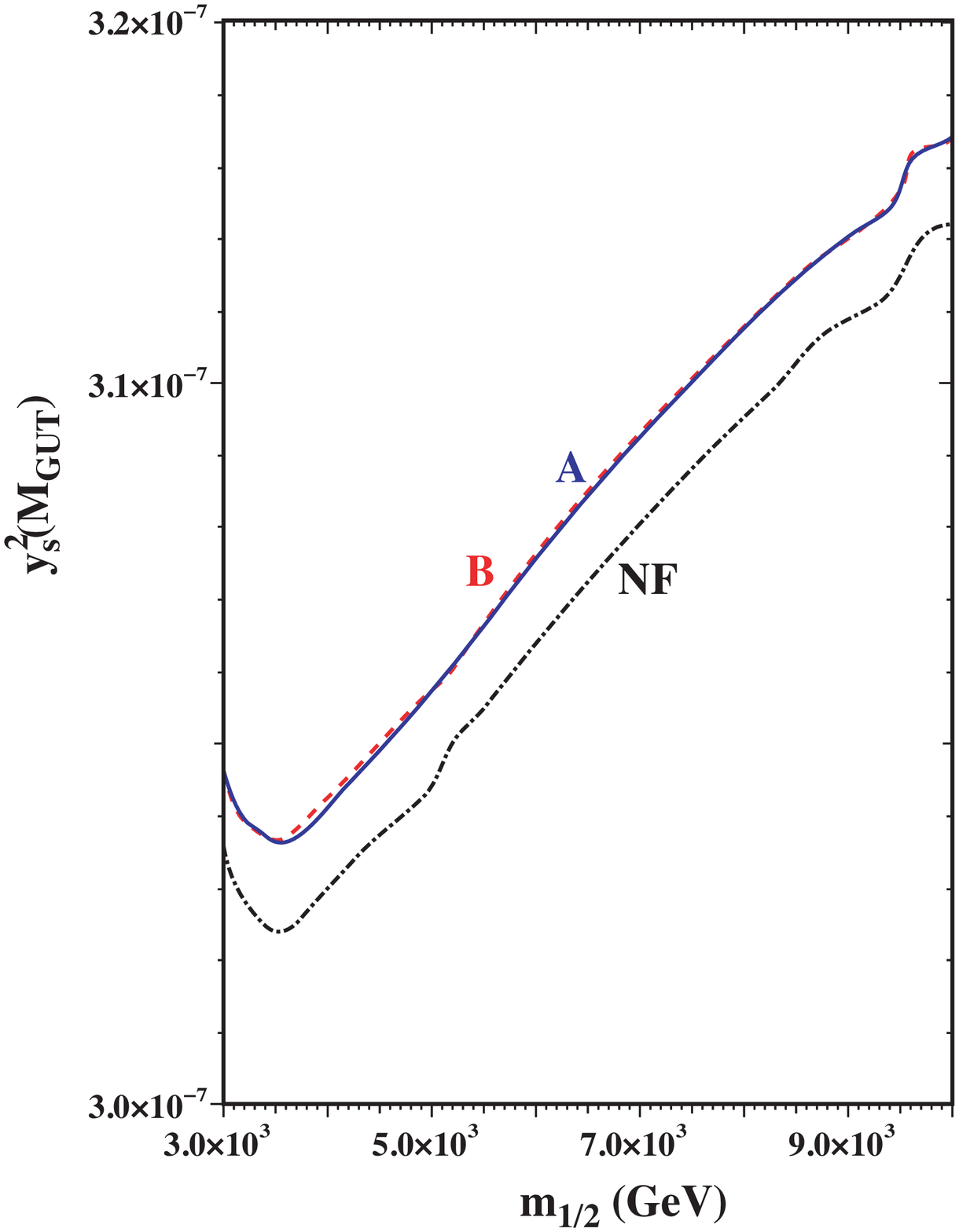} 
\includegraphics[width=5.2cm,height=5.2cm]{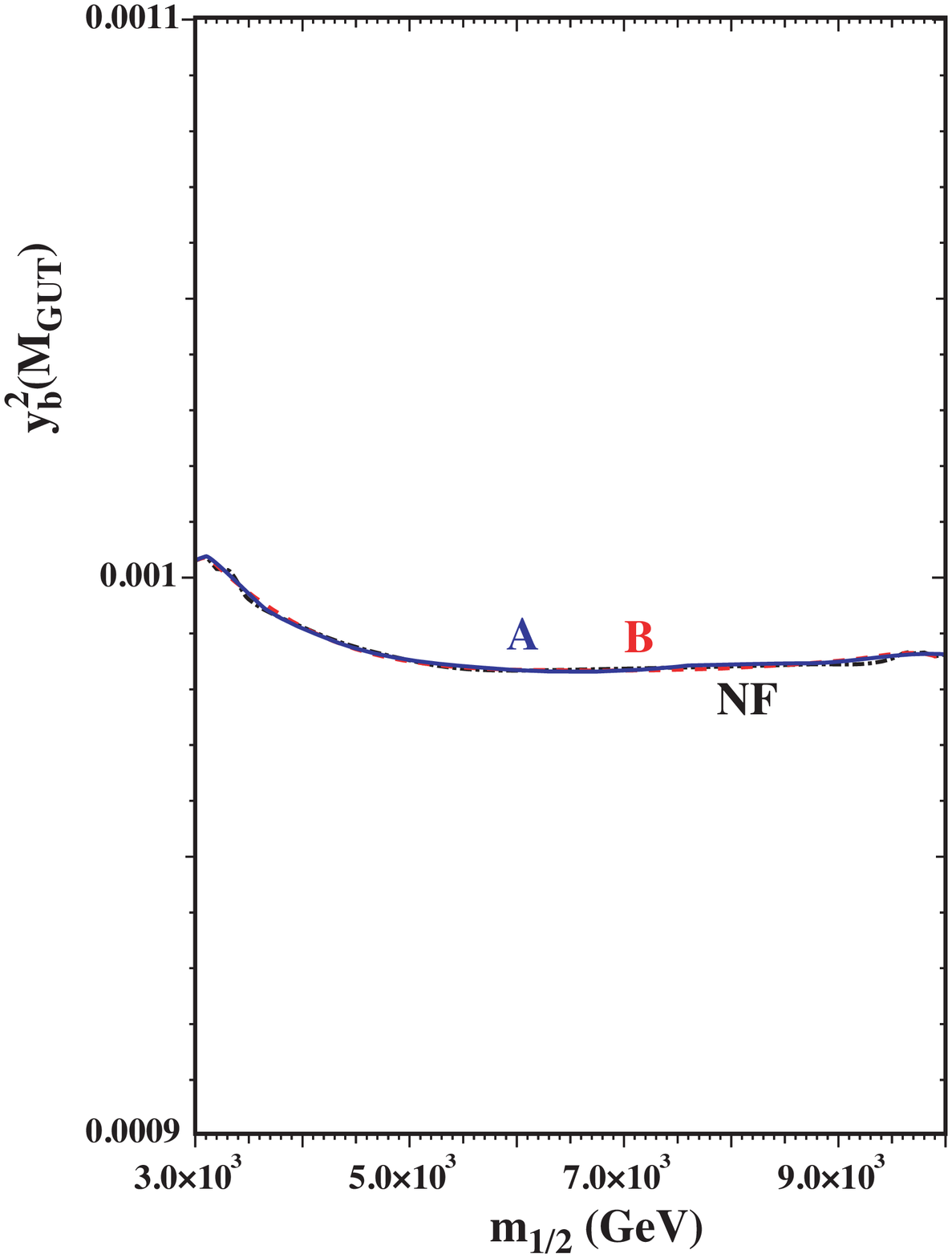} 
\caption{\it{Comparisons of the values of the Yukawa couplings at $M_{\rm GUT}$ for choices {\bf A} (blue solid line), {\bf B} (red dashed line) and {\bf NF} (black dot-dashed line). 
}
\label{fig:YukCoups_M1}}
\end{figure}
 
\subsubsection*{Hadronic Uncertainties} 
 
 In addition to the WCs, the proton decay amplitudes in Eq. (\ref{eq:ApKnue}) depend on hadronic matrix elements.
 As discussed in detail in \cite{Ellis:2019fwf},
 in order to apply the limit in Eq. (\ref{eq:lft_PDKnu}),
 one needs to know not only the central values of the matrix elements but also their uncertainties. 
The relevant systematic uncertainties of the form factors were taken into account for the first time in~\cite{Aoki:2013yxa}. The total uncertainties found in $K$ final states were  20\%-40\%, whereas they were 30\%-40\% for $\pi$ final states, which were reduced to 10\% -15\% in \cite{Aoki:2017puj}. The uncertainties in all of the matrix elements 
in \eq{eq:lft_PDKnu} must be taken into account in order to determine the region of parameter space for which 
$\tau\left(p\rightarrow K^+ \bar{\nu}\right) > 6.6 \times 10^{33}~{\rm{yrs}}$. For the matrix elements contributing to the relevant amplitude ${\cal A}(p\to K^+\bar{\nu}_\tau)$ in \eq{eq:ApKnue}, Ref.~\cite{Aoki:2017puj} found
\begin{align}
\label{sec:hadmatxunc}
 & \langle K^+\vert (us)_Ld_L\vert p\rangle= \, 0.041 \pm 0.006 \, , \nonumber \\  
 & \langle K^+\vert (ud)_Ls_L\vert p\rangle =\,  0.139 \pm 0.016 \, , \nonumber \\ 
 & \langle K^+\vert (us)_Rd_L\vert p\rangle = -0.049 \pm 0.006 \, , \nonumber\\ 
 & \langle  K^+\vert (ud)_Rs_L\vert p\rangle = -0.134 \pm 0.014 \, ,
\end{align}
where we have quoted the total error obtained by combining the statistical and systematic errors
in quadrature. We note that the matrix elements $\langle K^+\vert (us)_Ld_L\vert p\rangle$ and $\langle K^+\vert (ud)_Ls_L\vert p\rangle $ are the most relevant, since the $C_{LL}$ coefficients dominate over $C_{RL}$.

\subsection{Flavor Violation }
\subsubsection{$\mathbf{\muegD}$}

The embedding of the MSSM in SU(5), as in either Eq. (\ref{eq:orembedding}) or (\ref{eq:orembedding2}),
can make an important difference. In particular, the different embeddings for the 
SU(5) Yukawa matrices $h_{10}$ and $h_{5}$ lead to different effective mass matrices 
for $h_D$, $h_E$ and $h_U$, as we have seen in Section~\ref{sec:sectionModelFramework}. 
Minimal SU(5) corresponds to the embedding (\ref{eq:orembedding2}) (without the phases), 
where $h_D$ and $h_E$ are the transposes of each other. When $h_D$ involves the CKM  matrix, $h_E$
inevitably leads to large right-handed currents, enhancing the branching ratio $\Bmueg$,
which can be written as
\beq
\label{eq:BRmeglim}
\Bmueg \; = \; \frac{3\pi^2 e^2}{G_F^2 m^4_{\mu}} \left(|a_{\mu e\gamma L}|^2 +  |a_{\mu e\gamma R}|^2\right) \; \simeq \;
2.15\times 10^{15}  \left(|a_{\mu e\gamma L}|^2 +  |a_{\mu e\gamma R}|^2\right) \, , 
\eeq
where we use the notation in \cite{Ellis:2016qra} for the amplitude of the decay $\mu\rightarrow e \gamma$. 
The experimental upper limit $\BtauegE \le 4.2 \times 10^{-13}$~\cite{TheMEG:2016wtm} imposes the constraints
\beq
|a_{\mu e\gamma L}|, |a_{\mu e\gamma R}| \; \lesssim \; 10^{-14} \, .
\label{FVlimits}
\eeq
These limits on the coefficients $a_{\mu e\gamma L}$ and $a_{\mu e\gamma R}$ constrain the amount of
flavor violation mediated by charginos and neutralinos in the MSSM. 
We note that care must be taken in an 
analysis in terms of mass-insertion operators in the presence of off-diagonal entries in all the
soft supersymmetry-breaking sectors, because there are correlations among the elements of the 
matrices and some cancellations may occur.

In order to understand the order of magnitude of possible contributions to $a_{\mu e\gamma R}$ and $a_{\mu e\gamma L}$ that are consistent with the limits in \eq{FVlimits}, we consider  simplified formulae for the neutralino  
contributions. There are significant contributions coming from chargino exchange, but these are suppressed relative to the neutralino exchange contributions.

\begin{figure}[!ht]
\vspace{5mm}
\centering
\includegraphics[width=13cm]{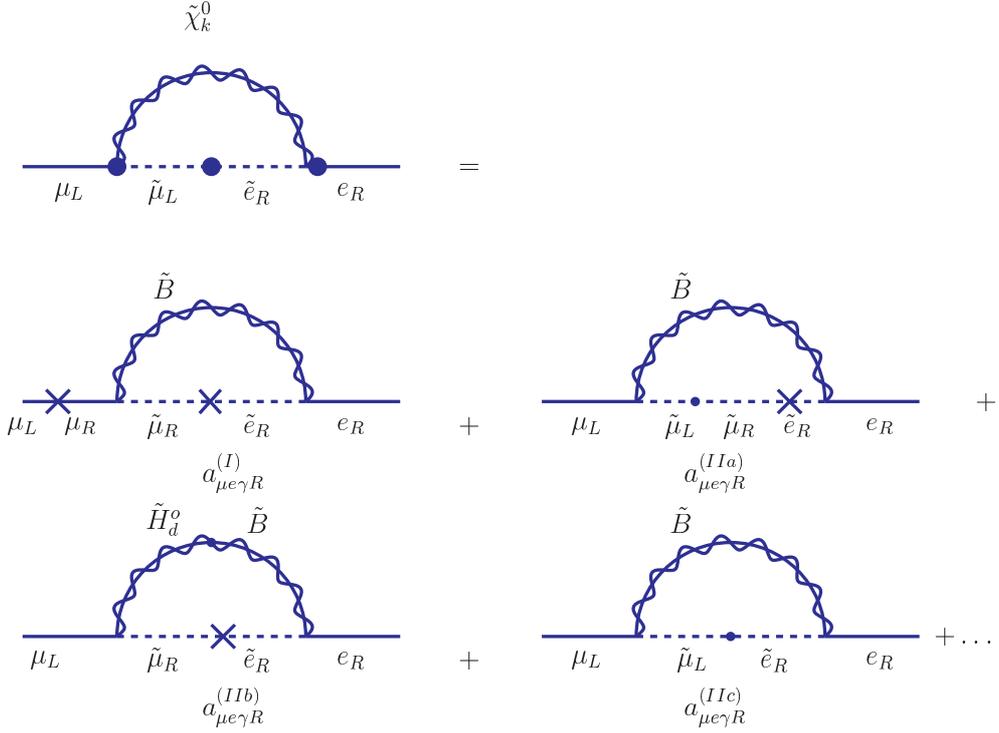}
\caption{\it Contributions to $\amuegR$.  
On the left hand side (of the equality), we depict the diagram in the mass eigenstate basis, and on the right hand side, the diagrams are split in the interaction basis. The external photon can couple to all charged-particle lines. The cross denotes the insertion of a
flavor-mixing term that does not change chirality, and the dot an insertion that changes chirality. 
}
\label{fig:plotamuR1}
\end{figure}

Fig.~\ref{fig:plotamuR1} displays the diagrams making the most important contributions to $a_{\mu e \gamma R}$. First the neutralino exchange diagram is shown in the mass-eigenstate basis, and then we identify
four main contributions in the interaction basis. 
The contributions from $a^{(I)}_{\mu e \gamma R}$, which requires a mass insertion outside the loop, can be approximated as
\bea
\label{eq:amuegRI}
a^{(I)}_{\mu e \gamma R}&\approx& -\frac{m^2_\mu}{96 \pi^2}  g_1^2  \frac{(m^2_{E})_{12}}{m^2_{\tilde{e}_R}m^2_{\tilde{\mu}_R}} \, .
\eea
As we will see when we consider specific models in Section \ref{mainmodels}, $(m^2_{E})_{12}$ is similar in both the cases {\bf A} and {\bf B}. Similarly 
 the contributions $a^{(IIa)}_{\mu e \gamma R}$, $a^{(IIb)}_{\mu e \gamma R}$ may be approximated by
\bea
\label{eq:amuegRIIab}
a^{(IIa)}_{\mu e \gamma R} &\approx& -\frac{m_\mu}{48 \pi^ 2}   \frac{\left(v {(a_E)}_{22} + m_\mu \mu \tan\beta\right)}{ m^2_{\tilde{\mu}_L}}         g_1^2  M_1 \frac{(m^2_{E})_{12}}{m^2_{\tilde{e}_R}m^2_{\tilde{\mu}_R}} \, , \nonumber\\
a^{(IIb)}_{\mu e \gamma R} &\approx& \frac{m_\mu }{16\pi^2} \frac{g_1 y_\mu}{3\sqrt{2}}\  {\rm{Re}}\left[ N_{11}^* N_{31}^* \right] M_1\
\frac{(m^2_{E})_{12}}{m^2_{\tilde{e}_R}m^2_{\tilde{\mu}_R}}\, .
\eea
Here $N_{11}$ and $N_{13}$ are mixing elements of the neutralinos, with $N_{11}\approx 1$ when the lightest neutralino is mainly bino, and  $N_{31}$ characterizes the mixing between the the Higgsino $\tilde H^0_d$ and the bino. 
These diagram factors are also proportional to $(m^2_{E})_{12}$, a common
factor between cases {\bf A} and {\bf B}.
Although the diagram $a^{(IIa)}$ also depends on ${(a_E)}_{22}$,
this quantity is also similar in cases {\bf A} and {\bf B}
for the models we consider below. 
In contrast, the diagram corresponding to
 $a^{(IIc)}_{\mu e \gamma R}$, which may be approximated by
\bea
\label{eq:amuegRIIc}
a^{(IIc)}_{\mu e \gamma R} &\approx& 
\frac{m_\mu }{16\pi^2}  \frac{g_1^2\ v \ {(a_E)}_{21} }{3 m_{\tilde \mu_L}^2} M_1 {\rm{Re}} [ N_{11}^*N_{31}^*  ]\frac{ \left( m_{\tilde e_R}^2 - m_{\tilde \mu_R}^2 \right)} { m_{\tilde e_R}^2 \  m_{\tilde \mu_R}^2}\, ,
\eea
is proportional to  $(a_E)_{21}$.  This mixing term is very different in cases {\bf A} and {\bf B} and can lead to differences in the total value of $a_{\mu e \gamma R}$ by an order of magnitude or more, as we see below.

There are similar contributions to $a_{\mu e \gamma L}$, but they are mediated by $(m^2_{L})_{12}$ instead of $(m^2_{E})_{12}$, and hence suppressed for these models, as we see below in Section \ref{mainmodels}.
The reason why  the right-handed contribution to $\Bmueg$, 
which is encoded in $|a_{\mu e \gamma R}|$ and  associated with $m^2_{E}$,
is significantly larger than $m^2_{L}$ is that  
$m^2_{E}$ is matched to $m^2_{10}$ at $\Mg$, and $m^2_{L}$ to $m^2_{5}$. Both $m^2_{10}$ and $m^2_{\bar{5}}$ start at zero at $\Mi$ (see \eq{eq:bcatMGUT}), but they evolve differently: 
\bea
\frac{d m^2_{\bar{5}}}  {dt}  \supset  -\frac{1}{16\pi^2}  \frac{96}{5} g^2_5 M^2_5 \, , \quad
\frac{d m^2_{10}}{dt}          \supset  -\frac{1}{16\pi^2}  \frac{144}{5} g^2_5 M^2_5 \, . 
\eea
Consequently, $(m^2_{10})_{ii}$ is typically twice as large as $(m^2_{\bar{5}})_{ii}$ at $\Mg$  
(see,  e.g., Fig.~2 of  \cite{Ellis:2016qra}).

We conclude this discussion by noting that the future MEG~II experiment is expected to be sensitive to $\Bmueg = 6 \times 10^{-14}$~\cite{Baldini:2018nnn}.

\subsubsection{$\mu \to eee$}
There are other proposals for future experiments that are sensitive to muon flavor violation, e.g., to the $\mu \to eee$ mode. The current experimental limit on this mode is provided by the SINDRUM experiment: ${\rm BR}(\mu \to eee) < 1.0 \times 10^{-12}$~\cite{Bellgardt:1987du},
and the Mu3e experiment aims at a sensitivity of $\sim 10^{-16}$ in the future~\cite{Blondel:2013ia}. 

The $\mu \to e \gamma$ dipole processes shown in 
Fig.~\ref{fig:plotamuR1} give the dominant contributions to $\mu \to eee$ in many supersymmetric models~\cite{Hisano:1995cp}.~\footnote{See also the detailed discussion of the related
$\tau \to 3 \mu$ process 
in Ref.~\cite{Arganda:2005ji}.}
In this case ${\rm BR}(\mu \to eee)$ is related to $\Bmueg$ by~\cite{Arganda:2005ji} 
\begin{equation}
    \frac{{\rm BR}(\mu \to eee)}{\Bmueg} = \frac{\alpha}{3\pi} 
    \biggl[ \ln \biggl(\frac{m_\mu^2}{m_e^2}\biggr)-\frac{11}{4} \biggr]
    \simeq 6 \times 10^{-3} ~. 
\end{equation}
This relation indicates that currently $\mu \to eee$ gives a much weaker limit on lepton flavor violation than $\mu \to e \gamma$, but will offer a better sensitivity in the future. 

\subsubsection{$\mu \to e$ conversion}
\label{sec:mueconversion}
Another promising process is $\mu \to e$ conversion on a nucleus. The tightest current experimental bound on the $\mu \to e$ conversion rate is provided for gold nuclei by the SINDRUM II collaboration: ${\rm BR}( \mu + {\rm Au} \to e +{\rm Au}) < 7 \times 10^{-13}$~\cite{Bertl:2006up}. In the future, COMET Phase II at J-PARC~\cite{Adamov:2018vin} ($\mu + {\rm Al} \to e +{\rm Al}$) and Mu2e at FNAL~\cite{Abusalma:2018xem} ($\mu + {\rm Al, Ti} \to e +{\rm Al, Ti}$) may offer sensitivity at the level of ${\cal O}(10^{-18})$~\cite{Strategy:2019vxc} and PRISM at J-PARC ($\mu + {\rm Pb, Au} \to e +{\rm Pb, Au}$)
at the level of ${\cal O}(10^{-19})$~\cite{Strategy:2019vxc}. 
Assuming again the dipole operator approximation for $\mu \to e$ conversion, 
there is a relation between $\Bmueg$ and ${\rm BR}( \mu + N \to e + N)$~\cite{Czarnecki:1998iz} that depends on the target nucleus $N$, {e.g.}, for $N = {\rm Al}$ we have ${\rm BR}( \mu + {\rm Al} \to e +{\rm Al}) \simeq 2.6 \times 10^{-3} \times  \Bmueg$ and for $N = {\rm Au}$ we estimate ${\rm BR}( \mu + {\rm Au} \to e +{\rm Au}) \simeq 2.7 \times 10^{-3} \times  \Bmueg$.
A sensitivity to $\mu \to e$ conversion at the level of $10^{-18} (10^{-19})$ would therefore correspond to
$\Bmueg \sim 4 \times 10^{-16} (4 \times 10^{-17})$.
We infer that $\mu \to e$ conversion processes may be more promising than $\mu \to e \gamma$ and $\mu \to eee$ in the future. 

\subsection{Electric dipole moments (EDMs) \label{subsec:EDM}}

The new limit on the electron EDM, $|d_e|< 1.1 \times 10^{-29}$~e.cm~\cite{Andreev:2018ayy} could in principle constrain parts of the parameter space that 
would otherwise be allowed if no flavor-violating terms in the soft terms were considered.

\begin{figure}[ht]
\centering
\includegraphics*[width=7.4cm,height=3.6cm]{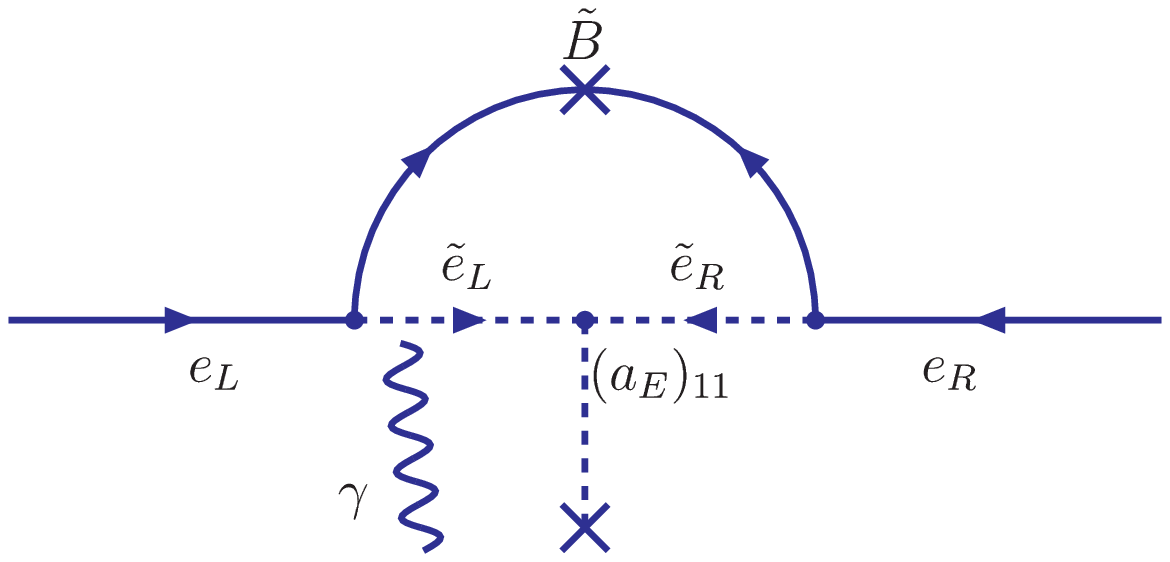}
\hspace*{1cm}
\includegraphics*[width=7.4cm,height=3.6cm]{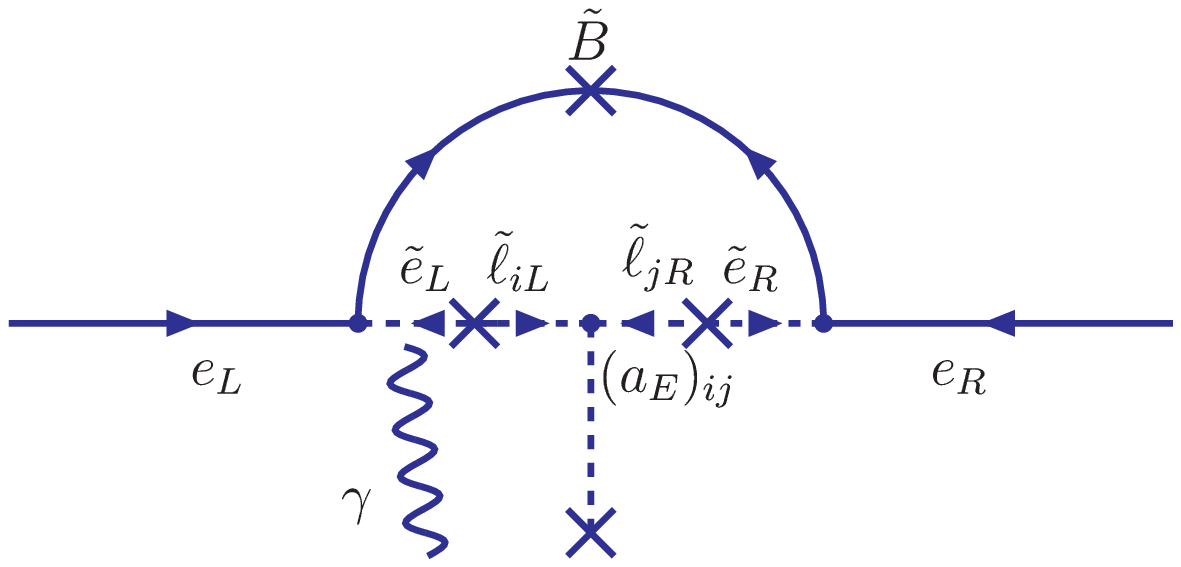}
\caption{\it{Contributions to the electron EDM mediated by the bino, without flavor violation (left diagram)
and with flavor violation (right diagram), respectively. The states $\ell$ are flavor eigenstates $\ell_1=e$, $\ell_2=\mu$, $\ell_3=\tau$.}
\label{fig:EEDM_FD} }
\end{figure}

At the one-loop level, there are supersymmetric contributions to the electron EDM
mediated by  charginos and neutralinos. 
A general expression is given by~\cite{Ibrahim:2007fb}
\bea
d_e \left(  m_{\tilde \chi^0} \right)&=&\frac{e\ \alpha_{\rm{EM}}}{4\pi \sin^2\theta_{\rm{W}}}
\ 
\sum_{k=1}^{2} \sum_{i=1}^{4} 
  {\rm{Im}}\left\{ \eta_{Eik}\right\}
\frac{ m_{\tilde \chi^0}  }{m^2_{\tilde e_k}} Q_{\tilde e}
 B\left( \frac{m_{\tilde \chi^0}^2}{m^2_{\tilde e_k}} \right),\label{eq:deX10}
\eea
where $Q_{\tilde e}=-1$, $B(x) \equiv 1/(2(1-x)^2) \left[ 1+ x+ 2 x \log x/ (1-x)\right]$, and
\bea
\eta_{Fik}&=&\left[-\sqrt{2} \tan\theta_W ( Q_f-T_{3f})\ N_{1i} (K_F)_{k, 1L}
-\sqrt{2} T_{3f} N_{2i} (K_{F})_ {k,1L}
+\frac{m_f}{\sqrt{2} M_W \cos\beta} N_{3i} (K_{F})_ {k,1R}\right]\times \nonumber\\
& & \left[\sqrt{2}\tan\theta_W Q_f N_{1i} (K^*_{F})_{k,1R}  
- \frac{m_f}{\sqrt{2} M_W \cos\beta}  N_{3i} (K^*_{F})_{k,1L}   \right].
\label{eq:etafijk}
\eea
In these expressions, the sfermion mass matrix, in the basis where Yukawa couplings are diagonal,  for each family is given by
\bea
\label{eq:effsfmssterms}
({\mathcal{M}}^2_{F})_{ij}&=&
\left[\begin{array}{cc}
(m^2_{F_L})_{ij}+(m^{2}_f)_{i} \delta_{ij}+D^f_{L} & 
-({a_F}_{ij}v_f+\mu^* \tan^s\beta \ (m_f)_{i} \delta_{ij}) \\
 -({a^{*}_F}_{ji}v_f+\mu \tan^s\beta \  (m_f)_{i} \delta_{ij})&
 (m^2_{F_R})_{ij}+(m^{2}_f)_{i} \delta_{ij}+D^f_{R}
\end{array}\right],
\nn\\
{\rm where} \quad D^f_{L,R} &=& \cos 2\beta M^2_Z(T^3_f-{Q}_{f_{L,R}} \sin^2\theta_W), \quad
s\ =\  \left\{\begin{array}{c} 1,\ f=d,e \\ -1,\ f=u
\end{array}\right.
\nn\\
& & F_L=Q, L, \quad F_R=D,E,
\eea
and $m_f$,  $f=d,e, u$,  are the masses of the fermions 
corresponding to the sfermions $F=D,E, U$.
Note that the indices  $1L$ and $1R$  of   $(K_{F})_{k,1L}$  and $(K^*_{F}) _{k,1R}$  in \eq{eq:etafijk} for $F=E$ correspond to $e_L$ and $e_R$, respectively,
which are the external lines in  Fig.~\ref{fig:EEDM_FD}. The index $k$ corresponds to the mass eigenstates from $k=1,\hdots, 6$, where
the sfermion mass eigenstates are defined by 
\bea
\label{eq:defeigenstates}
\left[\tilde f_1, \tilde f_2, \hdots, \tilde f_6\right]^T \equiv K^*_F\left[ \tilde f_{1L}, \tilde f_{1R}, \hdots, \tilde f_{3L}, \tilde f_{3R}
 \right]^T,
 \eea
such that $\widehat {\mathcal{M}}^{2}_F= K_F {\mathcal{M}}^2 K_F^\dagger$ is a diagonal matrix.

  In the models considered here, the lightest neutralino, ${\tilde \chi_1^0}$,  
  typically gives  the dominant
 contribution to the electron EDM in \eq{eq:deX10}. This term is proportional to
$N_{1i}K_{Ek1}$, so the most important contribution to $d_e$ comes from  
\bea
\eta_{E1k}&=& -2 \tan^2\theta_W \left( Q_e-T_{3e}\right)Q_e N_{11} N_{11} (K_{E})_{k,1L} (K^*_{E})_{k,1R}\nn\\
&=&
- \tan^2\theta_W Q_e N_{11} N_{11} (K_{E})_{k,1L} (K^*_{E})_{k,1R} \, ,
\eea
and
\bea
\label{eq:approxd_EDM}
d_e\approx -\frac{e\ \alpha_{\rm{EM}}}{4\pi \cos^2\theta_{\rm{W}}} \sum_{k=1}^{k=6}  {\rm{Im}} \left\{ N_{11} N_{11} (K_{E})_ {k,1L} (K^*_{E})_{k,1R} \right\}
\ 
\frac{m_{\tilde \chi_1^0}}{m^2_{e_k}}  B\left( \frac{m_{\tilde \chi_1^0}^2}{m^2_{e_k}} \right) \, ,
\eea
where the $m^2_{e_k}$ are the slepton mass eigenstates.  This contribution is depicted in the left Feynman diagram of Fig.~\ref{fig:EEDM_FD} in the flavor basis,
and we see that, in the absence of off-diagonal  and imaginary terms in $a_E$, the EDM is zero.
However, once the CKM matrix is introduced to seed flavor violation,
as in the flavor choices {\bf A} and {\bf B}
discussed earlier, imaginary parts appear in the soft squared-mass matrices
$(m^2_{L})_{1j}$, $({a_E})_{ij}$ 
and  $(m^2_{E})_{1j}$.  
We note that the function $B$ in \eq{eq:deX10} varies slowly over the range $\sim 0.2$ to $\sim 0.3$ for all of the spectra we consider and for all the indices $k$. Therefore the individual contributions in the terms of \eq{eq:deX10} depend mainly on the combination
\bea
 {\rm{Im}} \left\{ \eta_{E1k}\right\} \frac{1}{m^2_{e_k}}\propto 
 {\rm{Im}} \left\{ (K_{E})_{k,1L} (K^*_{E})_{k,1R}   \right\} \frac{1}{m^2_{e_k}}.
\eea
The imaginary part above can be easily understood in terms of the second diagram of Fig.~\ref{fig:EEDM_FD}, since
\bea
\label{eq:approxim}
\hspace{-2mm}
{\rm{Im}} \left\{ (K_{E})_{k,1L}(K_{E})^*_{k,1R}   \right\}  \sim v_d \  {\rm{Im}} \!\!\! \sum_{b,c=1,2, 3}\!\!  \left\{  \!
\frac{\left(m^2_L \right)_{1b} }{\sqrt{  (m^2_{L})_{11}   (m^2_{L})_{bb}    } } \frac{\left( a_{E} \right)_{bc} }{\sqrt{  (m^2_{L})_{bb}   (m^2_{E})_{cc}    } }   \frac{\left( m^2_{E} \right)_{c1}} {{\sqrt{  (m^2_{E})_{cc}   (m^2_{E})_{11}    } }} \! \right\} ,
\eea
where $b,c=1,2,3$. The imaginary parts of each of the contributions to the sum above can be written as 
\bea
{\rm{Im}} \left\{ (K_{E})_{k,1L}(K_{E})^*_{k,1R}   \right\}_{bc} \!\!  & \sim& \!\!
\frac{v_d}{ \sqrt{  (m^2_{L})_{11}   (m^2_{L})_{bb}    }    \sqrt{  (m^2_{L})_{bb}   (m^2_{E})_{cc}    }  \sqrt{  (m^2_{E})_{cc}   (m^2_{E})_{11}   }  }\nonumber\\
\times & & \hspace{-7mm}
\left[ {\rm Re} \left[  (m^2_E)_{c1} \right] \left[  
{\rm Im}   \left[  (m^2_L)_{1b} \right]  {\rm Re} \left[  (a_E)_{bc}    \right] +
 {\rm Re} \left[  (m^2_L)_{1b}  \right]  {\rm Im}    \left[  (a_E)_{bc} \right] 
\right] \right.\nonumber\\
\!\!\!\!\! + &  & \hspace{-7mm}
{\rm Im}  \left[  (m^2_E)_{c1} \right]  
\left. \left[  
{\rm Re}  \left[m^2_L   \right]_{1b} {\rm Re}  \left[ (a_E)_{bc}   \right]  
- {\rm Im}   \left[m^2_L   \right]_{1b}  {\rm Im}  \left[ (a_E)_{bc}   \right]  
\right] \right].
\label{eq:ImetaFlavordecomp}
\eea
We find that there are important contributions from the terms involving ${\rm Re} (a_E)_{33}$ but, depending on the model, contributions containing $(a_{E})_{11}$ can dominate for models with the flavor choice {\bf A}, and contributions containing $(a_{E})_{21}$ and $(a_{E})_{31}$ can also be important.

In \eq{eq:ImetaFlavordecomp} with $b = c = 3$, we find that in the models considered in Section \ref{sec:section_Analysis}
\bea
&& {\rm Re} \left[  (m^2_E)_{31}\right]   {\rm Im}    \left[  (m^2_L)_{13} \right]     > {\rm Re} \left[  (m^2_L)_{13}\right]
{\rm Im}  \left[  (m^2_E)_{31} \right],  \nonumber\\
&&{\rm Im}    \left[  (m^2_L)_{13} \right]  \sim {\rm Re}   \left[  (m^2_L)_{13} \right],  \nonumber\\
&& {\rm Im}   \left[ (a_E)_{33} \right] \ll  {\rm Re}  \left[ (a_E)_{33} \right].
\eea
Then the dominant term in Eqs.~(\ref{eq:approxim}) and (\ref{eq:ImetaFlavordecomp})  contains  ${\rm Re}(a_{E})_{33}$ and reduces to
\bea
\left[{\rm{Im}} \left\{ (K_{E})_{k,1L} (K_{E})^*_{k,1R}   \right\} \right]_{33}\sim  v_d\frac{  {\rm{Re}}\left[ (m^2_E)_{ 31}\right]    \rm{Re} \left[  (a_E)_{33}    \right]    \rm{Im}  \left[m^2_L   \right]_{13}       }{ (m^2_E)_{ 33}  (m^2_L)_{33}  \sqrt{ (m^2_E)_{ 33}  (m^2_L)_{33}  }   } \, .
\label{cont33}
\eea
In the models considered below, this contribution is similar in both of the choices {\bf A} and {\bf B}. 
However, for choice {\bf{B}}, the contribution from $(a_{E})_{31}$ can also be important, as we will see in Section~\ref{sec:section_Analysis}. 

It is relatively easy to understand how the contribution from $(a_{E})_{11}$ can dominate in models with flavor choice {\bf{A}} relative to choice {\bf{B}}. As seen in Eqs.~(\ref{eq:approxim}) and (\ref{eq:ImetaFlavordecomp}), when $b=c=1$ (and noting that the imaginary parts of $(m^2_E)_{11}$ and $(m^2_L)_{11}$ both vanish),
the term containing $(a_{E} )_{11}$ reduces to 
\bea
\left[{\rm{Im}} \left\{ K_{Ek1}K^*_{Ek2}   \right\} \right]_{11}=  v_d  \frac{{\rm{Im}}  \left\{  (a_{E})_{11}  \right\} }{  \sqrt{ (m^2_E)_{11} (m^2_L)_{11}    } }\, .
\label{cont11}
\eea
As we will see in Section \ref{sec:section_Analysis} below, ${\rm{Im}}  \left\{  (a_{E})_{11}  \right\}$ is typically four orders of magnitude larger in choice {\bf{A}}  than  in choice {\bf{B}}.
This can be traced to the matching condition in Eq.~(\ref{eq:trilinearpropaE}).
In fact, the contribution containing  $(a_{E})_{11} $ can be even larger than that containing  $(a_{E})_{33} $ in choice {\bf A}.

When one of the slepton  states dominates the contribution in \eq{eq:approxd_EDM}, we can write 
\bea
\label{eq:simplified_de}
|d_e|=1.1\times 10^{-29} \left[ \frac{ | {\rm{Im}} \left\{      \eta_{E1k} \right\} |    }{1 \times 10^{-8}}\right]\ 
\left[\frac{m_{\tilde \chi^0}}{1\times 10^3 \ \rm{GeV}} \right] \left[ \frac{[2 \times 10^3  \ \rm{GeV}]^2}{m^2_{e_k}} \right]  \ \left[ \frac{B(\frac{m_{\tilde \chi^0}^2}{m^2_{e_k}})}{0.29} \right]\ \rm{e.cm} \, .
\eea
When two or more contributions are important we can still use the formula above for each slepton, taking the signs of the ${\rm{Im}} \left\{ \eta_{E1k} \right\}$ into account.
Overall, therefore, we find that while the contributions from $b = c = 3$ for choices {\bf A} and {\bf B}
are similar, the contribution from $b = c = 1$ is much greater in choice {\bf A}, and we expect the EDM to be larger in choice {\bf A} than in choice {\bf B}.

\section{Analysis of Low-Energy Observables \label{sec:section_Analysis}}
\label{mainmodels}

\subsection{Models}
\label{Models}

In the continued absence of supersymmetry at the LHC, the allowed parameter space in constrained supersymmetric models has been pushed to ever higher mass scales~\cite{Ellis:2015rya,Ellis:2016tjc,Ellis:2017djk,Ellis:2018jyl,Bagnaschi:2018igf,Ellis:2019fwf,Bagnaschi:2015eha,Bechtle:2015nua}. For this reason, also in order to obtain a Higgs boson with mass consistent with the experimental value, $M_h \simeq 125$~GeV~\cite{lhch}, and a sufficiently long proton decay lifetime~\cite{Zyla:2020zbs}, supersymmetric mass scales in the range from 1 to 5 TeV are favored. Then, the requirement that the relic dark matter density agree with Planck results, $\Omega_\chi h^2 \simeq 0.12$~\cite{Planck},
imposes significant constraints on models and their parameters, as do the upper limits on dark matter scattering on matter~\cite{Aprile}.

It was found in the context of no-scale supergravity models that one or both of the MSSM Higgs fields 
must be twisted~\cite{Ellis:2017djk}, i.e., they must acquire masses
different from the universal masses for squarks and sleptons, which vanish at the input scale in no-scale models.~\footnote{Other models with non-universal Higgs masses include the NUHM1~\cite{nuhm1} and NUHM2~\cite{nuhm2,nuhm1},
which have been studied in~\cite{mc9,mc10,GAMBIT}.}
Models with universal input scalar masses suffer from tension between the Higgs mass measurement, proton decay limits and the cosmological relic density.  
With all fields untwisted it was possible to find parameters with a sufficiently large Higgs mass and acceptable relic density 
{\em or} long proton lifetime, but not both~\cite{Ellis:2017djk}.  

As discussed earlier, the trilinear and bilinear soft terms depend on the nature of the twisted Higgs fields, and on the assignments of the
modular weights that appear in the superpotential (see \eq{eq:boundarycond_Mi}).
We outline here the sample model classes
that we use for our analysis, which are adapted from some studied previously in~\cite{Ellis:2017djk}. The
models are distinguished by the parameters $p$ and $q$
that take values 0 or 1 depending on whether the $H$ 
and $\overline{H}$ fields are twisted or not, as well as the choices of modular weights. Here, we take $p=1$, and allow $q$ to take values 0 or 1. 
Once these are specified, the models have six free continuous parameters and one sign: 
\beq
 m_{1/2},\ m_{3/2},\ M_{in},\ \lambda,\ \lambda', \ \tan\beta, \ \rm{sign}(\mu).
\eeq
We recall that in the absence of the dimension-five coupling,
$c_5$, we cannot choose independently the two GUT couplings, $\lambda$ and $\lambda^\prime$.
In this case, typically the colored Higgs mass is low and proton decay is rapid. 
However, when $c_5 \ne 0$, the colored Higgs mass is
sufficiently large for small $\lambda^\prime$. As in previous work \cite{Ellis:2017djk,Ellis:2016tjc,Ellis:2019fwf}, we fix $\lambda^\prime = 10^{-5}$ in all of the models considered here in order to ensure a sufficiently large colored Higgs mass, $M_{H_C}$ and hence a sufficiently long nucleon lifetime. 
The lifetime for the dominant proton decay mode, $\TPDK$, 
increases with $\lambda$, for which we adopt either
$\lambda = 0.6$ or $\lambda = 1.0$. 
We take $\tan \beta = 7$ in all models except M1 where $\tan \beta = 6$ \footnote{In model M1, with $\tan \beta = 7$, the proton decay limit imposes a more stringent limit on $m_{1/2}$ such that $m_h = 125$ GeV is excluded along the relic density strip. Therefore we take $\tan \beta = 6$ for this case.}, 
and we choose
$\rm{sign}(\mu)>0$ in all models. We
consider two values of $\Mi$: $\Mi=10^{16.5}$~GeV, for which there is little RG running above $\Mg$, 
and $\Mi=10^{18}$~GeV, for which the RG running is more important.

We illustrate the effects of the choice of flavor structure using
a subset of the models considered previously in \cite{Ellis:2017djk}.
 As noted above, because of the 
restrictive nature of the untwisted no-scale boundary conditions,
we require that either one or both
of the Higgs five-plets are twisted in order to obtain simultaneously
the correct relic density, $\Omega_\chi h^2 = 0.12$ \cite{Planck}, 
and Higgs mass, $M_h = 125$ GeV \cite{lhch}, as well as $\TPDK$ consistent with the lower limit given in~\cite{Miura:2016krn}.
In Models M1 - M4 below, both Higgs multiplets are twisted,
whereas for models M5 and M6, 
only $H$ is twisted. The dark matter, $M_h$ and $\TPDK$
constraints can all be reconciled in these models. 
Ref.~\cite{Ellis:2017djk} also considered models in which only ${\overline H}$ is twisted. However, we find using {\tt FeynHiggs~2.16.0} \cite{FeynHiggs} a drop in the calculated Higgs mass of $\sim 2$ GeV, relative to previous versions, making it difficult to reconcile an acceptable relic density with $M_h \simeq 125$ GeV
and the $\TPDK$ constraint, and do not consider further
such models.
The models considered here are as follows:

\begin{enumerate}[label=(M{\arabic*})] 
\item In this model, we set $\Mi=10^{16.5}$ GeV, $\tan \beta = 6$, $\lambda = 0.6$, with $p=q=1$, and we take all modular weights $\alpha_F=\beta_S=0$. In this case,  $\aten=\afiv=m_{3/2}$,  $A_{\lambda}=2 m_{3/2}$, $A_{\lambda'}=0$, $B_H=2 m_{3/2}$, $B_{\Sigma}=0$. This model is similar to that considered in the left panel of Fig. 3 in \cite{Ellis:2017djk}.

\item  In this model,  we take $\Mi=10^{16.5}$ GeV, $\tan \beta = 7$, $\lambda = 0.6$, and $p=q=1$. However, in this case 
we fix $\alpten=\alpfiv=1$, $\alplam=2$,  $\alplamp=0$, $\betH$=2, $\betSg=0$, 
corresponding to $\aten=\afiv=0$,  $A_{\lambda}=A_{\lambda'}=0$, $B_H=B_{\Sigma}=0$. This is similar to the model considered in left panel of Fig. 4 of~\cite{Ellis:2017djk}.  

\item  In this model, we consider $\Mi=10^{18}$ GeV, $\tan \beta = 7$, $\lambda = 0.6$. We again take $p=q=1$, with the same modular weights as adopted in M2.  This model is similar to that considered in the right panel of Fig. 4 of~\cite{Ellis:2017djk}.  

\item  In this model, we consider $\Mi=10^{18}$ GeV, $\tan \beta = 7$, $\lambda = 1$. We again take $p=q=1$, with the same modular weights as adopted in M2.  This model is the same as that considered in the right panel of Fig. 4 of~\cite{Ellis:2017djk}.

\item In this case only $H$ is twisted, so that $p=1$ and $q=0$.  Once again, we take $\Mi = 10^{18}$ GeV, $\tan \beta = 7$, and $\lambda = 1$.  The modular weights are $\alpten=1$, $\alpfiv=0$, $\alplam=1$,  $\alplamp=0$, $\betH$=1, $\betSg=0$, which gives $\aten=\afiv=0$,  $A_{\lambda}=A_{\lambda'}=0$, $B_H=B_{\Sigma}=0$.
This model was considered in the left panel of Fig.~7 of~\cite{Ellis:2017djk}.

\item As in (M5), but in this case all modular weights are set to zero: $\alpha=\beta=0$ giving $\aten=m_{3/2}$, $\afiv=0$,  $A_{\lambda}=m_{3/2}$, so that $A_{\lambda'}=0$, $B_H=m_{3/2}$ and $B_{\Sigma}=0$. This model was studied in the right panel of Fig.~7 in~\cite{Ellis:2017djk}.

\end{enumerate}

For each of the models M1 - M6, we compute the proton decay lifetime, 
$\Bmueg$, and the induced electron EDM, comparing the flavor choices {\bf A} and {\bf B}, 
and also comparing the predictions of the {\bf NF} scenario for the proton lifetime.

\subsection{Both Higgs fields in twisted sectors}

Since both Higgs five-plets are twisted in models M1 and M2,
we must use $ p = q = 1$ in Eq. (\ref{eq:boundarycond_Mi}), yielding $m_H = m_{\bar H} = m_1 = m_2 = m_{3/2}$
at $\Mi$, whereas all the other scalar masses vanish there. Once the modular weights $\alpha$ and $\beta$
appearing in Eq. (\ref{eq:boundarycond_Mi}) are specified, all of the bi- and tri-linear terms are fixed relative to $m_{3/2}$, 
so the models are fully specified. 
In model M1, we take all modular weights to vanish, yielding the non-zero $A$-terms
$A_{\bf 10} = A_{\bf {\bar 5}}= m_{3/2}$ and $A_\lambda = 2m_{3/2}$, 
as well as a non-zero $B$-term for $B_H = 2 m_{3/2}$. In models M2 - M4, 
we take all $A$- and $B$-terms to vanish.  
In what follows, we display our results in
($m_{1/2}, m_1$) planes.

\begin{figure}[!ht]
\centering
\includegraphics[width=8cm]{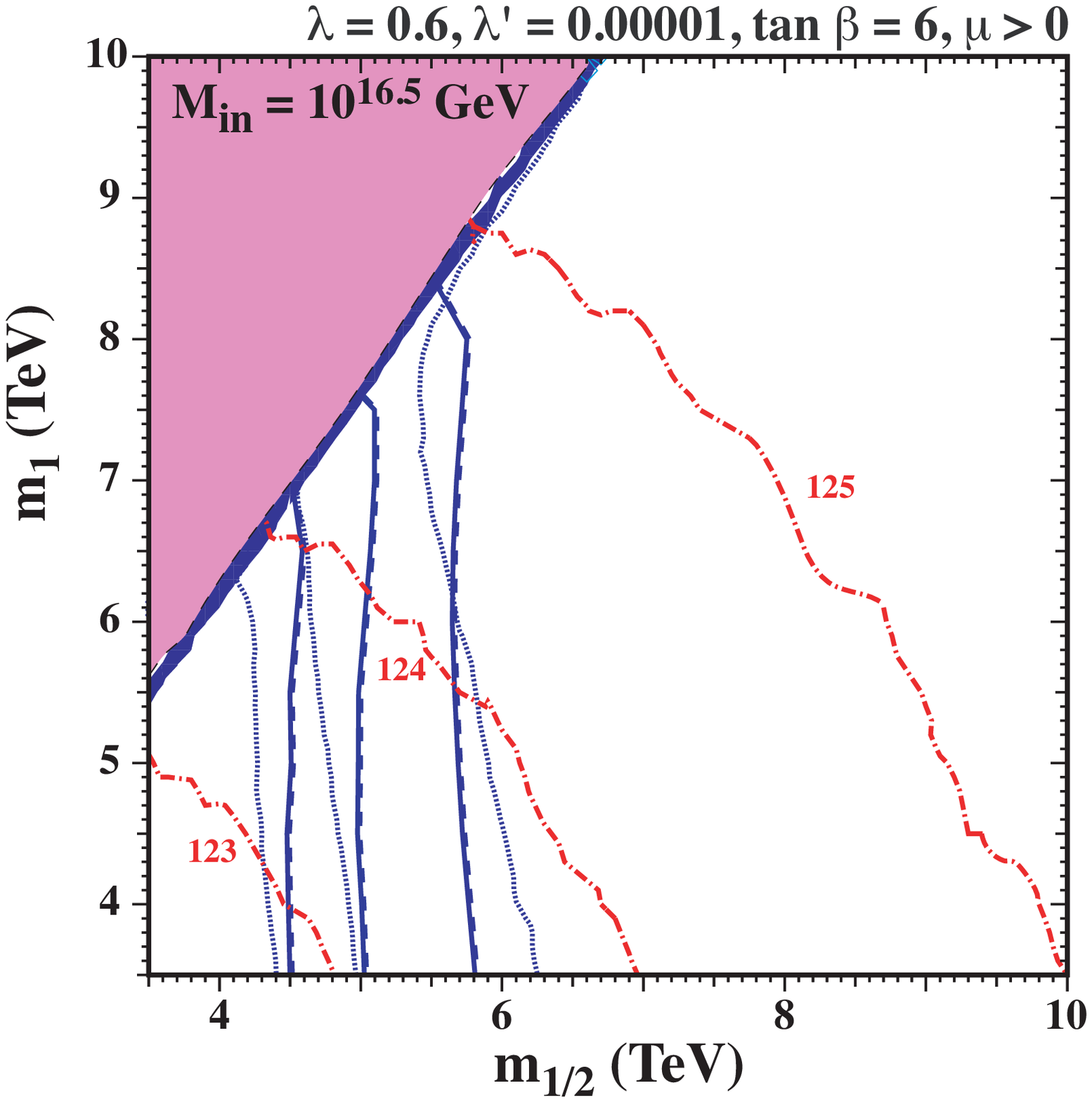}
\includegraphics[width=8cm]{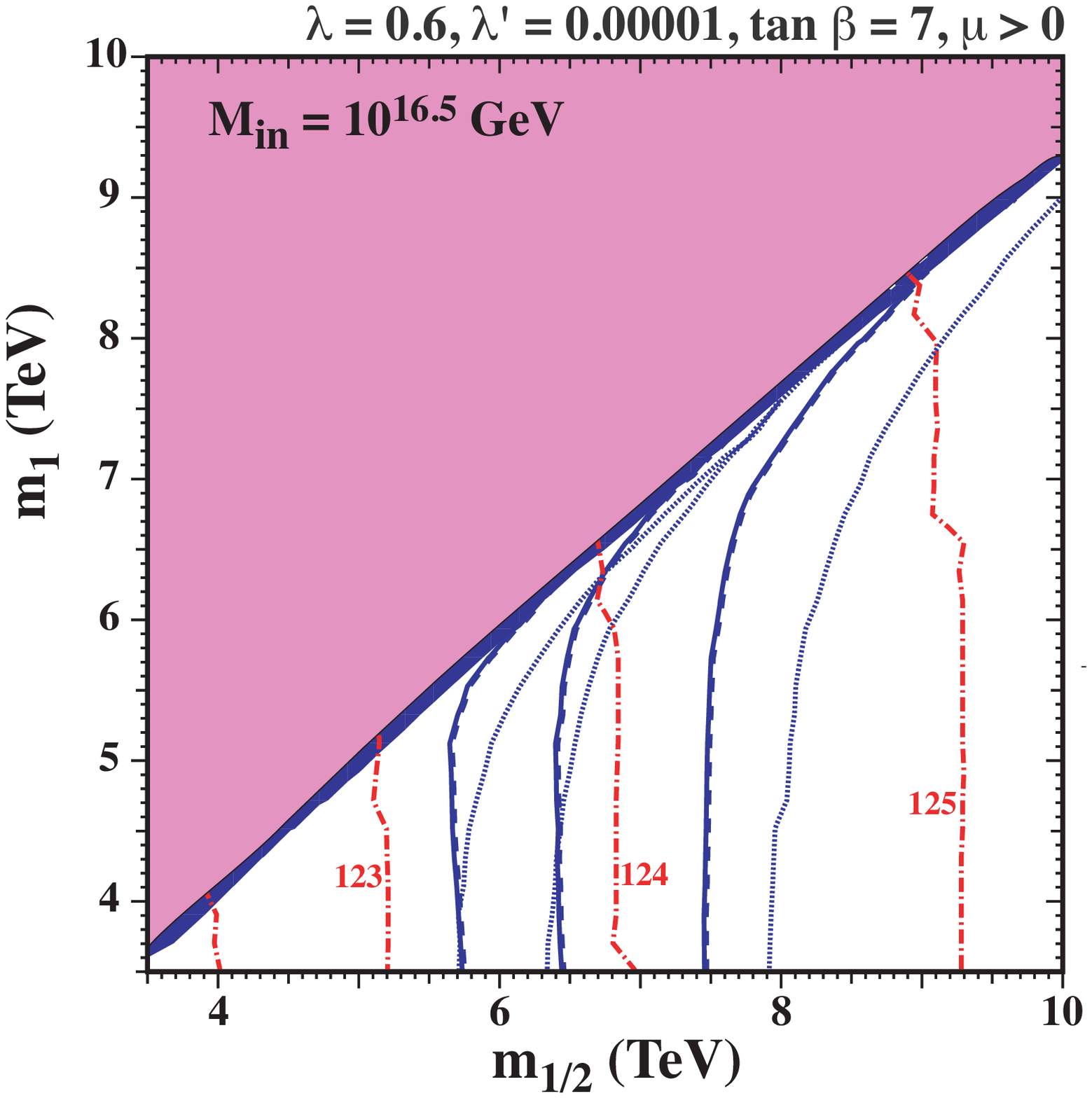}\\
\includegraphics[width=8cm]{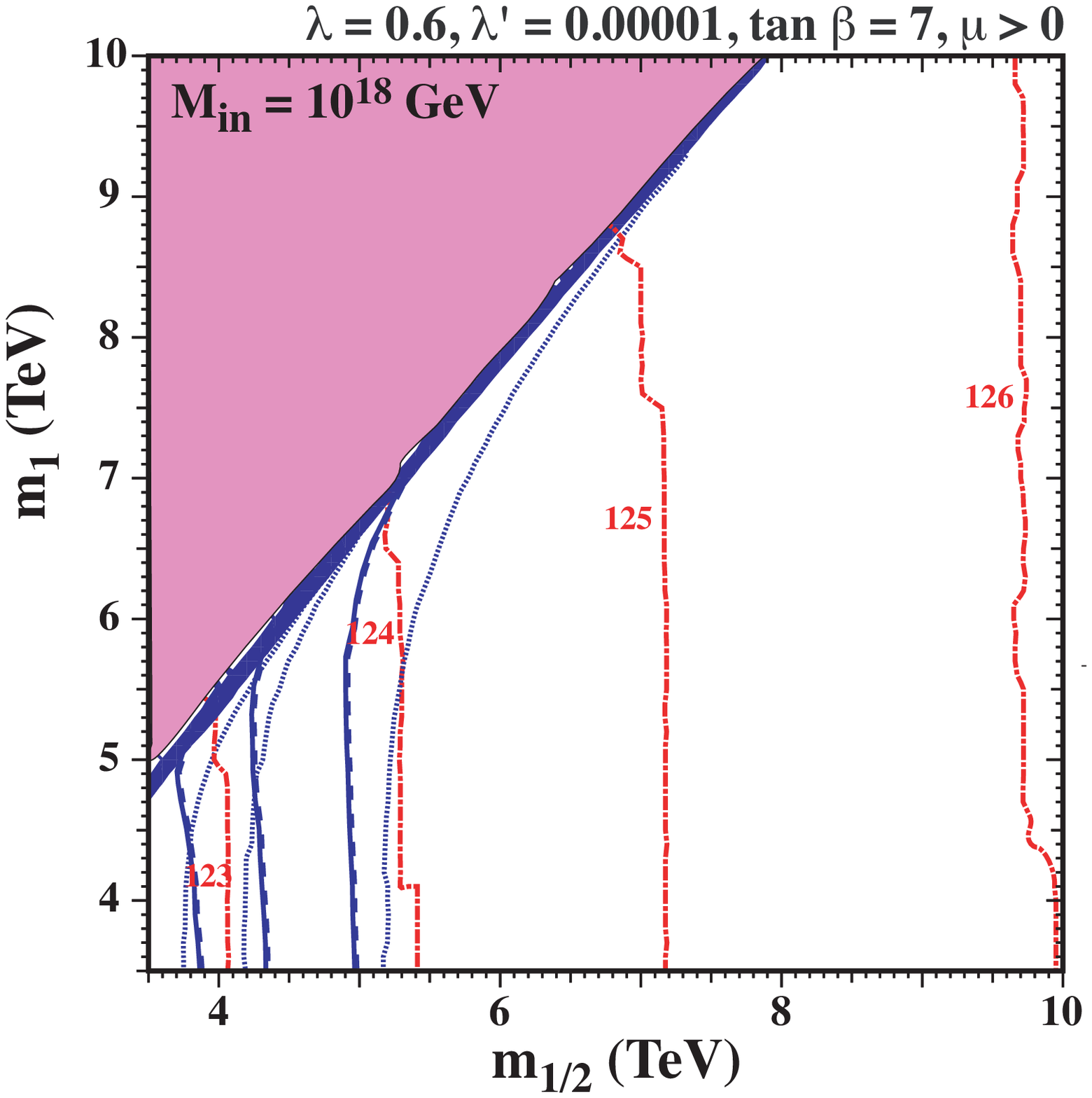}
\includegraphics[width=8cm]{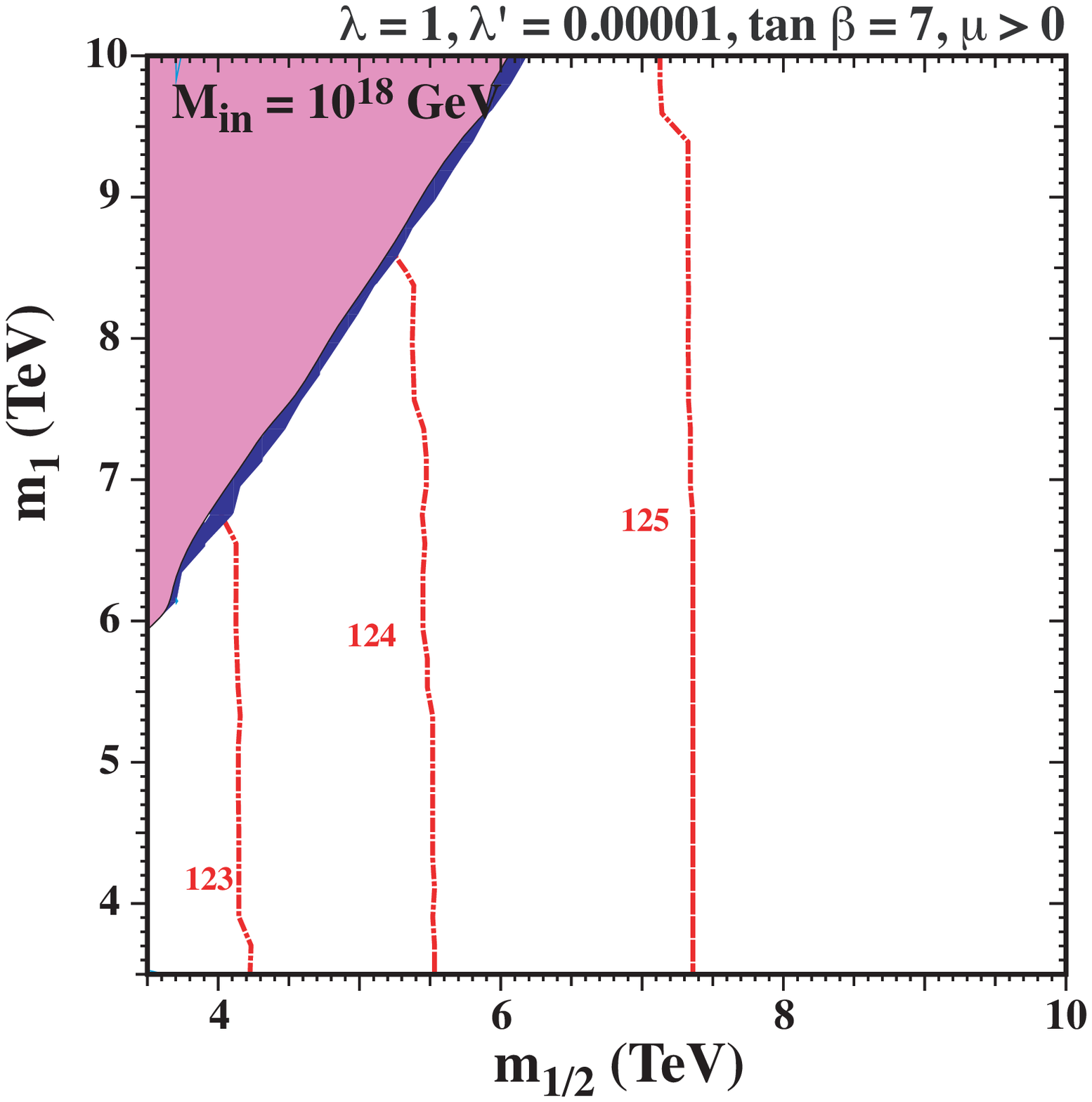}
\caption{\it Examples of ($m_{1/2}, m_1$) planes for Models M1 
with $\Mi = 10^{16.5}$~GeV, $\lambda = 0.6$, and $\tan \beta = 6$ (upper left),
M2 with $\Mi = 10^{16.5}$~GeV, $\lambda = 0.6$, and $\tan \beta = 7$ (upper right),
M3 with $\Mi = 10^{18}$~GeV, $\lambda = 0.6$, and $\tan \beta = 7$ (lower left) and
M4 with $\Mi = 10^{18}$~GeV, $\lambda = 1$, and $\tan \beta = 7$ (lower right).
We assume $\mu > 0$ in all panels, and the values of $\Mi$, $\tan \beta$, $\lambda$ and $\lambda'$ are
indicated in the legends. In the regions shaded pink there is no EWSB, and in the blue strips below these regions the relic density is in the range $0.06 < \Omega_\chi h^2 < 0.2$. The red dot-dashed curves are Higgs mass contours, with the masses labelled in GeV.  For each flavor choice, 
there are three contours for the proton lifetime,
$\TPDK$, corresponding to the central values and 
$1 \sigma$ variations in the hadronic matrix elements. The predictions of flavor
choices {\bf A} and {\bf B} are shown as the solid and dashed blue curves, respectively, and 
those of the {\bf NF} choice are shown as the blue dotted curves. 
\label{fig:M1_PD_YkA} }
\end{figure}

\paragraph{Model M1} In this model we fix $\Mi = 10^{16.5}$~GeV,
so there is little super-GUT running between $\Mi$ and $\Mg$, $\tan \beta = 6$ and $\mu > 0$. 
The chosen values of the couplings of the adjoint Higgs supermultiplets are $\lambda = 0.6$ and $\lambda' = 0.00001$.
We show in the upper left panel of Fig.~\ref{fig:M1_PD_YkA} the ($m_{1/2}, m_1$) plane for this model,
where we recall that $m_1 = m_{3/2}$ in this model. There is no EW symmetry breaking (EWSB)
in the triangular region shaded pink in the upper left corner, 
i.e., the solution for the MSSM $\mu$ parameter has $\mu^2 < 0$. 
The dark blue shaded strip just below the no-EWSB region corresponds to the focus point \cite{fp},
with the relic density taking values in the range $0.06 < \Omega_\chi h^2 < 0.2$. This is wider than the range determined by Planck \cite{Planck}, but we show an extended range in order to make it more visible on the scale of this figure. The red dot-dashed curves show contours of the Higgs mass as determined by {\tt FeynHiggs~2.16.0} \cite{FeynHiggs}.

\paragraph{Model M2} We show in the upper right panel of \Figref{fig:M1_PD_YkA}
the ($m_{1/2}, m_1$) plane for this model assuming $\Mi = 10^{16.5}$~GeV and $\mu > 0$,
and the same values of $\lambda$ and $\lambda'$ as in model M1, but $\tan \beta = 7$,
using the same shading and line conventions as in the left panel. As one might expect, 
since $A_0 = 0$ in this model, the region where there is no EWSB reaches down to 
lower values of $m_1$.~\footnote{We recall that  
the focus-point dark matter region disappears
for sufficiently large $A_0$ in the CMSSM.} As a result, the relic density takes
acceptable values at somewhat lower values of $m_1$ as well. \\

\paragraph{Model M3}
We exemplify the
importance of RG running between $\Mi$ and $\Mg$ in the lower left panel of \Figref{fig:M1_PD_YkA},
where we choose $\Mi = 10^{18}$~GeV  and  $\lambda = 0.6$.  Raising the value of $\Mi$ pushes the no-EWSB
boundary and the dark matter strip 
back to higher values of $m_1$.\\

\paragraph{Model M4}
We exemplify the
role of $\lambda$ in the lower right panel of \Figref{fig:M1_PD_YkA},
where we choose  $\lambda = 1$.  Raising the value of $\lambda$ also pushes the no-EWSB
boundary and the dark matter strip 
to higher values of $m_1$.\\

{\it Proton lifetime:} Also shown in Fig.~\ref{fig:M1_PD_YkA} are 
predictions for the proton lifetime, $\TPDK$.
For each case considered, we show 3 sets of 3 contours each,
corresponding to the current lower limit on $\TPDK$. 
The central contour in each set uses the central values 
for the hadronic matrix elements given in
\eq{sec:hadmatxunc}, and the outer contours to either side 
correspond to the $\pm 1 \sigma$ variations in these matrix elements indicated there, 
keeping the masses  $m_s(2 \ \rm{GeV})$ and $m_c(2 \ \rm{GeV})$
fixed at their central values. 
The sets of solid and dashed blue contours correspond to 
model choices {\bf A} and {\bf B}, respectively,
and we see that the choice between these flavor embeddings has very little effect the proton lifetime.
Along the dark matter strip, the lower limit on $\TPDK$ 
(assuming central values of the hadronic matrix elements)
corresponds to $m_{1/2} \gtrsim 5 (7) (4.5)$~TeV in model M1 (M2 with $\Mi = 10^{16.5}$~GeV)
(M3 with $\Mi = 10^{18}$~GeV and $\lambda = 0.6$), whereas the lower limit on $m_{1/2}$ from the 
proton lifetime is below 3.5~TeV for model M4 with $\Mi = 10^{18}$~GeV and $\lambda = 1$.

The predictions for $\TPDK$ with the {\bf A}
and {\bf B} flavor choices differ from those with the {\bf NF} flavor choice, 
which yield the blue dotted contours. Specifically, in the case of model M1 (upper left panel),
along the blue dark matter strip and
assuming the central values of the hadronic
matrix elements, the lower limit on $m_1 \ (m_{1/2})$ is {\it stronger} 
by about 700 (500)~GeV for model choices {\bf A} and {\bf B}
than for the {\bf NF} choice. On the other hand, in model M2 with $\Mi = 10^{16.5}$~GeV
(upper right panel) and $\Mi = 10^{18}$~GeV, $\lambda = 0.6$ (lower left panel), 
the lower limits on $m_1$ and $m_{1/2}$ are {\it weaker} 
by about 1~TeV for model choices {\bf A} and {\bf B} than for the {\bf NF} choice.
In model M3, the limits for choices {\bf A} and {\bf B} are about 900 (700) GeV {\it weaker} than choice {\bf NF}.
Finally, in model M4 with $\Mi = 10^{18}$~GeV and $\lambda = 1$ (lower right panel),
$\TPDK$ exceeds the current lower limit everywhere in the regions of the ($m_{1/2}, m_1$) planes displayed.\\

\begin{figure}[!ht]
\centering
\includegraphics[width=8cm]{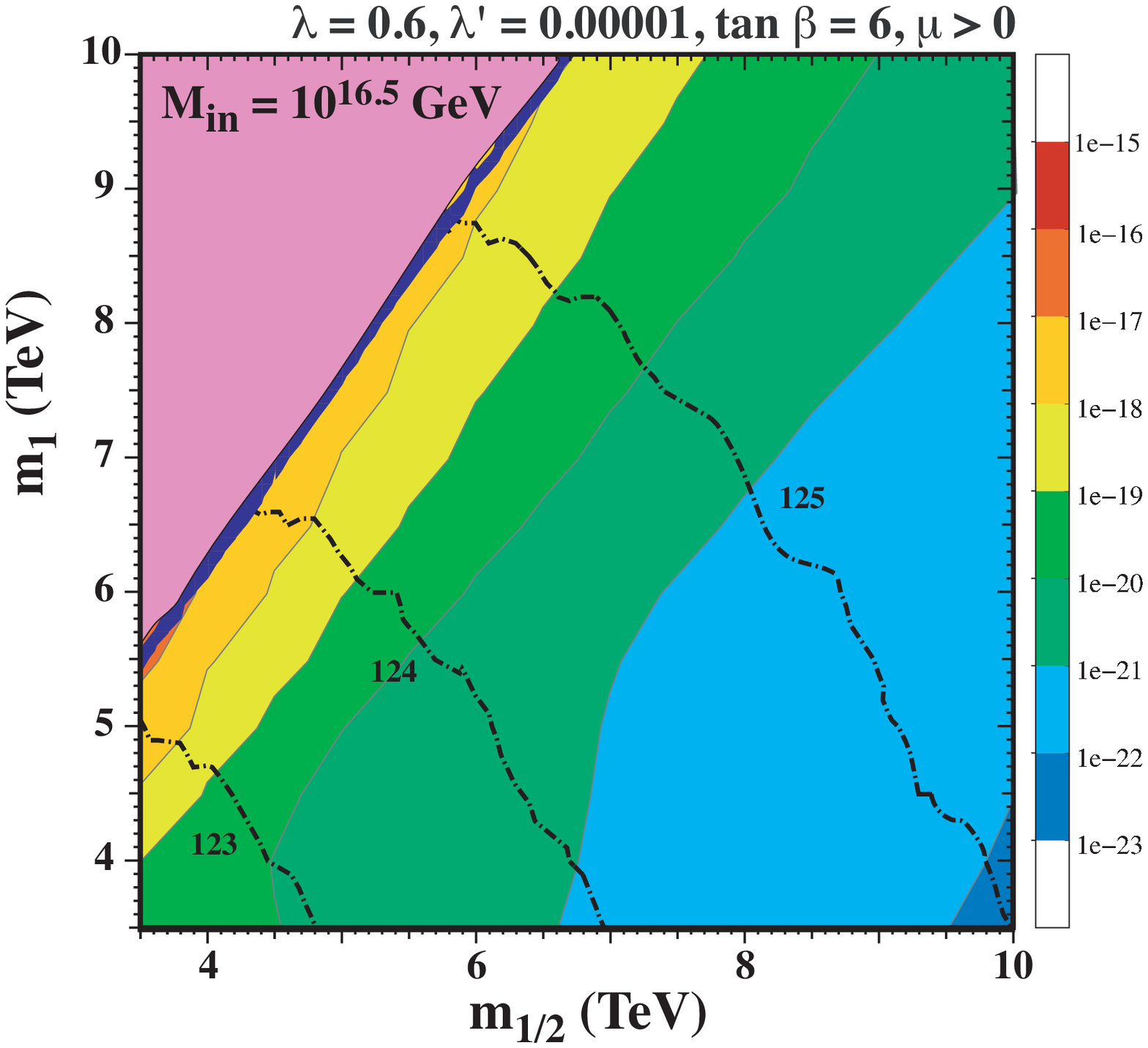}
\includegraphics[width=8cm]{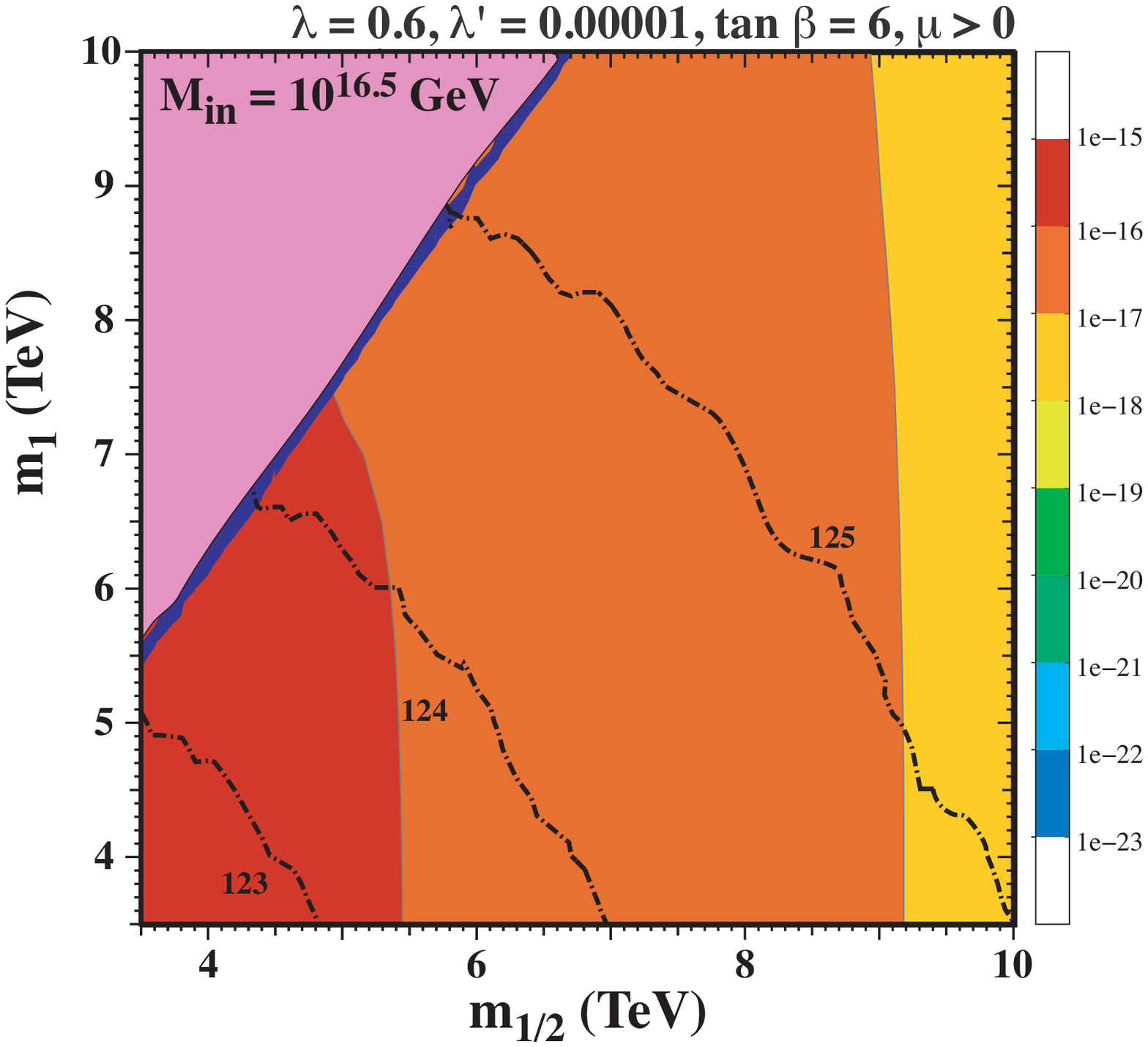}\\
\includegraphics[width=8cm]{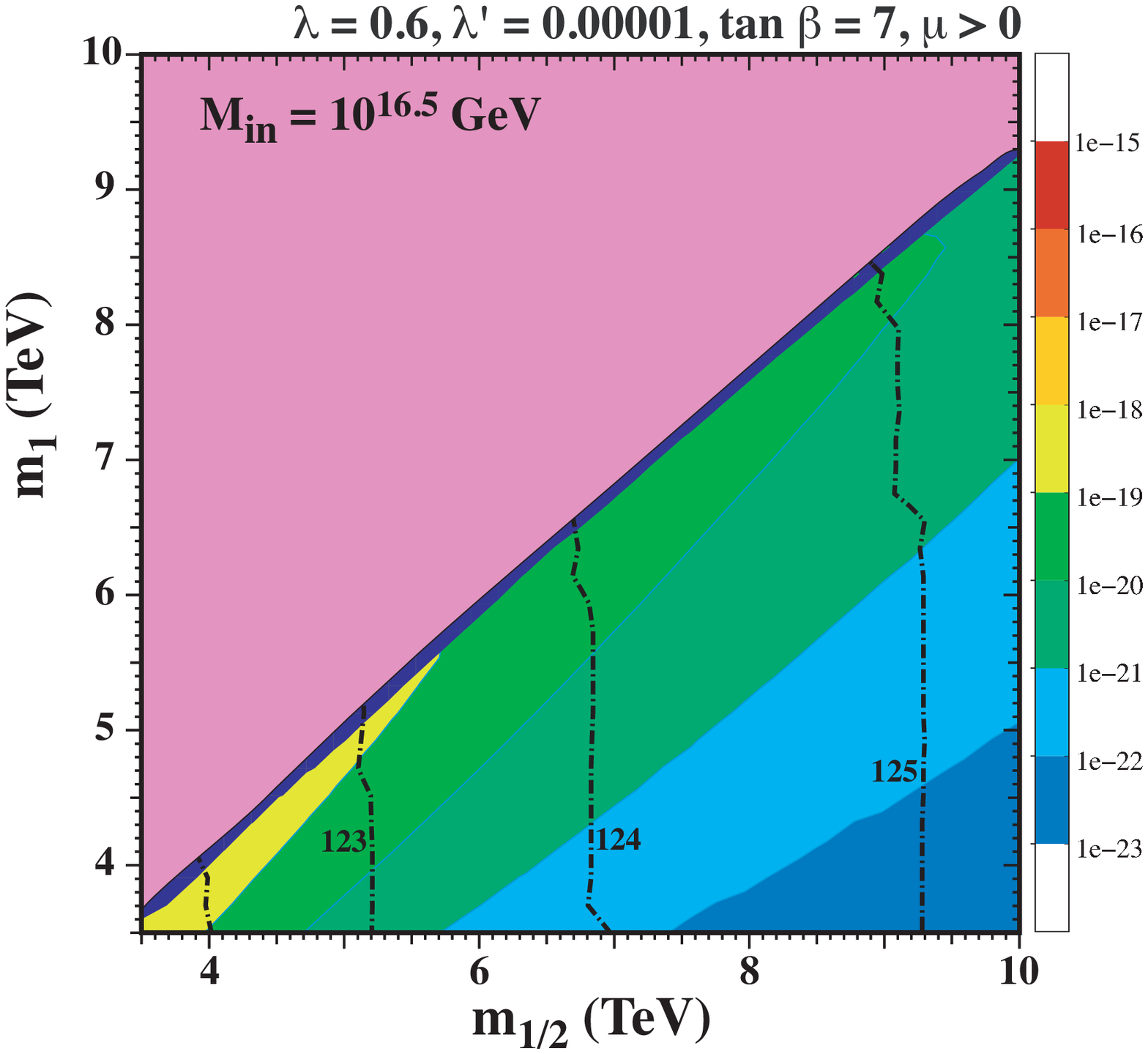}
\includegraphics[width=8cm]{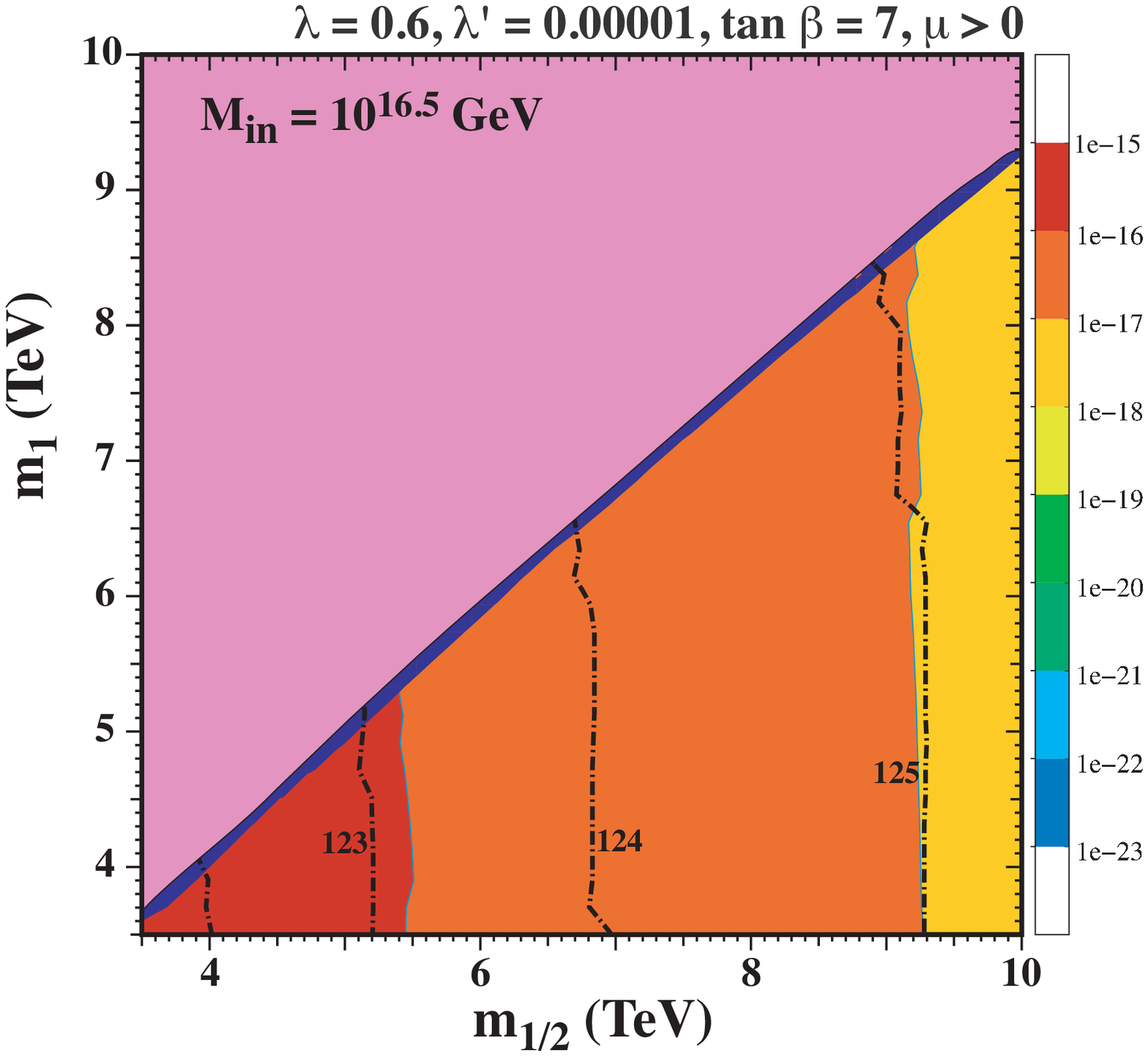}
\caption{\it As in Fig.~\ref{fig:M1_PD_YkA}, showing values of $\Bmueg$ for the flavor
choices {\bf A} (left) and {\bf B} (right) in model M1 with $\Mi = 10^{16.5}$~GeV
and $\tan \beta = 6$ (upper panels) and in model M2
with $\Mi = 10^{16.5}$~GeV, $\tan \beta = 7$ and the indicated values
of $\lambda$ and $\lambda'$ (lower panels). 
The color-coding for $\Bmueg$
is indicated in the bars beside the panels.}
\label{fig:M1_MUEG}
\end{figure}

{\it Flavor violation:} We show in the upper panels of Fig.~\ref{fig:M1_MUEG} ($m_{1/2}, m_1$) planes with values of
BR($\mu \to e \gamma$) for model M1, which has $\Mi = 10^{16.5}$~GeV
and $\tan \beta = 6$, and the flavor choices {\bf A} (left)  and  {\bf B} (right).   
As  in  Fig.~\ref{fig:M1_PD_YkA},  the  region  where  there  is  no EWSB is shaded pink and 
$0.06 < \Omega_\chi h^2 < 0.2$ in the dark blue strip. The contours where the Higgs mass is 123, 124 and 125~GeV
are shown here as black dot-dashed lines. The lower panels of Fig.~\ref{fig:M1_MUEG} are the
corresponding ($m_{1/2}, m_1$) planes for model M2 with $\Mi = 10^{16.5}$~GeV, $\tan \beta = 7$.
In all the panels $\lambda = 0.6$ and $\lambda' = 0.00001$.

For choice {\bf A} $\Bmueg$ is always below the current experimental upper
limit of $4.2\times 10^{-13}$~\cite{TheMEG:2016wtm}. In the region of greatest interest
along the blue relic density strip, the branching ratio
may exceed $10^{-17}$, but a small portion at low ($m_{1/2}, m_1$) where it reaches $10^{-16}$ is excluded by
the proton decay limit. Moreover, $\Bmueg$ decreases significantly below the strip and at larger masses. 
The low values for the branching ratio arise primarily from the choice of an embedding in 
which $h_E$ is diagonal at the EW scale (see Eq.~(\ref{eq:YukDiagMEW})). 

In contrast, for choice {\bf B} the lepton Yukawa couplings are not diagonal at the EW scale
and we see in the right panel of Fig.~\ref{fig:M1_MUEG} that $\Bmueg$ is significantly larger,
with values above $10^{-16}$ becoming consistent with $\TPDK$, $M_h$ and the relic dark matter density. 
Indeed, $\Bmueg$ is larger than $10^{-18}$ even at very large gaugino masses $> 10$~TeV.
We note that in this case that the dependence of $\Bmueg$ on $m_{1/2}$ is much stronger than that
on $m_1$. Nevertheless, there is a stretch of the focus-point strip with $4.5 \, {\rm TeV} \lesssim m_{1/2} \lesssim 6 \, {\rm TeV}$, compatible with the present limit on $\TPDK$ and the Higgs mass, where $\mu \to e$ conversion may be accessible to the PRISM experiment~\cite{Strategy:2019vxc}.~\footnote{We note that our analysis ignores the possible effects of neutrino couplings, which are not constrained in the SU(5) GUT. There is freedom in selecting how to incorporate them in the SU(5) theory, and they could potentially increase $\Bmueg$.}

In order to understand this behavior, we analyze a benchmark point in model M1 lying on
the relic density strip with $m_{1/2}=6000$ GeV, which corresponds to  $m_1=9070$~GeV, 
and a Higgs mass of 
$M_h=125.2 \pm  0.9$~GeV according to {\tt FeynHiggs~2.16.0}. We show in Table~\ref{numbers12}  the relevant mass parameter values in model M1 that are used to extract the approximate
values for $\TPDK$, $\Bmueg$ and the electron EDM.
As one can see, there is essentially no difference in the selectron masses between cases {\bf{A}} and {\bf{B}} and only a 2\% difference in the the smuon masses. As we discussed earlier, $(m^2_{L})_{12} \ll (m^2_{E})_{12}$ so that $a_{\mu e \gamma L}$ is suppressed.
We also see that in all models (M1-M4), $(m^2_{E})_{12}$ is within a factor of two and $(a_E)_{22}$ is nearly identical between cases {\bf{A}} and {\bf{B}}.  The difference seen in Fig.~\ref{fig:M1_MUEG} between
cases {\bf{A}} and {\bf{B}} is a result of 
$a^{(IIc)}_{\mu e \gamma R}$ (see Eq.~(\ref{eq:amuegRIIc})),
which is proportional to $(a_E)_{21}$ and is more than a factor 
of $10^3$ times larger in case {\bf B} due to the choice of 
$U^E_R = V^*_{\rm CKM}$ as opposed to $U^E_R = ${\bf 1} in case {\bf A}.
We note that the predictions for $\TPDK$ are similar for flavor choices {\bf A} and {\bf B},
beyond the current limit but well within the projected reach of Hyper-Kamiokande~\cite{HK}.
While the predictions for $\Bmueg$ and the electron EDM differ for flavor choices {\bf A} and {\bf B},
they lie significantly below the current limits and prospective experimental sensitivities.

\begin{table}
\begin{center}
\label{tbl:MassesFV}
\begin{tabular}{|r|c|c|}
\hline 
\multicolumn{3}{|c|}{{\bf M1}: $\Mi= 10^{16.5}$~GeV, $\lambda=0.6$}\\
\hline
Parameter & {\bf A} & {\bf B}\\
\hline
$\mu$ [GeV] & \multicolumn{2}{|c|}{1022}\\
$M_1$  [GeV]  & \multicolumn{2}{|c|}{2010} \\
$M_2$ [GeV] &  \multicolumn{2}{|c|}{3983}\\
\hline
$m_{\tilde e_L}$ [GeV]  & 3493&  3493 \\
$m_{\tilde e_R}$ [GeV]  & 2866 & 2866 \\
$m_{\tilde \mu_{L}}$ [GeV]  & 3494  & 3554\\
$m_{\tilde \mu_{R}}$ [GeV]& 2829& 2873\\
$(m^2_{E})_{12}$ [GeV]$^2$    &   $598 \, e^{i \, 0.36}$ &  $832 \, \e^{-i \, 0.40}$  \\ 
$(m^2_{E})_{31}$  [GeV]$^2$ &  $\phNF{1.7\times 10^{4}}{2.8}$   &    $\phNF{2.3\times 10^{4}}{3.0}$  \\
$(m^2_{L})_{12}$ [GeV]$^2$    &  $1.7 \, \e^{-i \, 0.87}$   &  $1.7 \, \e^{-i \, 0.87}$  \\ 
$(m^2_{L})_{13}$  [GeV]$^2$ &   $\phPF{39}{1.9}$   &  $\phPF{39}{2.4}$ \\
$(a_E)_{11}$ [GeV]  &   $\phNF{0.42}{6.4 \times 10^{-6}}$ &
$\phNF{0.42}{6.9 \times 10^{-10}}$\\
$(a_E)_{21}$ [GeV]  &  $4.7 \times 10^{-5} \   \e^{-i \, 0.21}$ & $\phNF{0.04}{0.002}$\\ 
$(a_E)_{22}$ [GeV]  &   16   $\, \e^{-i \, 1.3 \times 10^{-5}}$  &  16   $\, \e^{-i \, 1.4 \times 10^{-6}}$\\ 
$(a_E)_{33}$  [GeV] & $\phNF{640}{3.1 \times 10^{-10}}$ & $\phPF{640}{1.1 \times 10^{-10}}$  \\
\hline
$\TPDK$ [yrs] & $8.9 \times 10^{33} $ &   $8.9\times 10^{33} $  \\
$\Bmueg$ & $1.4 \times 10^{-18}$  &  $3.9\times 10^{-17} $ \\
$d_e$ [e.cm] & $4.0 \times 10^{-33}$ & $-5.9\times 10^{-34}$  \\

\hline \hline
\multicolumn{3}{|c|}{{\bf M2}:  $\Mi= 10^{16.5}$~GeV, $\lambda=0.6$}\\
\hline
Parameter & {\bf A} & {\bf B}\\
\hline
$\mu$  [GeV]& \multicolumn{2}{|c|}{  1016}\\
$M_1$   [GeV]& \multicolumn{2}{|c|}{1993 }\\ 
$M_2$  [GeV]&\multicolumn{2}{|c|}{3964}\\
\hline
$m_{\tilde e_L}$  [GeV]& 3504& 3504  \\
$m_{\tilde e_R}$ [GeV] & 2831 & 2831 \\
$m_{\tilde \mu_{L}}$ [GeV] & 3504 & 3571\\
$m_{\tilde \mu_{R}}$  [GeV] &  2829 &  2854\\
$(m^2_{E})_{12}$  [GeV]$^2$ & $\phPF{140}{0.36}$  &  $\phNF{255}{0.40}$ \\
$(m^2_{E})_{31}$  [GeV]$^2$ & $\phNF{3.9\times 10^{3}}{2.8}$ &   $\phNF{6.1\times 10^{3}}{3.1}$  \\
$(m^2_{L})_{12}$  [GeV]$^2$ & $\phPF{0.58}{2.28}$  & $\phPF{0.58}{2.28}$ \\
$(m^2_{L})_{13}$  [GeV]$^2$ & $\phPF{13}{1.9}$  &  $\phPF{13}{2.3}$    \\
$(a_E)_{11}$  [GeV] & $\phNF{0.094}{3.7\times 10^{-6}}$  & $\phNF{0.094}{4.8\times 10^{-10}}$ \\
$(a_E)_{21}$  [GeV] &  4.2 $\times10^{-6}   \, \e^{-i \,   0.16} $ & $0.048   \, \e^{-i \, 4.0 \times 10^{-4}} $\\
$(a_E)_{22}$ [GeV]  &   $\phPF{11}{7.8  \times 10^{-6}} $ &   $\phNF{11}{3.3  \times 10^{-6}} $   \\
$(a_E)_{33}$  [GeV] & $\phNF{0.024}{2.3\times10^{-10}}$   &   $\phNF{0.024}{7.8\times10^{-6}}$   \\
\hline
$\TPDK$ [yrs] &  $5.0 \times 10^{33}$  &   $5.0 \times 10^{33}$ \\
$\Bmueg$ &  $7.9 \times 10^{-20}$ &   $6.2   \times 10^{-17}$ \\
$d_e$ [e.cm] &   $4.8  \times 10^{-34}$ &   $2.2 \times 10^{-35}$ \\
\hline
\end{tabular}
\end{center}
\caption{\it  Benchmark points in models M1 and M2 with $m_{1/2}=6000$~GeV. For M1,  $m_1=9070$~GeV, and for M2,  $m_1=5950$~GeV. We list values of the parameters relevant for $\Bmueg$ and the electron EDM
obtained with flavor choices {\bf{A}} and {\bf{B}}, as well as the corresponding
predictions for $\TPDK, \Bmueg$ and the electron EDM.  \label{numbers12}}
\end{table}

Similar behavior is found for model M2, shown in the lower two
panels of Fig.~\ref{fig:M1_MUEG}. Along the relic density strip (now at lower $m_1$ relative to M1), the branching ratio ranges from $10^{-20}$ to a few $\times 10^{-19}$ for flavor choice {\bf A}.
As we saw for model M1, the branching ratio is considerably larger
for flavor choice {\bf B} and may be as large $\mathcal{O}(10^{-16})$ while remaining consistent with proton decay limits. A representative benchmark point along the relic density strip at $m_{1/2} = 6000$ GeV for M2 is also given in Table \ref{numbers12}, with $m_1 = 5950$ GeV. For this point $M_h = 123.6 \pm 0.7$ GeV, which is consistent within the uncertainties with the experimental value.
In this case, we again see that the dominant difference in the branching ratio between
choices {\bf A} and {\bf B} is due to $(a_E)_{12}$. 
The predictions for $\TPDK$ are again similar for flavor choices {\bf A} and {\bf B},
and are consistent with the current limit within the current matrix element uncertainties.
As in the case of model M1, the predictions for $\Bmueg$ and the electron EDM 
again differ for flavor choices {\bf A} and {\bf B},
while lying significantly below the current limits.

\begin{figure}[!ht]
\centering
\includegraphics[width=8cm]{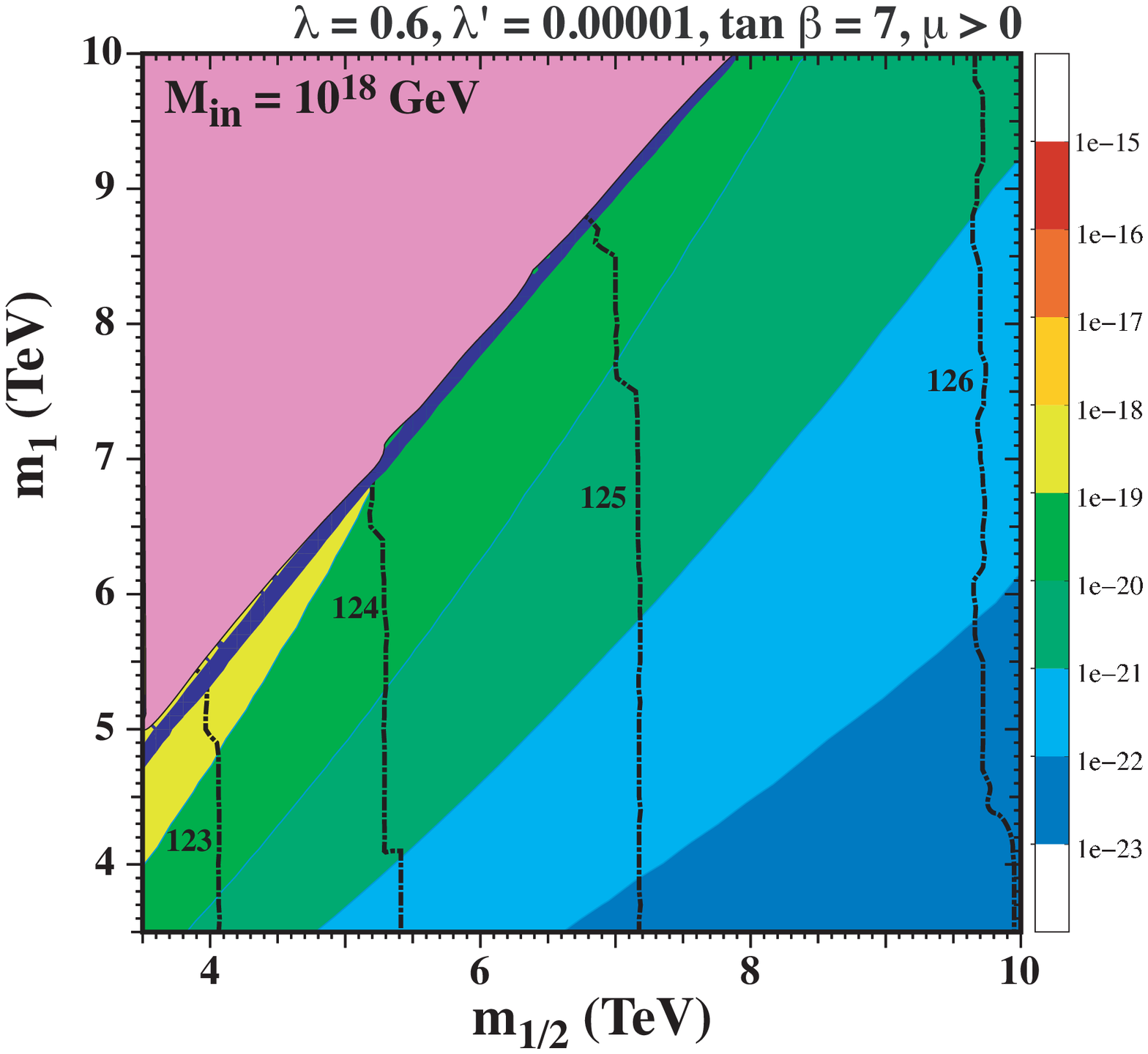}
\includegraphics[width=8cm]{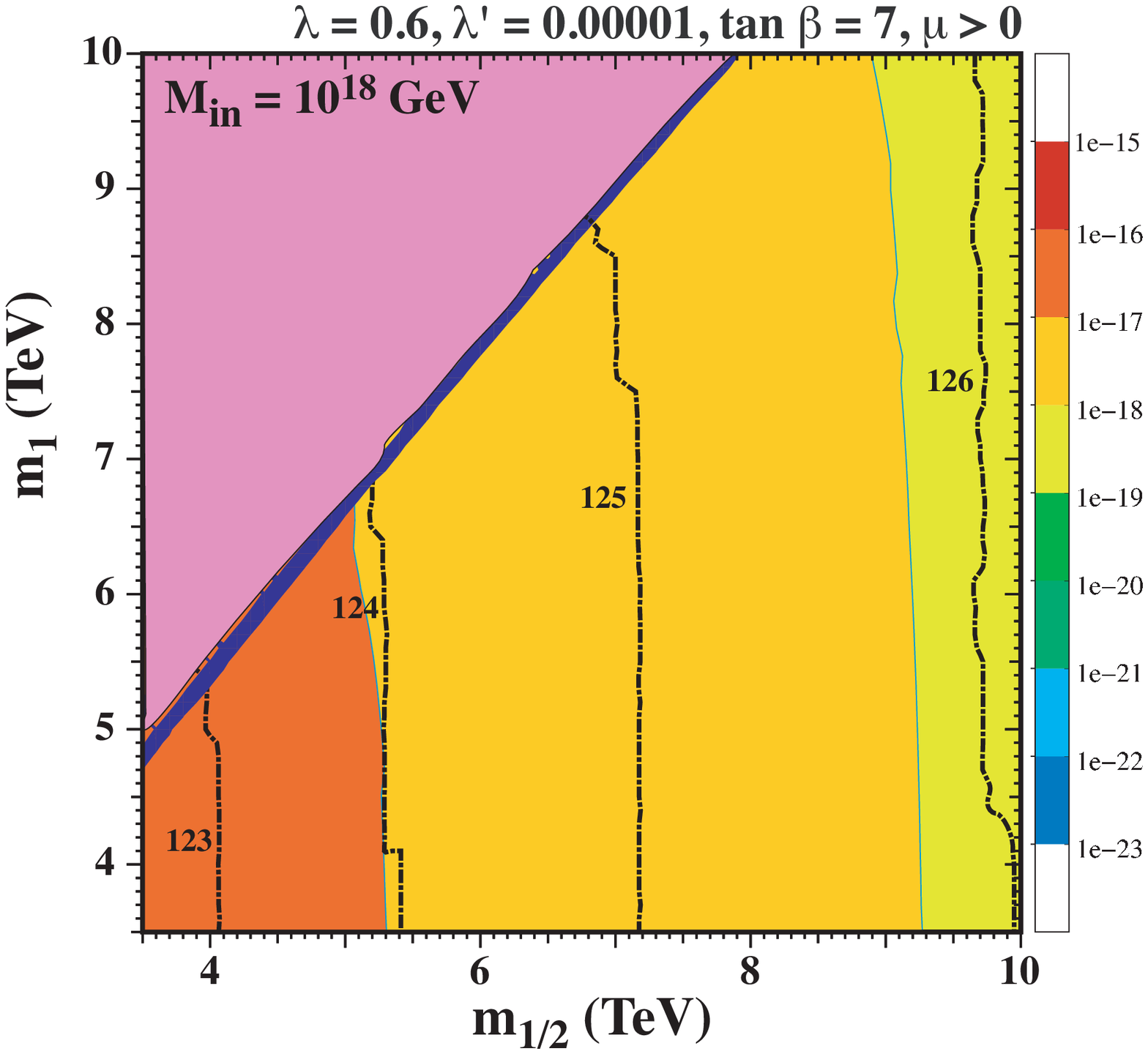}\\
\includegraphics[width=8cm]{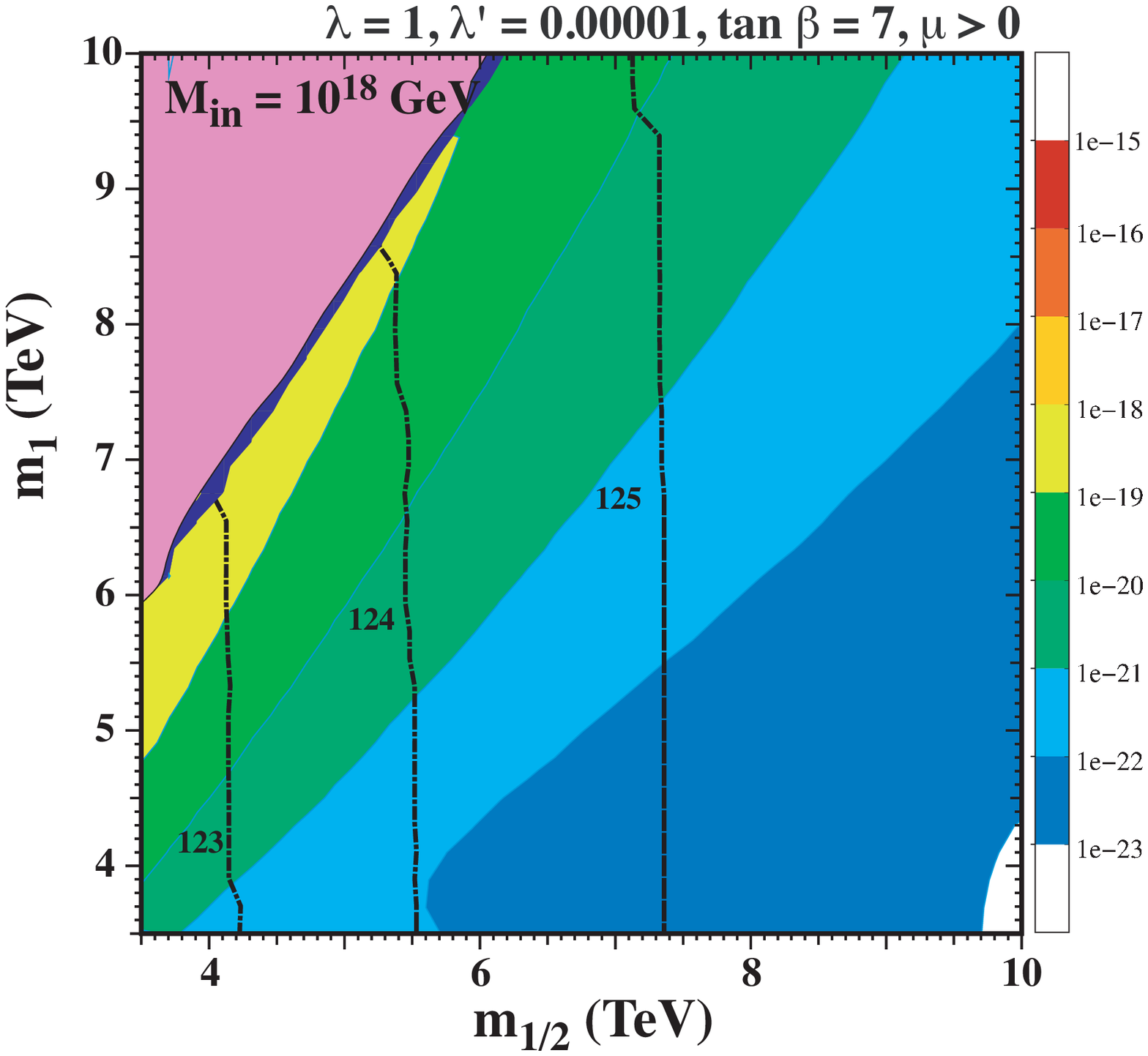}
\includegraphics[width=8cm]{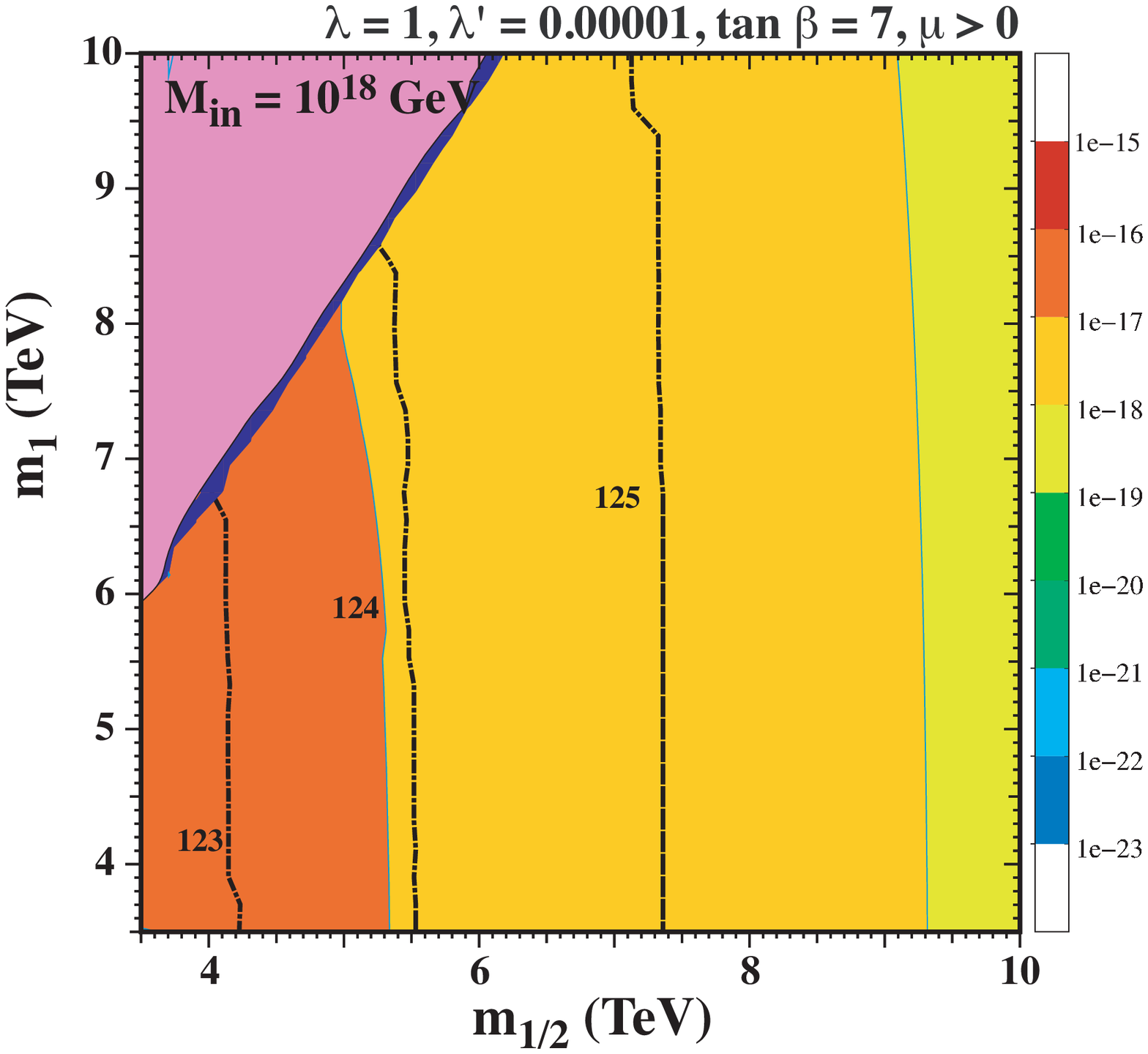}
\caption{\it As in Fig.~\ref{fig:M1_MUEG}, showing values of $\Bmueg$ for the flavor
choices {\bf A} (left) and {\bf B} (right) in model M3 with $\Mi = 10^{18}$~GeV,
$\tan \beta = 7$, $\lambda' = 0.00001$ and $\lambda = 0.6$ (upper panels), and model M4 with $\lambda = 1$ (lower panels). 
The color-coding for $\Bmueg$
is indicated in the bars beside the panels.}
\label{fig:M2a_MUEG}
\end{figure}

Fig.~\ref{fig:M2a_MUEG} shows the values of $\Bmueg$ found in models M3 (upper panels) and M4 (lower panels),
in flavor choice {\bf A} (left panels) and {\bf B} (right panels). In flavor choice {\bf A}, values of
$\Bmueg > 10^{-18}$ are compatible with the dark matter, $M_h$ and $\TPDK$ constraints in both models M3 (barely) and M4
(comfortably). In flavor choice {\bf B}, $\Bmueg$ reaches higher values along the dark matter strip,
with values $> 10^{-16}$ being compatible with both the $M_h$ and $\TPDK$ constraints. In general, with
flavor choice {\bf A} the values of $\Bmueg$ decrease away from the dark matter strip, whereas with
flavor choice {\bf B} the values of $\Bmueg$ depend primarily on $m_{1/2}$, with much less dependence on $m_1$.

In Table \ref{numbers34}, we show representative benchmark
points for models M3 and M4 along the relic density strip with $m_{1/2} = 6000$ GeV. 
For M3, $m_1 = 7850$ GeV, giving $M_h = 124.5 \pm 0.7$ GeV and 
for M4, $m_1 = 9780$ GeV, with $M_h = 124.4 \pm 0.7$ GeV.
We again see that the large increase in $(a_E)_{12}$ in choice
{\bf B} relative to {\bf A} accounts for the increase in $\Bmueg$.
In both cases, $\TPDK$ should be within reach of the Hyper-Kamiokande experiment~\cite{HK},
but the predictions for $\Bmueg$ and the electron EDM are below the projected experimental
sensitivities for both flavor choices.

\begin{table}
\begin{center}
\label{tbl:MassesFV2}
\begin{tabular}{|r|c|c|}
\hline 
\multicolumn{3}{|c|}{{\bf M3}:  $\Mi = 10^{18}$~GeV, $\lambda=0.6$}\\
\hline
Parameter & {\bf A} & {\bf B}\\
\hline
$\mu$  [GeV] & \multicolumn{2}{|c|}{1076 }\\
$M_1$  [GeV]  & \multicolumn{2}{|c|}{2181 } \\
$M_2$  [GeV] &  \multicolumn{2}{|c|}{4332}\\
\hline
$m_{\tilde e_L}$  [GeV] & 4844& 4844 \\
$m_{\tilde e_R}$   [GeV] & 4846 & 4846\\
$m_{\tilde \mu_{L}}$  [GeV] & 4844   & 4872 \\
$m_{\tilde \mu_{R}}$  [GeV]  &4846  & 4936 \\
$(m^2_{E})_{12}$  [GeV]$^2$   & $\phPF{908}{0.36}$ &  $\phNF{1395}{-0.40}$ \\
$(m^2_{E})_{31}$  [GeV]$^2$ &     $\phNF{2.6\times10^4}{2.8}$ & $\phNF{3.3\times10^4}{3.1}$ \\
$(m^2_{L})_{12}$  [GeV]$^2$   & $\phPF{3.7}{2.3}$ & $\phPF{3.7}{2.3}$ \\
$(m^2_{L})_{13}$  [GeV]$^2$ &  $\phPF{82}{1.9}$ &   $\phPF{ 82}{2.2}$  \\
$(a_E)_{11}$  [GeV] & $\phNF{0.21}{6.2\times10^{-6}}$ &  $\phNF{0.21}{5.7\times10^{-10}}$ \\
$(a_E)_{21}$  [GeV]  &$\phNF{0.00003}{0.27}$ &   $\phNF{0.05} {0.0006}$  \\
$(a_E)_{22}$  [GeV] & $\phPF{14} {3\times 10^{-6}}$  &  $\phNF{14} {9.2 \times 10^{-7}}$ \\
$(a_E)_{33}$  [GeV] & $\phPF{400}{1.0\times 10^{-11}}$ &  $\phPF{400}{1.6\times 10^{-9}}$ \\
\hline
$\TPDK$ [yrs] &   $9.9\times 10^{33}$   &  $9.9\times 10^{33}$  \\
$\Bmueg$ &  $5.5 \times 10^{-20}$ &  $4.8 \times 10^{-18}$  \\
$d_e$ [e.cm] &  $7.1  \times 10^{-34}$ &  $1.8  \times10^{-34}$ \\
\hline \hline
\multicolumn{3}{|c|}{\bf{M4}:  $\Mi = 10^{18}$~GeV, $\lambda=1$}\\
\hline
Parameter & {\bf A} & {\bf B}\\
\hline
$\mu$  [GeV]  & \multicolumn{2}{|c|}{1071  }\\
$M_1$  [GeV]  & \multicolumn{2}{|c|}{2184  }\\
$M_2$  [GeV] &  \multicolumn{2}{|c|}{4337}\\
\hline
$m_{\tilde e_L}$  [GeV]  & 4842 &  4842 \\
$m_{\tilde e_R}$  [GeV]  &  4845 & 4845\\
$m_{\tilde \mu_{L}}$  [GeV] & 4845  & 4933 \\
$m_{\tilde \mu_{R}}$ [GeV]  & 4842  & 4867\\
$(m^2_{E})_{12}$  [GeV]$^2$  &  $\phPF{1123}{0.36}$  &  $\phNF{1691}{-0.40}$  \\
$(m^2_{E})_{31}$  [GeV]$^2$ & $\phNF{3.2\times10^4}{2.8}$  & $\phNF{4.6\times10^4}{2.7}$     \\
$(m^2_{L})_{12}$  [GeV]$^2$  &  $\phPF{4.6}{2.3}$    &  $\phPF{4.6}{2.3}$  \\
$(m^2_{L})_{13}$  [GeV]$^2$ & $\phPF{100}{1.9}$  &  $\phPF{100}{2.3}$  \\
$(a_E)_{11}$ [GeV]  & $\phNF{0.27}{4.4\times10^{-6}}$ & $\phNF{0.27}{3.7\times10^{-10}}$  \\
$(a_E)_{21}$ [GeV]  &  $\phNF{0.00003}{0.28}$ &  $\phNF{0.051}{0.0005}$ \\
$(a_E)_{22}$  [GeV] &  $\phPF{14} {   2.7 \times 10^{-6}}$  &  $\phNF{14} {8.3 \times 10^{-7}}$ \\
$(a_E)_{33}$  [GeV] & $\phPF{380}{1.1\times10^{-11}}$ &  $\phPF{380}{1.6\times10^{-9}}$ \\
\hline
$\TPDK$ [yrs] & $2.7 \times 10^{34}$   & $2.7 \times 10^{34}$ \\
$\Bmueg$ & $9.2 \times 10^{-20}$ &   $4.4  \times 10^{-18}$ \\
$d_e$ [e.cm] & $8.2 \times 10^{-34}$  &    $3.4 \times 10^{-34}$ \\
\hline
\end{tabular}
\end{center}
\caption{\it Benchmark points in models M3 and M4 with $m_{1/2}=6000$~GeV. For M3,  $m_1=7850$~GeV, and for M4,   $m_1=9780$~GeV. We list values of the parameters relevant for $\Bmueg$ and the electron EDM
obtained with flavor choices {\bf{A}} and {\bf{B}}, as well as the corresponding
predictions for $\TPDK, \Bmueg$ and the electron EDM.
\label{numbers34} }
\end{table}

\begin{figure}[!ht]
\centering
\includegraphics[width=8cm]{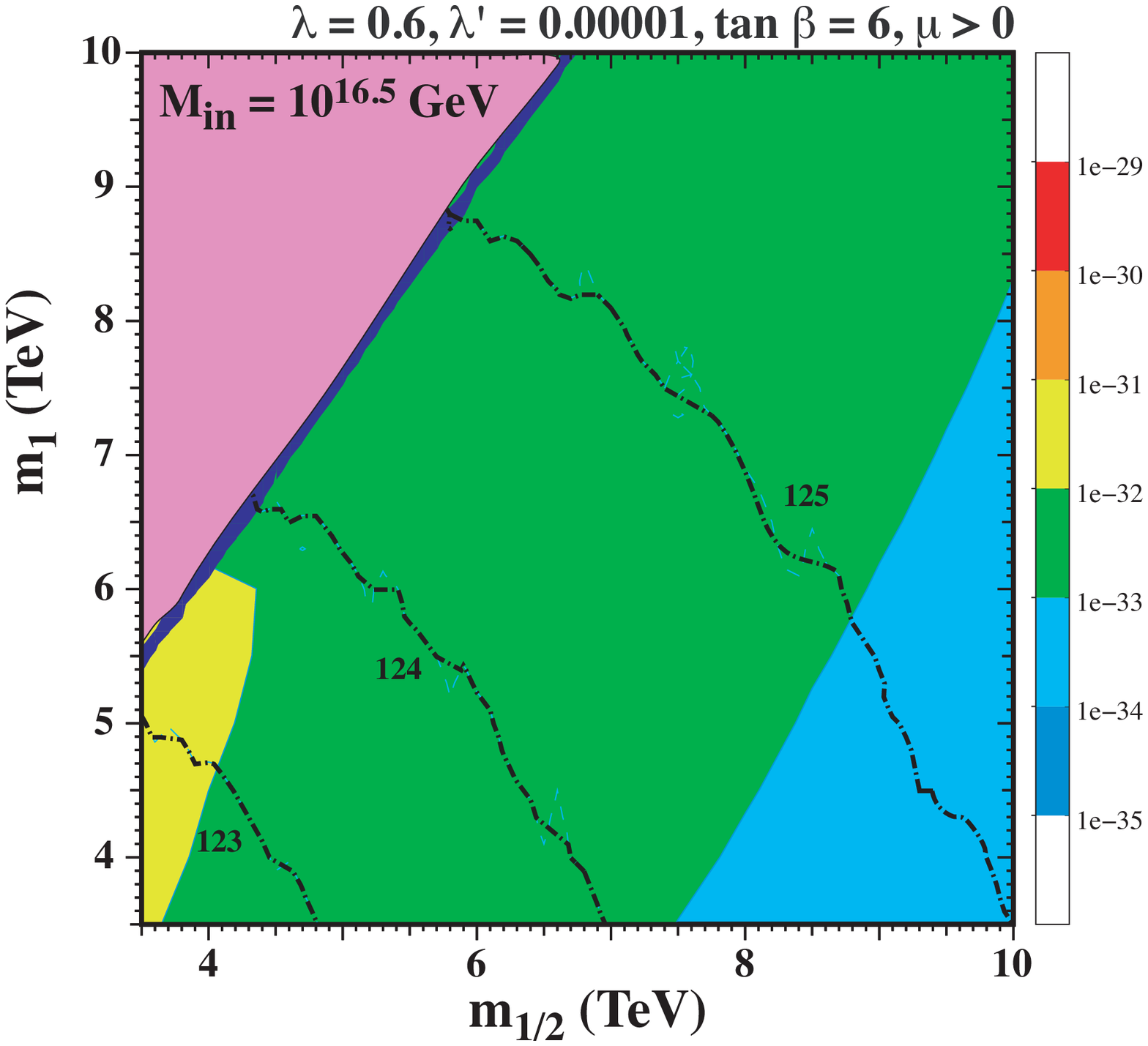}
\includegraphics[width=8cm]{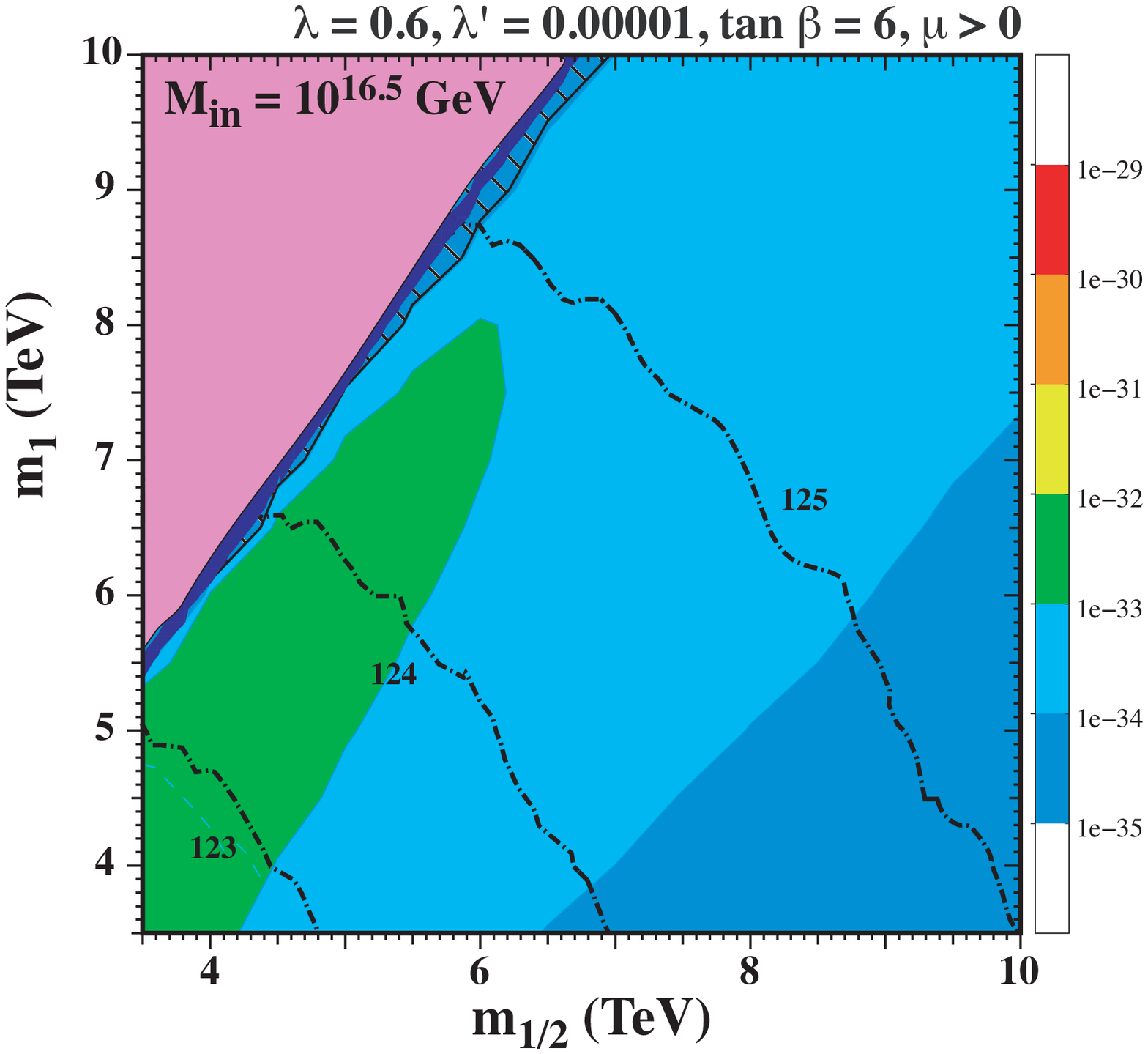}\\
\includegraphics[width=8cm]{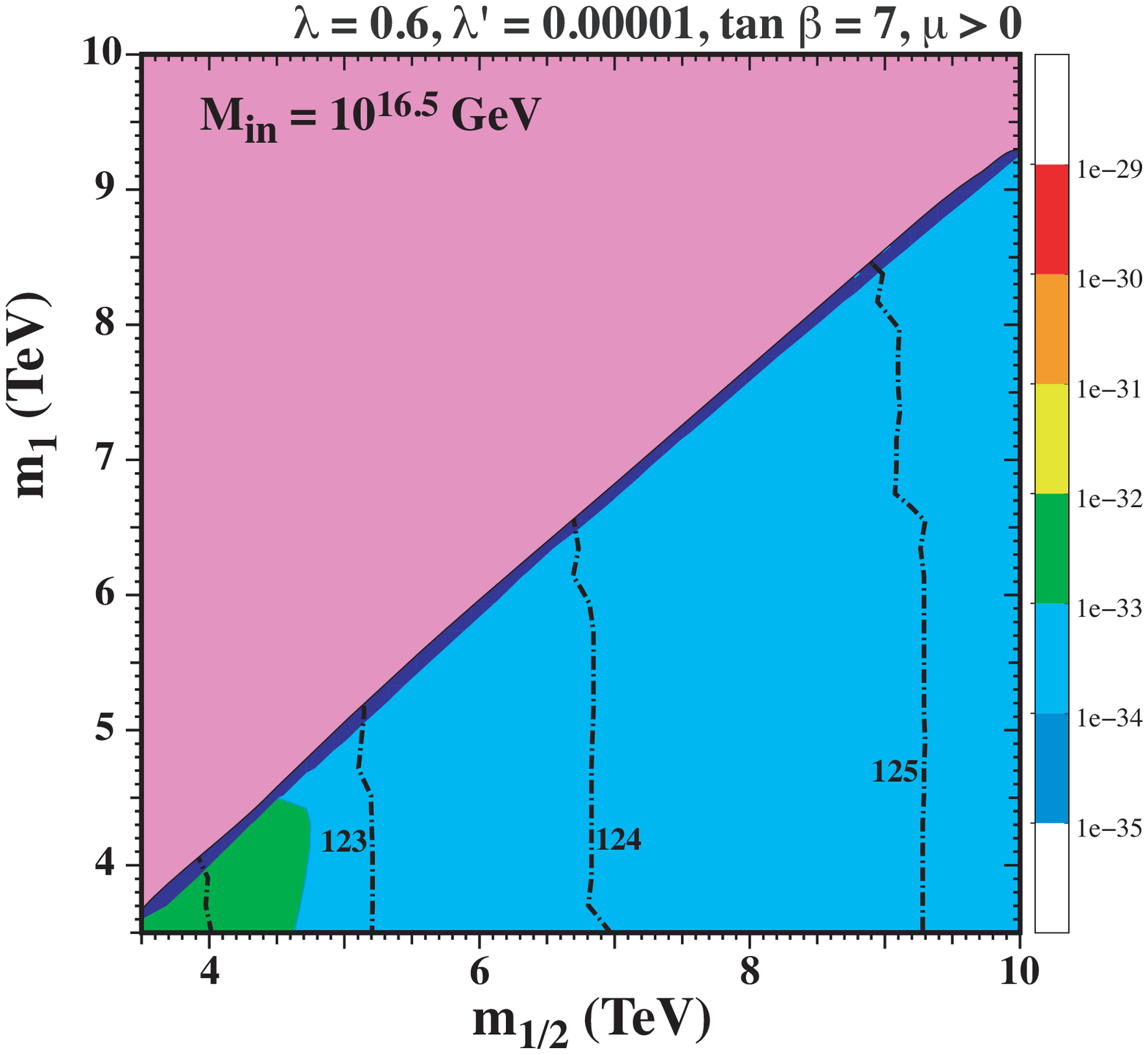}
\includegraphics[width=8cm]{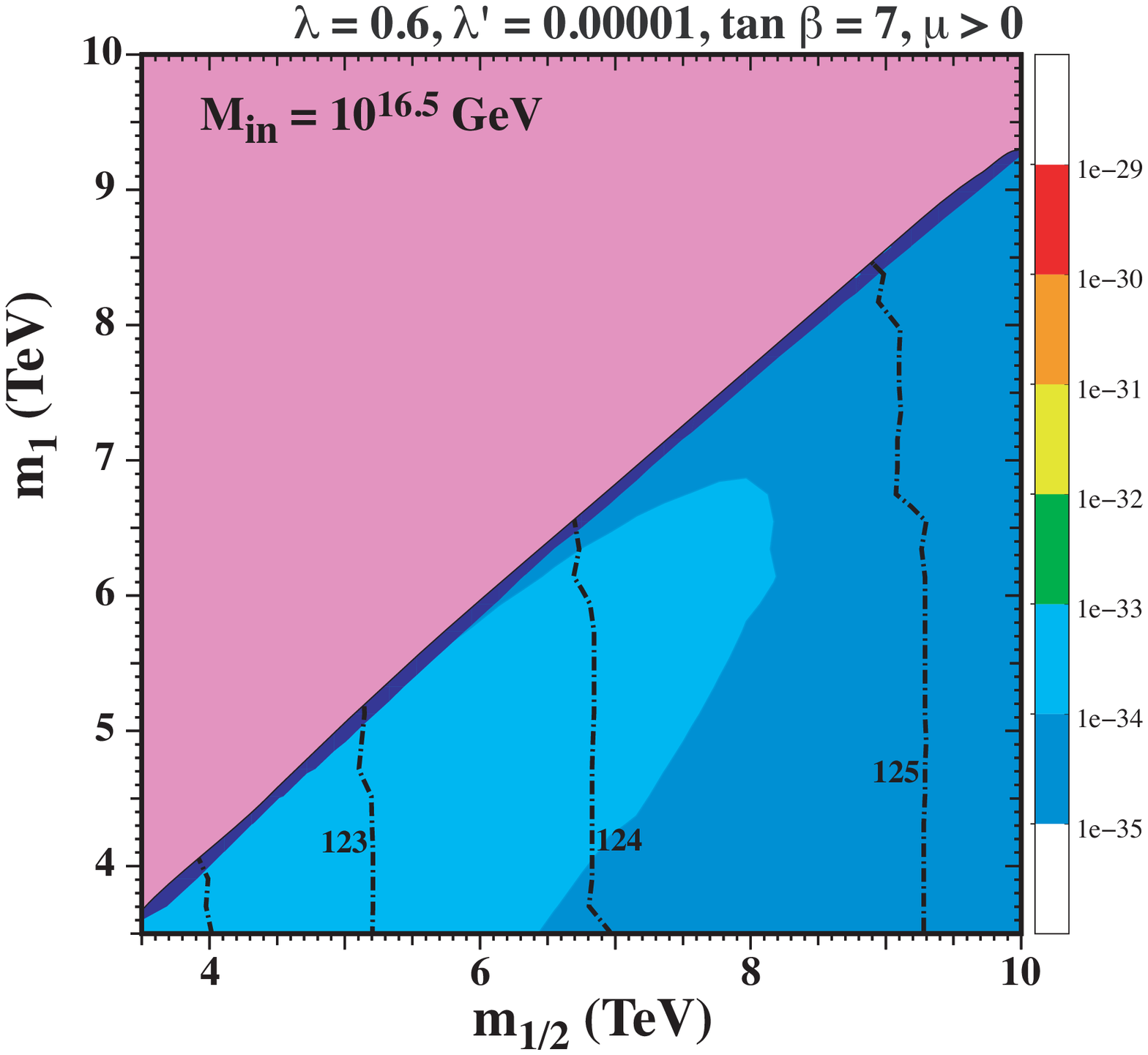}
\caption{\it As in Fig.~\ref{fig:M1_PD_YkA}, showing values of the electron EDM for the flavor
choices {\bf A} (left) and {\bf B} (right) in model M1 with $\Mi = 10^{16.5}$~GeV
and $\tan \beta = 6$ (upper panels) and in model M2
with $\Mi = 10^{16.5}$~GeV, $\tan \beta = 7$ and the indicated values
of $\lambda$ and $\lambda'$ (lower panels). 
The color-coding for the electron EDM
is indicated in the bars beside the panels.
\label{fig:M1_EDMS} }
\end{figure}

{\it Electron EDM:}
In the upper panels of Fig.~\ref{fig:M1_EDMS} we show the values of the electron EDM (eEDM), $d_e$, calculated 
using {\tt SUSY\_FLAVOR}~\cite{susyflavor} in model M1, presented
in the corresponding ($m_{1/2}, m_1$) plane used in Figs.~\ref{fig:M1_PD_YkA} and \ref{fig:M1_MUEG}. 
In the absence of flavor effects and in the absence of complex phases in the supersymmetric 
parameters (which we do not consider here), the EDM would be zero. Once the CKM matrix is
introduced as a seed of flavor and CP violation, the CKM phase propagates in all of the spectra,
generating a non-zero eEDM. 
The values of $d_e$ displayed in
Fig.~\ref{fig:M1_EDMS} are for the flavor choices {\bf A} (left) and {\bf B} (right). 

We see that the eEDM is generally larger for choice {\bf A}
reaching $ \sim 10^{-32}$ e.cm in the portion of the dark matter strip that is consistent
with $\TPDK$ and $M_h$. 
We also see that the eEDM is roughly a factor of 10 larger for case {\bf A} than it is for case {\bf B}.
As one can see from Table \ref{numbers12}, the contribution from Eq.~(\ref{cont33}) in both cases {\bf A} and {\bf B} are similar. However since ${\rm Im} (a_E)_{11}$ is about four orders of magnitude larger for case {\bf A} relative to case {\bf B}, the contribution from Eq.~(\ref{cont11})
boosts the eEDM in case {\bf A}. 
 For M1 and case {\bf A}, indeed the most important contributions come from $(a_{E})_{11}$ and $(a_{E})_{33}$, where 
\bea
\left[{\rm{Im}} \left\{ (K_{E})_{k,1L}(K_{E})^*_{k,1R}   \right\} \right]_{11} \sim 0. 5 \left[{\rm{Im}} \left\{ (K_{E})_{k,1L}(K_{E})^*_{k,1R}   \right\} \right]_{33}  \sim - 3 \times 10^{-13} \, .
\eea
Due to the overall sign in Eq.~(\ref{eq:approxd_EDM}), this is a positive contribution to the eEDM.
However, for the choice {\bf B} the contribution containing $(a_E)_{11}$ is negligible due to the smallness of ${\rm{Im}}\left\{ (a_E)_{11} \right\}$, but the contribution containing  $(a_E)_{31}$ becomes important, and we have instead~\footnote{The ratio of the (31) component to the (33) component using the approximation in Eq.~(\ref{eq:approxim}) is -0.3, however, in the full numerical computatation it is slightly great than -1.}
 \bea
 \label{relativesign}
-\left[{\rm{Im}} \left\{ (K_{E})_{k,1L}(K_{E})^*_{k,1R}    \right\} \right]_{31} \sim  \left[{\rm{Im}} \left\{ (K_{E})_{k,1L}(K_{E})^*_{k,1R}    \right\} \right]_{33}  \sim - 2 \times 10^{-13} \, .
\eea
Although all the other contributions in the cases {\bf A} and {\bf B} above are small, we keep them in \eq{eq:approxim}
and get with  \eq{eq:simplified_de}
\bea
|d_e|^{\bf{A}} \sim 2.3 \times 10^{-33} \, {\rm e.cm} \, , \quad
|d_e|^{\bf{B}} \sim 1.4 \times 10^{-33} \, {\rm e.cm} \, .
\eea
The reader should keep in mind that the approximation of  \eq{eq:simplified_de} should give the right order of magnitude, but the exact numerical factor is difficult to obtain with this approximation, due to the detailed structure of the complete $6\times 6 $ diagonalization matrices $K_E$. 

This range of eEDM values is well below
the experimental limit, and with flavor choice {\bf B} the eEDM 
remains below $10^{-33}$ e.cm along all the dark matter strip.
Indeed, the eEDM falls precipitously as the relic density strip is approached,
changing sign as it passes through zero in the thin cross-hatched region,
where its magnitude is below $10^{-34}$~e.cm. 
We see in (\ref{relativesign})
that $\left[{\rm{Im}} \left\{ (K_{E})_{k,1L}(K_{E})^*_{k,1R}    \right\} \right]_{31} > 0$ in choice {\bf B}, and so the opposite signs in (\ref{relativesign}) would explain the change in sign in $d_e$ with respect to choice {\bf A} if in the exact diagonalization we had
\beq
\left|\left[{\rm{Im}} \left\{ (K_{E})_{k,1L}(K_{E})^*_{k,1R}    \right\} \right]_{31}\right| > \left|\left[{\rm{Im}} \left\{ (K_{E})_{k,1L}(K_{E})^*_{k,1R}    \right\} \right]_{33} \right| \, ,
\eeq
as is the case in the full numerical calculation, 
causing $d_e$ to become negative.
The value of $d_e$ is reduced in choice {\bf B}, with
respect to {\bf A}, due to a cancellation. 
We find that the values of 
$\left[{\rm{Im}} \left\{ (K_{E})_{k,1L}(K_{E})^*_{k,1R}    \right\} \right]_{33}$ at the benchmark point
are similar in the two flavor choices,
being equal to $-6.1 \times 10^{-13}$ and $-6.2 \times 10^{-13}$ for choices {\bf A} and {\bf B}, respectively. This accounts for the cancellation in {\bf B}, but not in {\bf A}, and we find that
the sign of the eEDM at our benchmark point is indeed opposite in choices {\bf A} and {\bf B}. 

The lower panels of Fig.~\ref{fig:M1_EDMS} show the values of the eEDM in model M2
with the flavor choices {\bf A} (left) and {\bf B} (right), presented
in the corresponding ($m_{1/2}, m_1$) planes displayed in Figs.~\ref{fig:M1_PD_YkA} and \ref{fig:M1_MUEG}. 
We see again that larger values
of the eEDM are found with flavor choice {\bf A} than with choice {\bf B}: $< 10^{-33}$~e.cm
compared with $\lesssim 10^{-34}$~e.cm.
For M2, the contribution  from 
$(a_{E})_{11}$ dominates over the contribution from $(a_{E})_{33}$:
\bea
|\left[{\rm{Im}} \left\{ (K_{E})_{k,1L}(K_{E})^*_{k,1R}   \right\} \right]_{11}\sim 3.3 |\left[{\rm{Im}} \left\{ (K_{E})_{k,1L}(K_{E})^*_{k,1R}   \right\} \right]_{33} \sim 3.6 \times 10^{-14},
\eea 
whereas for choice {\bf{B}}
\bea
|\left[{\rm{Im}} \left\{ (K_{E})_{k,1L}(K_{E})^*_{k,1R}   \right\} \right]_{31}\sim 0.4 |\left[{\rm{Im}} \left\{ (K_{E})_{k,1L}(K_{E})^*_{k,1R}    \right\} \right]_{33} \sim 7 \times 10^{-15},
\eea 
and we obtain from \eq{eq:simplified_de}
\bea
|d_e|^{\bf{A}} \sim 9.8 \times 10^{-35} \, {\rm e.cm}, \quad
|d_e|^{\bf{B}} \sim 2.2 \times 10^{-35} \, {\rm e.cm}.
\eea
Thus this analytic approximation accounts for an enhancement by a factor of roughly 4.5, the full numerical ratio between {\bf A} and {\bf B} being about 22.

\begin{figure}[!ht]
\centering
\includegraphics[width=8cm]{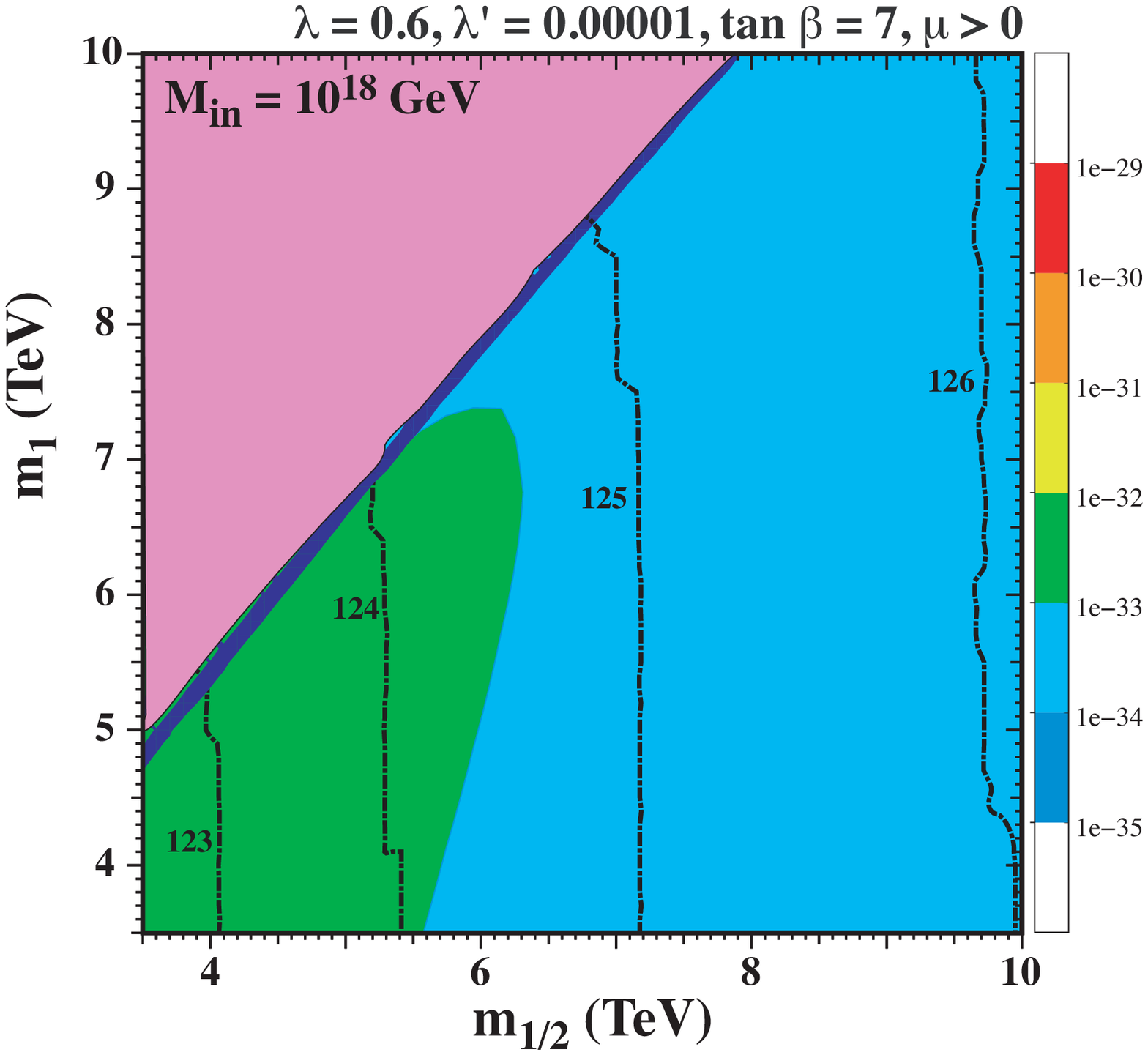}
\includegraphics[width=8cm]{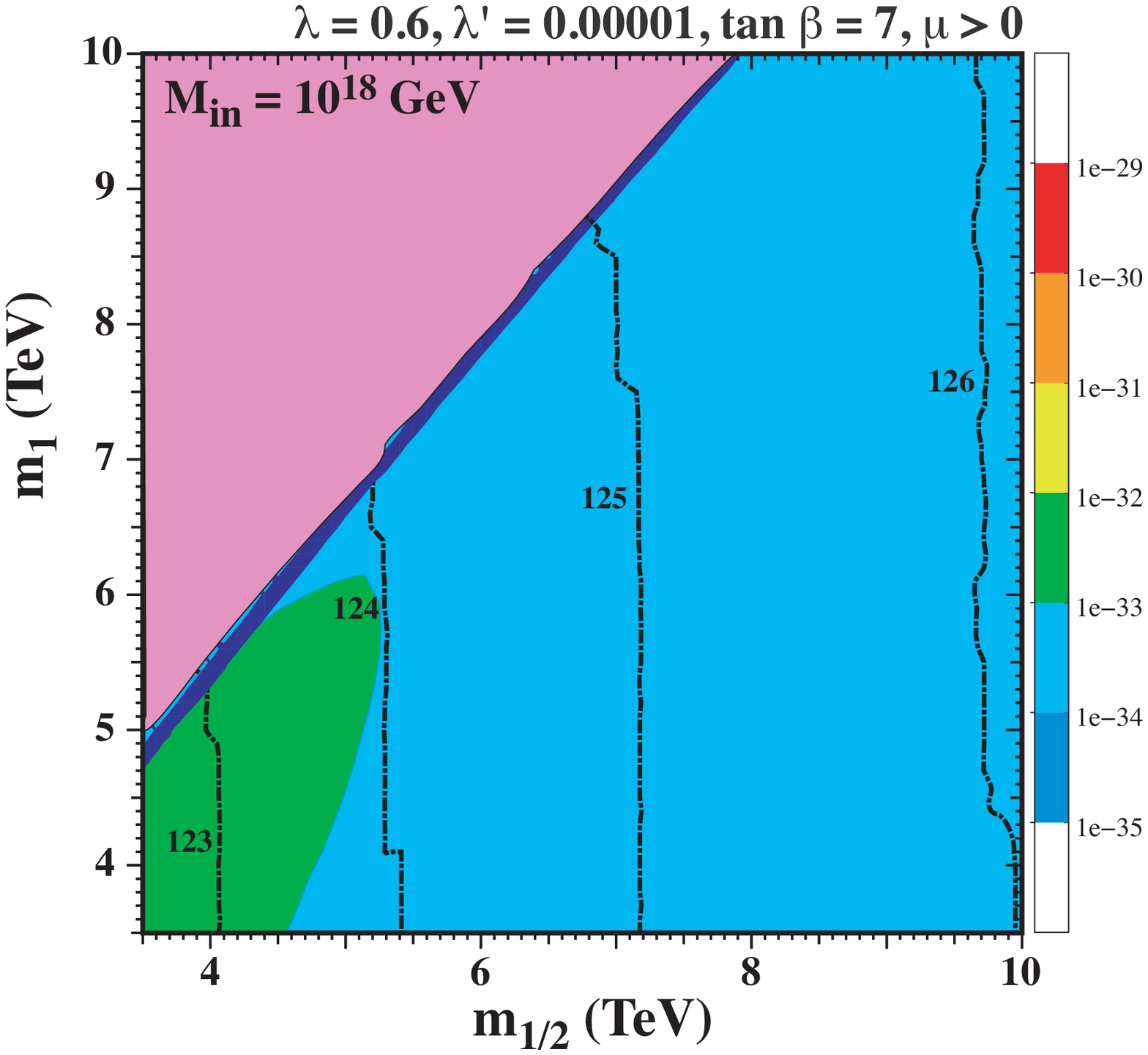}\\
\includegraphics[width=8cm]{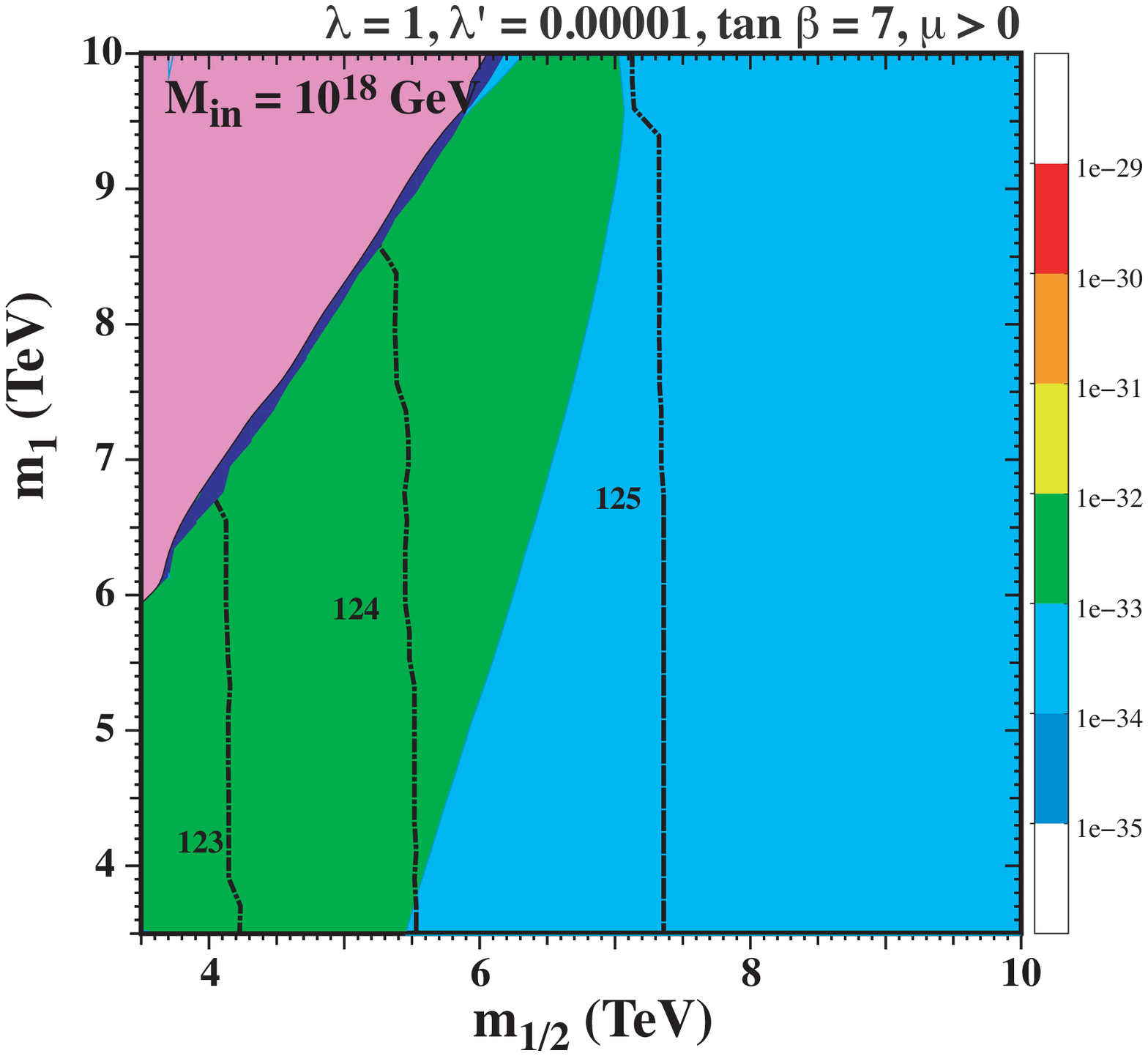}
\includegraphics[width=8cm]{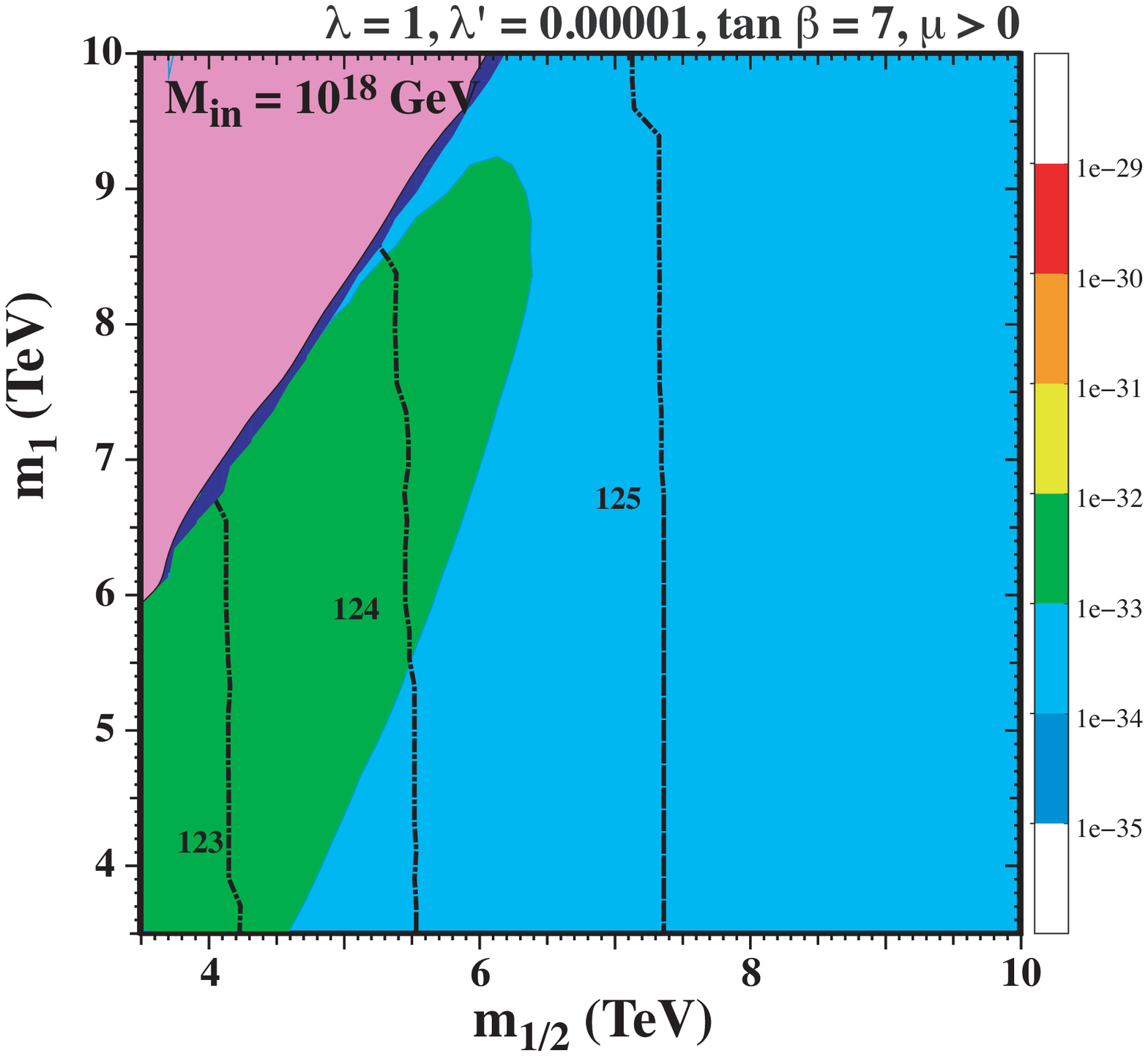}
\caption{\it As in Fig.~\ref{fig:M1_PD_YkA}, showing values of of the electron EDM for the choices {\bf A} (left) and {\bf B} (right) in model M3 with $\Mi = 10^{18}$ GeV, $\tan \beta = 7$, 
$\lambda' = 0.00001$ and $\lambda = 0.6$ (upper panels) or model M4 with $\lambda = 1$ (lower panels). 
The color-coding for the electron EDM
is indicated in the bars beside the panels.
\label{fig:M2a_EDMS} }
\end{figure}

In Fig.~\ref{fig:M2a_EDMS} we show the values of the electron EDM in models M3 (upper panels) and M4 (lower panels),
on the corresponding ($m_{1/2}, m_1$) planes displayed in Figs. \ref{fig:M1_PD_YkA} and \ref{fig:M2a_MUEG}, for flavor choices
{\bf A} (left panels) and {\bf B} (right panels). We see that the eEDM is generally larger in model M3
than in model M4, and larger with flavor choice {\bf A} than with flavor choice {\bf B}.
However, along the dark matter strips the $\TPDK$ constraint generally imposes $d_e < 10^{-33}$~e.cm,
except in the case of model M4 with choice {\bf A}, for which $d_e$ may reach a few $\times 10^{-33}$~e.cm.
Note that M4, for both cases {\bf A} and {\bf B}, all the elements $(a_{E})_{11}$, $(a_{E})_{31}$ and $(a_{E})_{33}$ are important. 

\subsection{Models in which only $\bar H$ is twisted}

As mentioned above, unless ${\bar H}$ is twisted we find 
no solutions for which the relic density, Higgs mass and proton
lifetime are consistent with experiment. In this Section,
we consider models in which {\it only} ${\bar H}$ twisted, i.e., we leave 
$H$ untwisted so that $m_0 = m_1 = 0$, whereas $m_2 = m_{3/2}$. 
As in the previous Section, we consider two choices for the
modular weights, one in which the modular weights take values such that 
all tri- and bi-linear terms vanish,
and another in which the weights all vanish,
leaving some of the tri- and bi-linear terms non-zero. 
These are labelled models M5 and M6, respectively.

\begin{figure}[!ht]
\centering
\includegraphics[width=8cm]{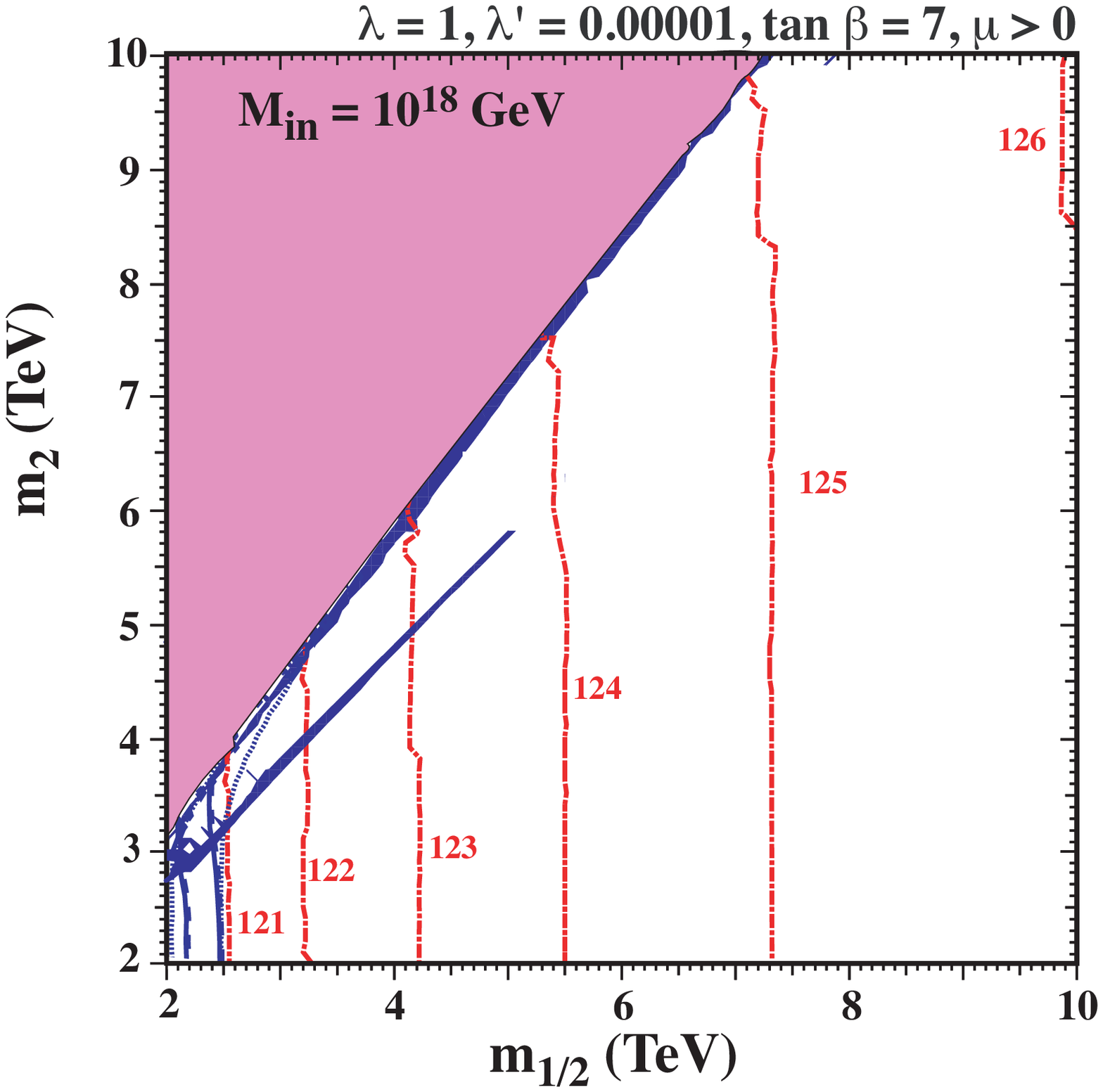}
\includegraphics[width=8cm]{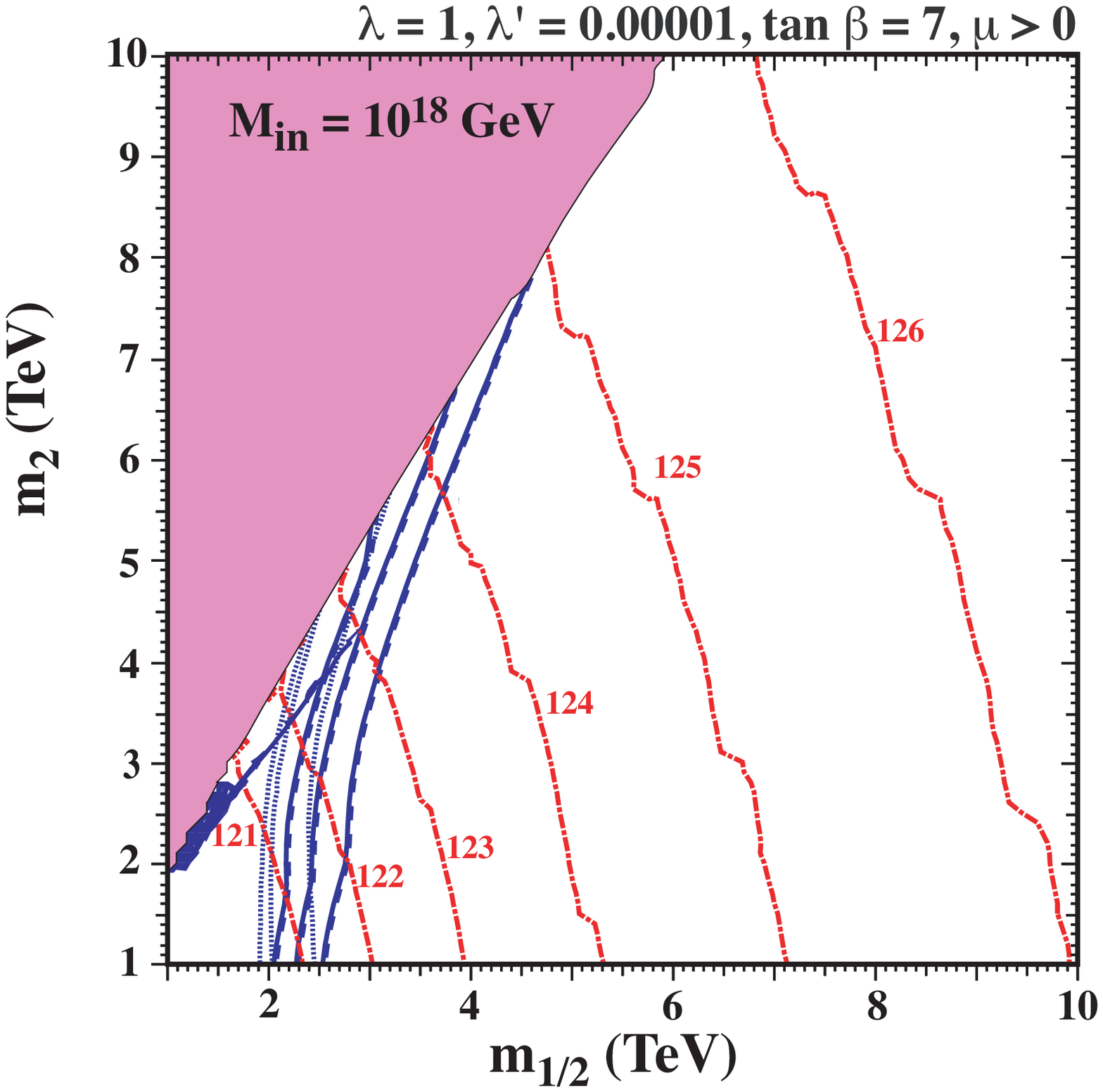}
\caption{\it Examples of ($m_{1/2}, m_2$) planes for Models M5 and M6
with $\Mi = 10^{18}$~GeV, $\lambda = 1$, and $\tan \beta = 7$.
We assume $\mu > 0$ in both panels. As in Fig.~\ref{fig:M1_PD_YkA}, in the regions shaded pink there is no EWSB, and in the blue strips below these regions the relic density is in the range $0.06 < \Omega_\chi h^2 < 0.2$. The red dot-dashed curves are Higgs mass contours, with the masses labelled in GeV.  For each flavor choice, in the left panel
there are two contours for the proton lifetime,
$\TPDK$, corresponding to the central values and 
$-1 \sigma$ variations in the hadronic matrix elements, with the $+1 \sigma$ curves 
invisible at
lower values of $m_{1/2}$. In the right panel there are three contours for the proton lifetime,
corresponding to the central values and $\pm 1 \sigma$ variations in the 
hadronic matrix elements. The predictions of flavor
choices {\bf A} and {\bf B} are shown as the solid and dashed blue curves, respectively, and 
those of the {\bf NF} choice are shown as the blue dotted curves. 
\label{fig:M56} }
\end{figure}

\paragraph{Model M5} In this model we fix $\Mi = 10^{18}$~GeV,
 $\tan \beta = 7$, and $\mu > 0$. 
The chosen values of the couplings of the adjoint 
Higgs supermultiplets are $\lambda = 1$ and $\lambda' = 0.00001$.
We show in the left panel of Fig.~\ref{fig:M56} the 
($m_{1/2}, m_2$) plane for this model,
where we recall that $m_2 = m_{3/2}$ when only $\bar H$ is twisted. 
There is no EW symmetry breaking (EWSB)
in the triangular region shaded pink in the upper left corner, 
i.e., the solution for the MSSM $\mu$ parameter has $\mu^2 < 0$.
As in Fig.~\ref{fig:M1_PD_YkA}, the red dot-dashed curves show contours 
of the Higgs mass as calculated using {\tt FeynHiggs~2.16.0} \cite{FeynHiggs},
and there is a dark blue shaded strip just below the 
no-EWSB region, corresponding to the focus point \cite{fp},
where the relic density taking values in the range $0.06 < \Omega_\chi h^2< 0.2$. 
In addition to this strip, there is a band at lower $m_2$, which corresponds
to a funnel where rapid annihilation via direct-channel $H/A$ poles 
when $m_\chi \simeq M_{H/A}/2$ brings the relic density into this range. 
This band actually consists of two unresolved narrow strips with $m_\chi >$ 
and $m_\chi < M_{H/A}/2$, between which the relic density takes lower values. 
We note that this funnel strip ends when $M_h < 124$ GeV.
Beyond this endpoint, the suppression in the annihilation cross-section
due to the large value of $m_{1/2}$ is strong enough that the relic density
always exceeds the observed value, i.e., $\Omega_\chi h^2 > 0.12$, even on the $H/A$ poles.

\paragraph{Model M6} The corresponding results for model M6 are shown in the right panel of Fig.~\ref{fig:M56}. In this case,
$A_{10} = A_\lambda = B_H = m_2$, and the other bi- and tri-linear terms vanish
at $\Mi$. The dependence of the $A$-terms on $m_2$ induces a weak dependence 
of $M_h$ on $m_2$, as is readily seen by comparing 
the Higgs mass contours in the two panels of Fig.~\ref{fig:M56}.
Importantly, in this case we do not find the focus-point strip
along the boundary of the no-EWSB region. Indeed, in this case $m_A^2 < 0$, 
where $m_A$ is the pseudoscalar Higgs mass, everywhere in the pink shaded region. 
There is, nevertheless, a funnel strip at lower $m_2$, which extends only as far as
$m_{1/2} \sim 3$ TeV, where $M_h \sim 123$ GeV in this case.\\

{\it Proton lifetime:} The proton lifetime limits for M5 are weaker than those in M6. 
In both cases, there is little flavor dependence
and the proton lifetimes for cases {\bf A} and {\bf B} are nearly identical 
and also similar to the {\bf NF} case.  In the case of M5, we 
only see two sets of lines, as the $-1 \sigma$ variations in the hadronic matrix
elements, which increase the proton lifetime for fixed supersymmetric model parameters, push the contour for 
$\tau_p = 6.6 \times 10^{33}$ yrs to low values of $m_{1/2} < 2$ TeV, below its displayed range
and where $M_h$ is too small.
There is considerable parameter space in model M5 where the relic density
$\Omega_\chi h^2 \approx 0.12$ and the proton lifetime constraint is satisfied. 
In contrast, in model M6 only the portion of the funnel strip
between $m_{1/2} = 2.5$ and 3~TeV satisfies the proton decay constraint
when the matrix elements are varied by $\pm 1 \sigma$.

\begin{figure}[!ht]
\centering
\includegraphics[width=8cm]{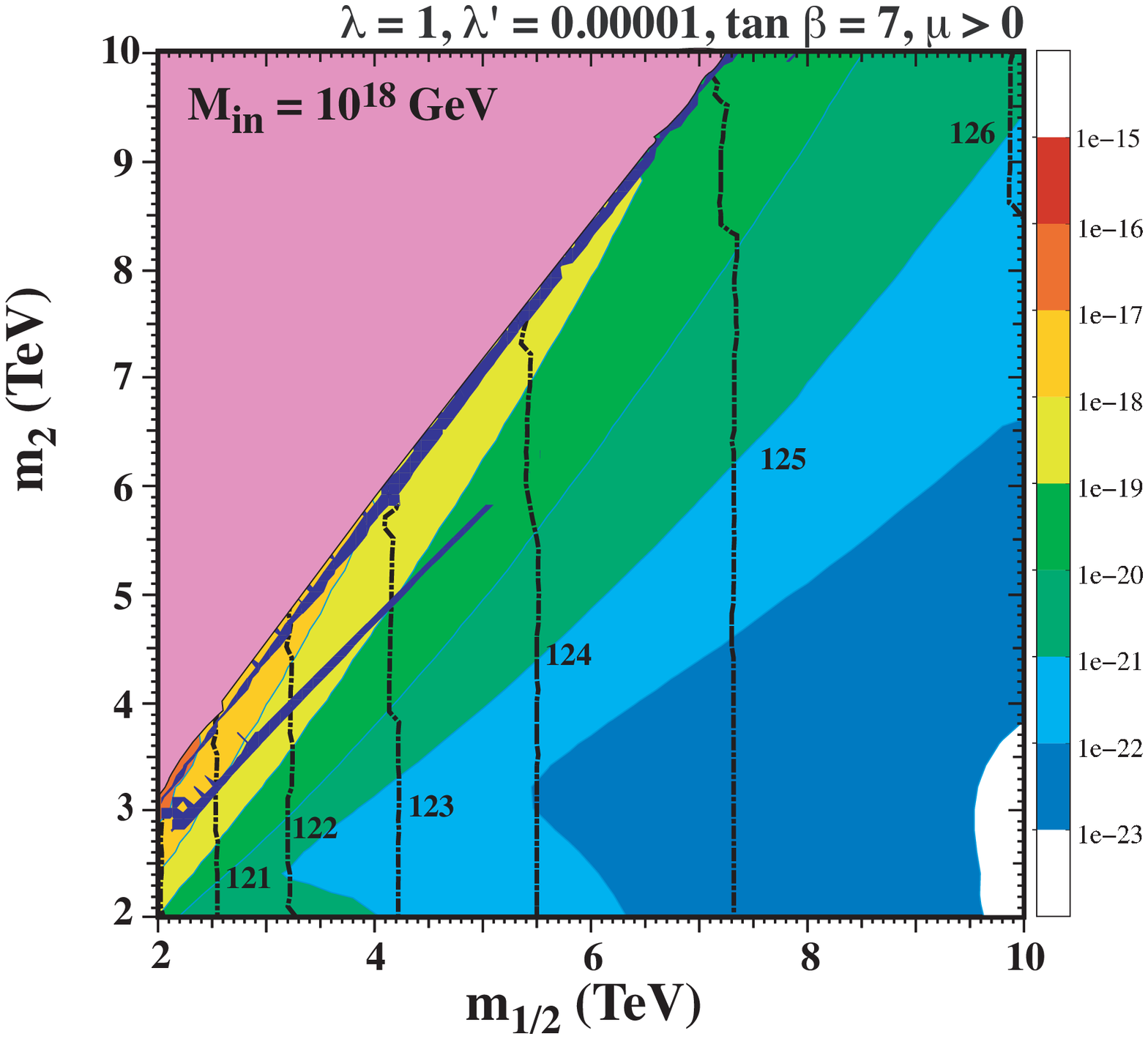}
\includegraphics[width=8cm]{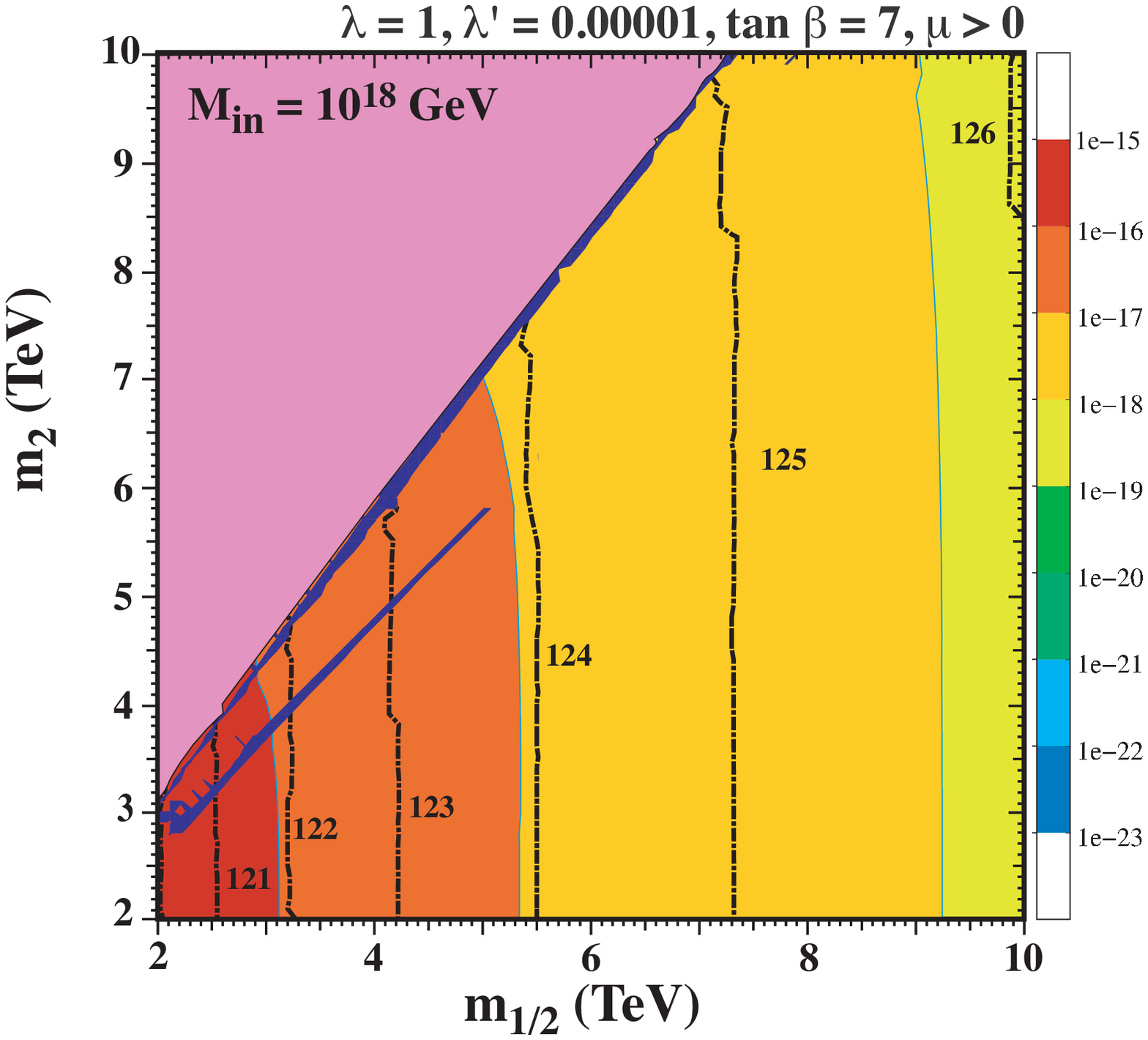}\\
\includegraphics[width=8cm]{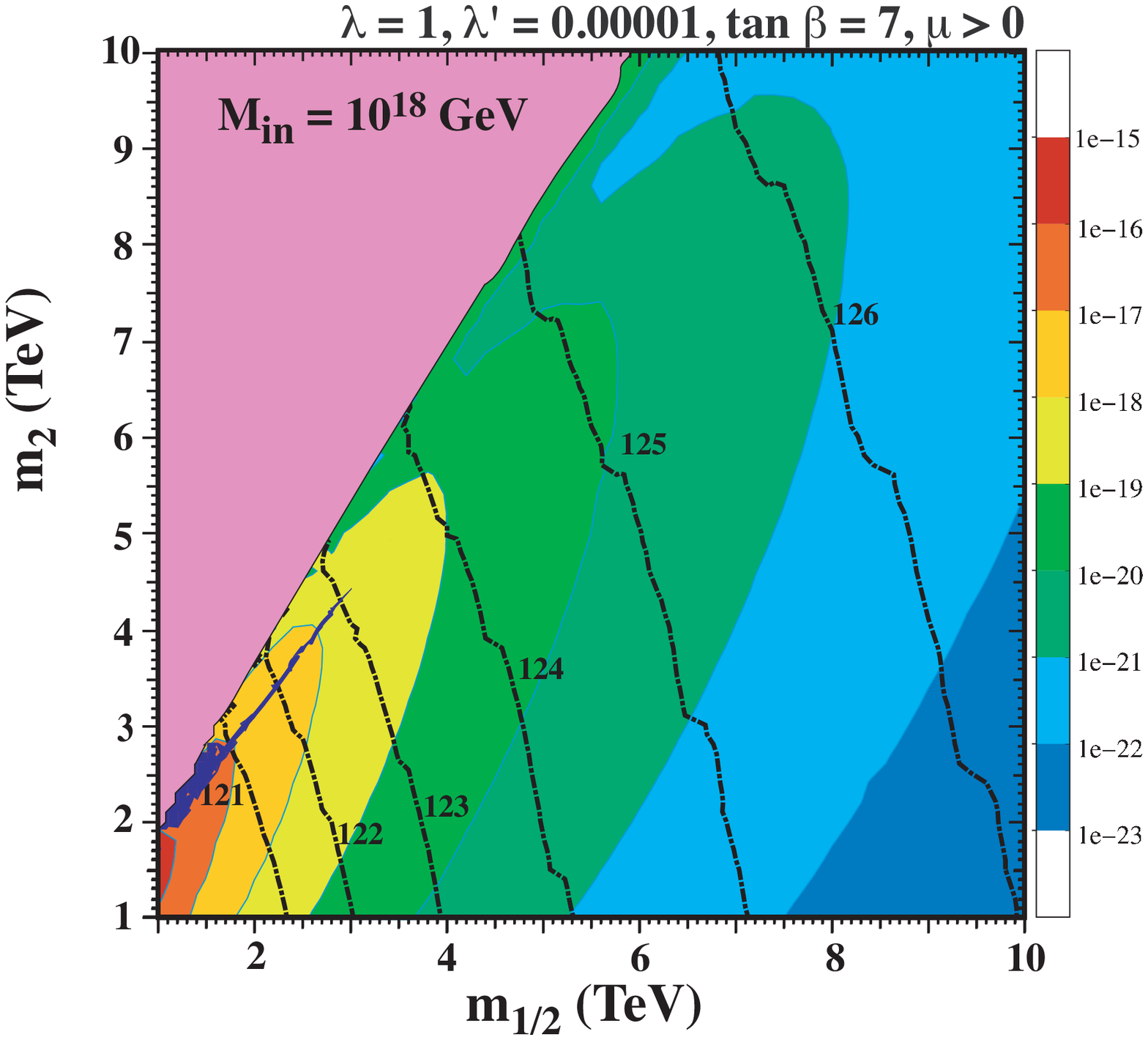}
\includegraphics[width=8cm]{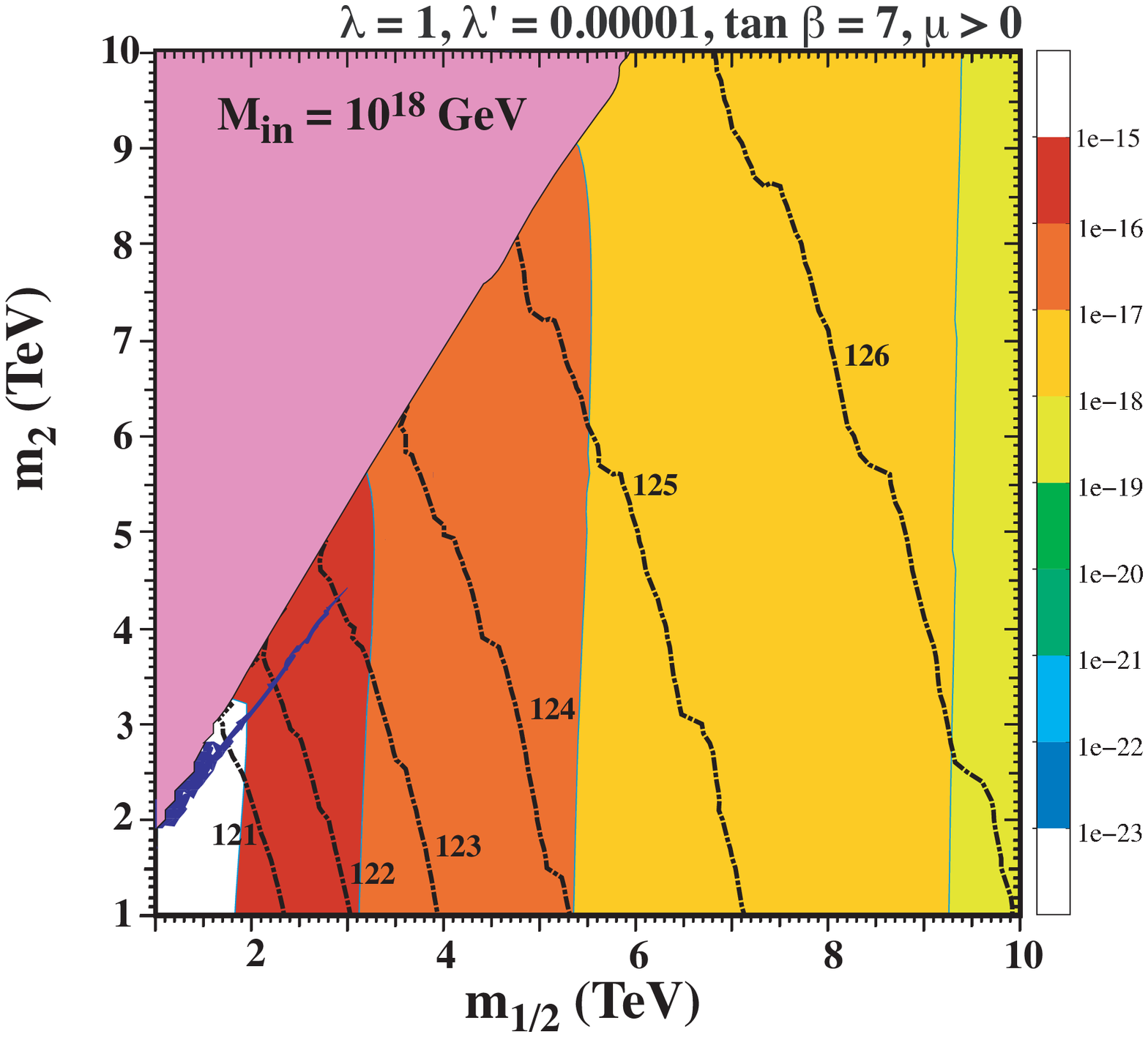}
\caption{\it As in Fig.~\ref{fig:M1_MUEG}, showing values of $\Bmueg$ 
in the $(m_{1/2}, m_2)$ planes for the flavor
choices {\bf A} (left) and {\bf B} (right) in model M5 (upper panels) 
and model M6 (lower panels) both with $\Mi = 10^{18}$~GeV,
$\tan \beta = 7$, $\lambda' = 0.00001$ and $\lambda = 1$. 
The color-coding for $\Bmueg$
is indicated in the bars beside the panels.}
\label{fig:M56f}
\end{figure}

{\it Flavor violation:} Values of the branching ratio $\Bmueg$ for models M5 and M6
are shown in Fig.~\ref{fig:M56f} (upper and lower panels, respectively) 
for flavor choices {\bf A} and {\bf B}
(left and right panels, respectively). As was the case in model M1, 
for flavor choice {\bf A}, the branching ratio in model M5
exceeds $10^{-17}$ only at very low $m_{1/2}$, 
in this case for $m_{1/2} < 1.4$ TeV, for
which $M_h < 121$ GeV. For the portion of the focus-point strip with $M_h > 123$ GeV, $\Bmueg < 10^{-18}$. 
Furthermore, the branching ratio is over an order of magnitude smaller in the funnel strip 
than it is in the focus-point strip. In contrast, for choice {\bf B}, there is little 
difference in $\Bmueg$ between the two relic density strips. We again see that overall 
the branching ratio for choice {\bf B} is significantly larger than for choice {\bf A}.  
We provide in Table~\ref{numbers56} the parameters of 
a benchmark point in model M5 lying on the relic density strip with $m_{1/2} = 6000$ TeV,
$m_2 = 8385$ GeV and $M_h = 124.4 \pm 0.7$ GeV.
Once again, we see that $(a_E)_{21}$ is significantly larger for choice {\bf B} than for choice {\bf A}, leading to the increased branching ratio for 
$\mu \to e \gamma$. 

In the case of model M6, $\Bmueg$ is generally smaller than in M5 for flavor choice {\bf A}, though it does exceed $10^{-16}$ for $m_{1/2} < 1.2$ TeV. For flavor choice {\bf B}, 
the branching ratio exceeds $10^{-15}$ for $m_{1/2} < 2$ TeV,
in the unshaded the region. At the tip of the funnel strip,
the branching ratio exceeds $10^{-16}$. In a portion of this strip that is compatible with the present limit on $\TPDK$ and the Higgs mass, $\mu \to e$ conversion may be accessible to the PRISM experiment.~\cite{Strategy:2019vxc} In this case, 
because of the lack of a focus-point strip, we
provide also in Table~\ref{numbers56} the parameters of 
a benchmark point on the funnel strip with $m_{1/2} = 3000$~TeV
and $m_2 = 4470$ GeV, corresponding to $M_h = 123.2 \pm 0.8$ GeV. The differences between the branching ratios in choices {\bf A} and {\bf B} can again be attributed to the increase in $(a_E)_{21}$ for choice {\bf B} seen in the Table.


\begin{table}
\begin{center}
\label{tbl:MassesFV3}
\begin{tabular}{|r|c|c|}
\hline 
\multicolumn{3}{|c|}{{\bf M5}:  $\Mi=10^{18}$~GeV, $\lambda=1$}\\
\hline
Parameter & {\bf A} & {\bf B}\\
\hline
$\mu$  [GeV] & \multicolumn{2}{|c|}{1013}\\
$M_1$  [GeV]  & \multicolumn{2}{|c|}{2184  } \\
$M_2$  [GeV] &  \multicolumn{2}{|c|}{4336 }\\
\hline
$m_{\tilde e_L}$  [GeV] & 5024& 5024\\
$m_{\tilde e_R}$   [GeV] & 4485 & 4485\\
$m_{\tilde \mu_{L}}$  [GeV] & 5024 & 5125  \\
$m_{\tilde \mu_{R}}$  [GeV]  &4485 & 4505 \\
$(m^2_{E})_{12}$  [GeV]$^2$   & $\phPF{1012}{0.36}$ &  $\phNF{1462}{0.40}$ \\
$(m^2_{E})_{31}$  [GeV]$^2$ & $\phNF{2.9\times10^4}{2.8}$  &  $\phPF{3.5\times 10^4}{3.1} $ \\
$(m^2_{L})_{12}$  [GeV]$^2$   & $\phNF{0.71}{2.3}$  & $\phNF{0.71}{2.3}$ \\
$(m^2_{L})_{13}$  [GeV]$^2$ & $\phPF{16}{1.9}$  &   $\phPF{12}{1.6}$   \\
$(a_E)_{11}$ [GeV] &  $\phNF{0.2}{6\times10^{-6}}$ &  $\phNF{0.2}{5.5\times10^{-10}} $  \\
$(a_E)_{21}$  [GeV] &  $\phNF{0.00003} {0.28}$ &  $\phNF{0.052}{0.0005}$  \\
$(a_E)_{22}$  [GeV] & $\phPF{14}{2.7 \times 10^{-6}}$ & $\phNF{14}{8.6 \times 10^{-7}}$  \\
$(a_E)_{33}$  [GeV] &  $\phPF{380}{1.2\times10^{-11}}$ &    $\phPF{380}{1.6\times10^{-9}}$  \\
\hline
$\TPDK$  & $2.8  \times 10^{34}$ & $2.8  \times 10^{34}$ \\
$\Bmueg$& $1.5  \times 10^{-19}$ &  $4.7 \times 10^{-18}$ \\
$d_e$ [e.cm] &  $4.8 \times 10^{-34}$ &  $5.2 \times 10^{-35}$  \\
\hline \hline
\multicolumn{3}{|c|}{{\bf M6}:  $\Mi = 10^{18}$~GeV, $\lambda=1$}\\
\hline
Parameter & {\bf A} & {\bf B}\\
\hline
$\mu$  [GeV]  & \multicolumn{2}{|c|}{2679 }\\
$M_1$  [GeV]  & \multicolumn{2}{|c|}{1073} \\
$M_2$  [GeV] &  \multicolumn{2}{|c|}{2130}\\
\hline
$m_{\tilde e_L}$  [GeV]  & 2525&   2525  \\
$m_{\tilde e_R}$  [GeV]  &  2221& 2221 \\
$m_{\tilde \mu_{L}}$  [GeV] & 2525  &  2577 \\
$m_{\tilde \mu_{R}}$ [GeV]  & 2222   &  2226\\
$(m^2_{E})_{12}$  [GeV]$^2$  &  $\phPF{581}{0.36}$  &  $\phNF{800}{0.40}$   \\
$(m^2_{E})_{31}$  [GeV]$^2$ & $\phNF{1.6\times10^4}{2.8}$  &  $\phPF{1.9\times10^4}{3.1}$  \\
$(m^2_{L})_{12}$  [GeV]$^2$  &  $\phPF{0.0019}{3.0}$    &  $\phPF{0.0019}{3.0}$   \\
$(m^2_{L})_{13}$  [GeV]$^2$ &   $\phPF{9.4\times10^{-3}}{0.79}$ & $\phPF{7.9\times10^{-3}}{0.59}$  \\
$(a_E)_{21}$ [GeV]  &  $\phNF{0.00006}{0.36}$ &  $\phNF{0.027}{0.0007}$ \\
$(a_E)_{11}$  [GeV] &  $\phNF{5.9 \times 10^{-2}}{4.7\times 10^{-6}}$  &  $\phPF{5.9\times 10 ^{-2}}{10^{-9}} $   \\
$(a_E)_{22}$  [GeV] &  $\phPF{6.3} { 4\times 10^{-7}}$ &  $\phNF{6.3} {3 \times 10^{-7}}$  \\
$(a_E)_{33}$  [GeV] & $\phPF{130}{1.5\times10^{-10}}$   &  $\phPF{130}{7.7\times10^{-10}}$  \\
\hline
$\TPDK$ & $6.96 \times 10^{33}$  & $6.96 \times 10^{33}$  \\
$\Bmueg$ &$5.1 \times 10^{-19}$  & $1.43 \times 10^{-16}$  \\
$d_e$ [e.cm] & $7.6 \times 10^{-34}$ &  $5.8  \times 10^{-35}$ \\
\hline
\end{tabular}
\end{center}
\caption{\it Benchmark points in model M5 with $m_{1/2}=6000$~GeV and in model M6 with
$m_{1/2}$= 3000~GeV. For M5,  $m_2=8385$~GeV, and for M6,  $m_2=4470$~GeV. We list values of the parameters relevant for $\Bmueg$ and the electron EDM
obtained with flavor choices {\bf{A}} and {\bf{B}}, as well as the corresponding
predictions for $\TPDK, \Bmueg$ and the electron EDM.}
\label{numbers56}
\end{table}

\begin{figure}[!ht]
\centering
\includegraphics[width=8cm]{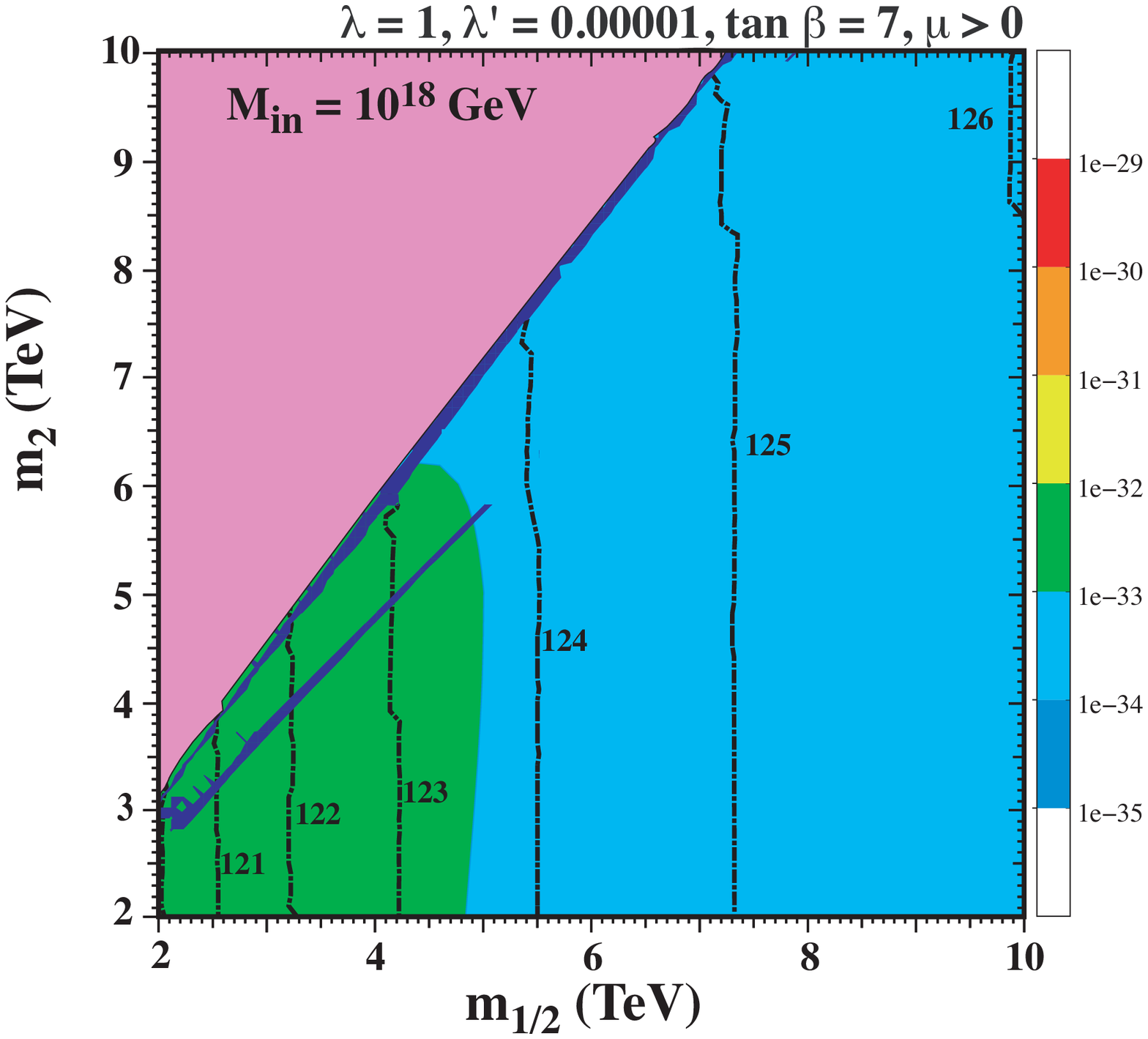}
\includegraphics[width=8cm]{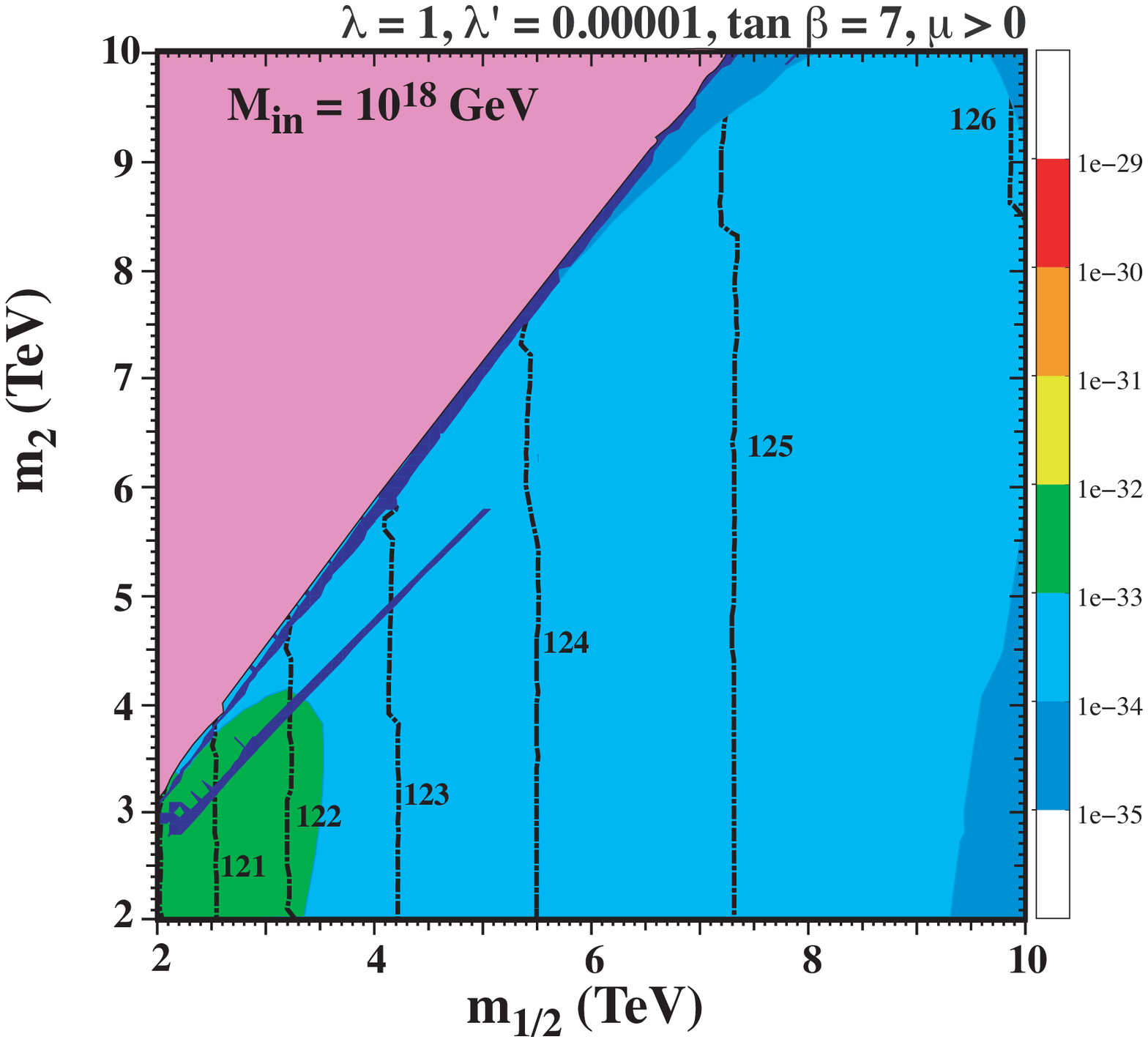}\\
\includegraphics[width=8cm]{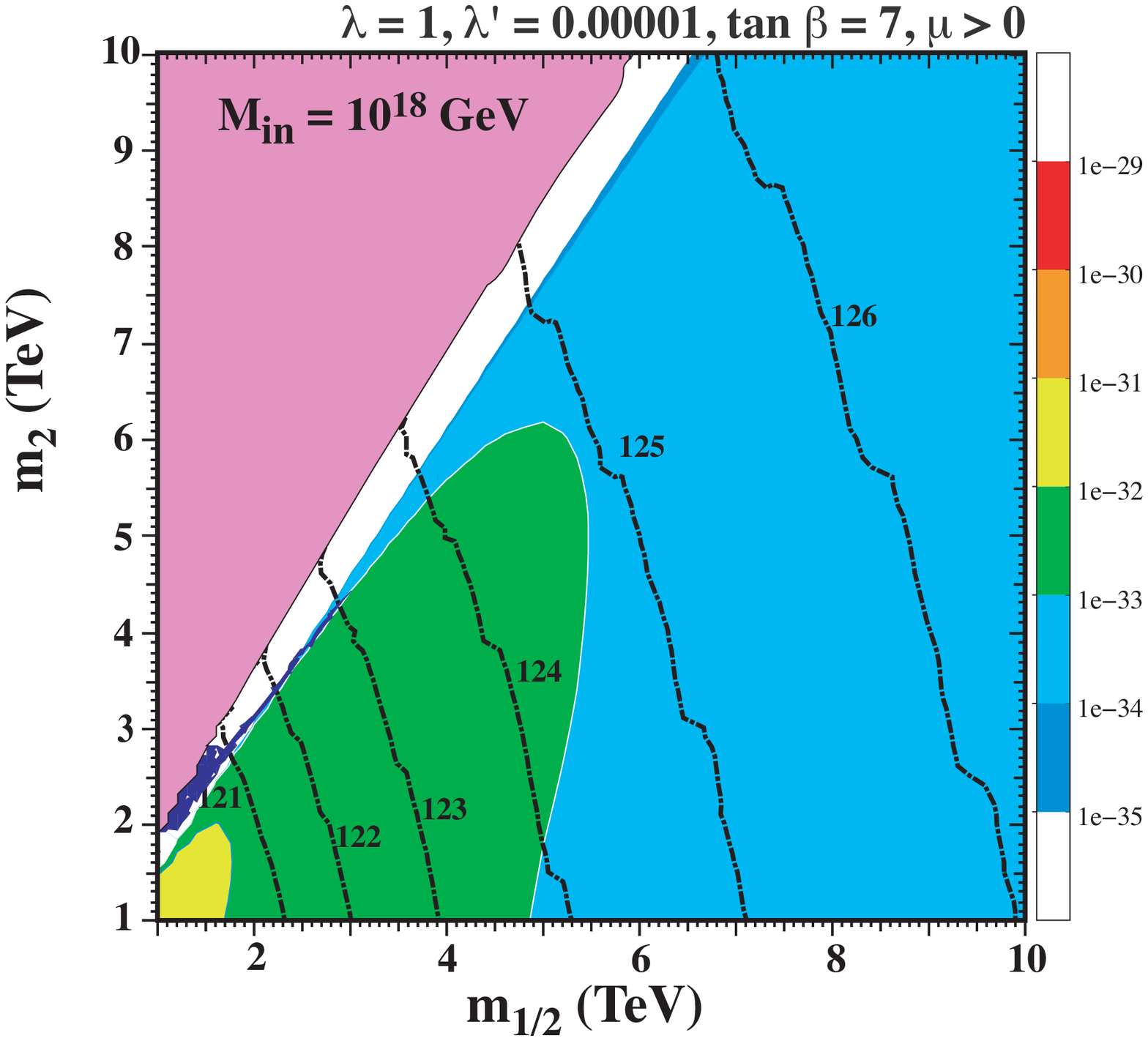}
\includegraphics[width=8cm]{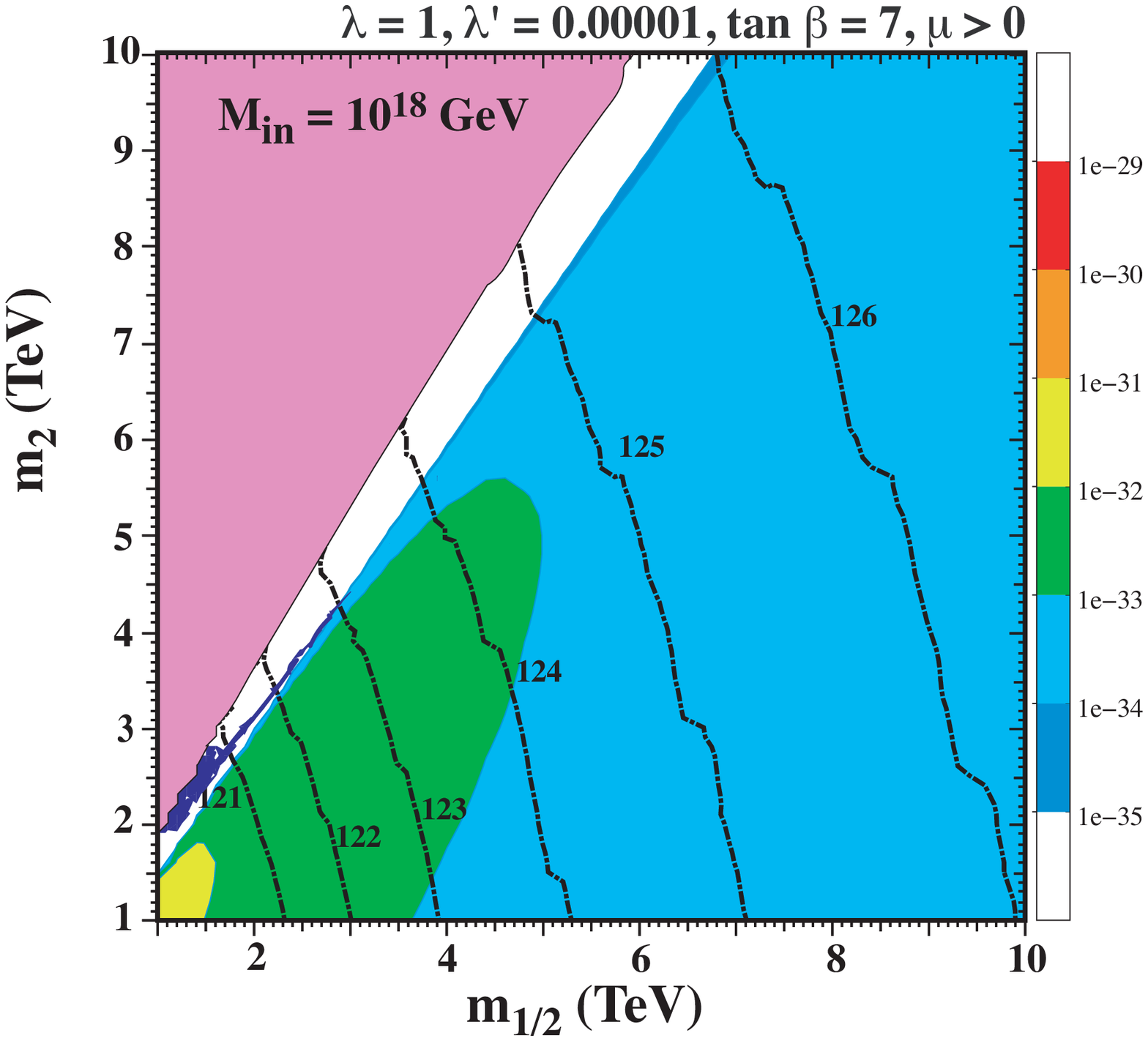}
\caption{\it As in Fig.~\ref{fig:M2a_EDMS}, showing values of of the 
electron EDM in the $(m_{1/2}, m_2)$ planes for the choices {\bf A} (left) and {\bf B} (right) 
in model M5 (upper panels) and in model M6 (lower panels) 
with $\Mi = 10^{18}$ GeV, $\tan \beta = 7$, 
$\lambda' = 0.00001$ and $\lambda = 1$. 
The color-coding for the electron EDM
is indicated in the bars beside the panels.
\label{fig:M56_EDMS} }
\end{figure}

{\it Electron EDM:} Predictions for the electron EDM in models M5 (upper panels)
and M6 (lower panels) are shown in Fig.~\ref{fig:M56_EDMS}, again with flavor choice
{\bf A} in the left panels and flavor choice {\bf B} in the right panels. Predictions 
are everywhere significantly below the present experimental sensitivity. Overall,
we see that the predicted values are somewhat smaller in model M6 than in model M5,
and somewhat larger with choice {\bf A} than with choice {\bf B}. In the most
favorable case, namely model M5 with flavor choice {\bf A}, the electron EDM
varies between $10^{-32}$~e.cm and $10^{-34}$~e.cm along the focus-point strip, 
and between $10^{-32}$~e.cm and $10^{-33}$~e.cm along the rapid-annihilation strip.
In the least favorable case, namely model M6 with flavor choice {\bf B}, 
the electron EDM is below $10^{-35}$~e.cm along all the rapid-annihilation strip.
In the case of M6, the approximation for $|d_e|$
we use in Eq.~(\ref{eq:approxim}) gives the
correct order of magnitude for choice {\bf{A}}, 
but falls short for choice {\bf B} by an about an order of magnitude, as the contributions of other elements in the
matrix $(a_{E})_{bc}$ must be taken into account when
determining the total value of $d_e$. 

As for the previous benchmarks, in both models M5 and M6 $\TPDK$ is within reach of
Hyper-Kamiokande~\cite{HK}, whereas $\Bmueg$ and the electron EDM
lie below the prospective future experimental reaches.

\section{Overview and Conclusions \label{sec:section_Conclusions}}

We have studied in this paper the phenomenological scope for SU(5) super-GUTs,
in which variants of no-scale boundary conditions are imposed on
the soft supersymmetry-breaking parameters at some input scale $\Mi > \Mg$.
Specifically, the soft supersymmetry-breaking scalar masses for the
squarks and sleptons vanish at $\Mi$, whereas those for the ${\mathbf{5}}$
and  $\overline{\mathbf{5}}$ Higgs supermultiplets depend whether they have twisted boundary conditions
at $\Mi$, as seen in (\ref{eq:boundarycond_Mi}), where other details of the
boundary conditions such as modular weights can be found. 

In addition to these input conditions,
the low-energy phenomenology of such models depends on the magnitude of
the hierarchy between $\Mg$ and $\Mi$, for which we consider the illustrative
values of $10^{16.5}$ and $10^{18}$~GeV.
We consider the constraints on such no-scale SU(5) models that are imposed by
the cosmological density of cold dark matter, $\TPDK$ and $M_h$.
We find that the Higgs field responsible for the charge-2/3 quark masses
must be twisted, while twisting the Higgs responsible for the charge-1/3 
and charged-lepton masses is optional.

Within this general framework, we have considered six specific choices for
the input boundary conditions. In addition to $\Mi$, modular weights and
GUT Higgs trilinear couplings - see (\ref{W5}) - these include
possible dimension-5 effects on GUT unification.
We emphasize also that the super-GUT running between $\Mi$ and $\Mg$ is
sensitive to the way in which the MSSM matter fields are embedded into
GUT supermultiplets, and specifically the underlying origin of CKM flavor mixing.
For each of these six no-scale models, we have considered two choices for flavor mixing, which yield
predictions for $\Bmueg$ and the electron EDM that are quite different, but less so for $\TPDK$. We also contrast
their predictions for $\TPDK$ with those made when neglecting off-diagonal entries
in the Yukawa coupling matrices. Though the differences in $\TPDK$ between the two
flavor choices are small, the differences from when the mixing is neglected may be larger
than the uncertainties associated with hadronic matrix elements in some cases.
We note that the ranges of sparticle mass parameters favored by the dark matter density and
$M_h$ (as calculated using {\tt FeynHiggs~2.16.0}) are generally beyond the current
$\TPDK$ limit as well as the reach of the LHC.

As can be seen in the various panels of Figs.~\ref{fig:M1_MUEG}, \ref{fig:M2a_MUEG} and \ref{fig:M56f},
the predictions for $\Bmueg$ are strongly dependent on the flavor choice as well as the choice of
no-scale model. However, in all cases except portions of the dark matter strips in models M1 and M6 with flavor choice {\bf B}, the value of $\Bmueg$ lies significantly below the current and
projected experimental sensitivities. The electron EDM is also below the current and projected experimental sensitivities, 
as can be seen in Figs.~\ref{fig:M1_EDMS}, \ref{fig:M2a_EDMS} and \ref{fig:M56_EDMS}.
On the other hand, there are significant regions of parameter space
for all models where $\TPDK$ is within reach of the Hyper-Kamiokande experiment.
As seen in the Tables, this is in particular the case for all the benchmark points
highlighted there.

These examples demonstrate explicitly that {\it there is no supersymmetric flavor problem in
no-scale models}, the reasons being that the no-scale boundary condition that every
soft supersymmetry-breaking matter scalar mass vanishes at the input scale $\Mi$ is flavor-universal, and that the
leading-order renormalization by gauge interactions is also flavor-universal.
Nevertheless, $\TPDK$ may well be within reach.

\section*{Acknowledgements}

The work of J.E.~was supported partly by the United Kingdom STFC Grant ST/P000258/1 
and partly by the Estonian Research Council via a Mobilitas Pluss grant. 
The work of K.A.O. was supported partly
by the DOE grant DE-SC0011842 at the University of Minnesota and he
 acknowledges support by the Director, Office of Science, Office of High Energy Physics of the U.S. Department of Energy under the Contract No. DE-AC02-05CH11231.
 L.~V. acknowledges hospitality and financial support from the Fine Theoretical Physics Institute at the University of Minnesota and from the Abdus Salam International Centre for Theoretical Physics, Italy, during various stages of this project, as well as the Fundamental Research Program at 
 the Korea Institute for Advanced Study.
The work of N.N. was supported by the Grant-in-Aid for Scientific Research B (No.20H01897), Young Scientists B (No.17K14270), and Innovative Areas (No.18H05542).


\begin{thebibliography}{10}
 
 \bibitem{ATLAS20}
M.~Aaboud {\it et al.} [ATLAS Collaboration],
  JHEP {\bf 1806}, 107 (2018)
  [arXiv:1711.01901 [hep-ex]];
  M.~Aaboud {\it et al.} [ATLAS Collaboration],
  Phys.\ Rev.\ D {\bf 97}, no. 11, 112001 (2018)
  [arXiv:1712.02332 [hep-ex]];
  ATLAS Collaboration, \url{https://twiki.cern.ch/twiki/bin/view/AtlasPublic/ SupersymmetryPublicResults}.
  
  \bibitem{CMS20}
M.~Aaboud {\it et al.} [ATLAS Collaboration],
  JHEP {\bf 1806}, 107 (2018)
  [arXiv:1711.01901 [hep-ex]];
  M.~Aaboud {\it et al.} [ATLAS Collaboration],
  Phys.\ Rev.\ D {\bf 97}, no. 11, 112001 (2018)
  [arXiv:1712.02332 [hep-ex]];
CMS Collaboration, \url{ https://twiki.cern.ch/twiki/bin/view/CMSPublic/PhysicsResultsSUS}.

\bibitem{lhch}
G.~Aad {\it et al.}  [ATLAS Collaboration],
  Phys.\ Lett.\ B {\bf 716}, 1 (2012)
  [arXiv:1207.7214 [hep-ex]];
   S.~Chatrchyan {\it et al.}  [CMS Collaboration],
  Phys.\ Lett.\ B {\bf 716}, 30 (2012)
  [arXiv:1207.7235 [hep-ex]].



\bibitem{Maiani:1979cx}
L.~Maiani,
in Proceedings, Gif-sur-Yvette Summer School On Particle Physics,
  1979, 1-52;
Gerard 't~Hooft and others (eds.),
{\it Recent Developments in Gauge Theories, Proceedings of the Nato Advanced
  Study Institute, Cargese, France, August 26 - September 8, 1979},
Plenum press, New York, USA, 1980, Nato Advanced Study Institutes
  Series: Series B, Physics, 59.;
Edward Witten,
{\em Phys. Lett.} B105, 267, 1981.

\bibitem{Ellis:1990zq}
John~R. Ellis, S.~Kelley and Dimitri~V. Nanopoulos,
{\em Phys. Lett.} B249, 441, 1990;
John~R. Ellis, S.~Kelley and Dimitri~V. Nanopoulos,
{\em Phys. Lett.} B260, 131, 1991;
Ugo Amaldi, Wim de~Boer, and Hermann Furstenau.
\newblock {\em Phys. Lett.}, B260, 447, 1991;
Paul Langacker and Ming-xing Luo,
{\em Phys. Rev.} D44, 817, 1991;
C.~Giunti, C.~W. Kim and U.~W. Lee,
{\em Mod. Phys. Lett.} A6, 1745, 1991.

 \bibitem{ehnos}        
  		H.~Goldberg,
                Phys.\ Rev.\ Lett.\ {\bf 50} (1983) 1419;
                J.~Ellis, J.~Hagelin, D.~Nanopoulos, K.~Olive and M.~Srednicki,
                Nucl.\ Phys.\ B {\bf 238} (1984) 453.

 \bibitem{mh}
J.~R.~Ellis, G.~Ridolfi and F.~Zwirner,
  Phys.\ Lett.\ B {\bf 257} (1991) 83;
  Phys.\ Lett.\ B {\bf 262} (1991) 477;
Y.~Okada, M.~Yamaguchi and T.~Yanagida,
  Prog.\ Theor.\ Phys.\  {\bf 85}, 1 (1991);
  A.~Yamada,
  Phys.\ Lett.\ B {\bf 263} (1991) 233;
  Howard~E. Haber and Ralf Hempfling,
{Phys. Rev. Lett.} 66 (1991) 1815;
M.~Drees and M.~M.~Nojiri,
  Phys.\ Rev.\ D {\bf 45} (1992) 2482;
  P.~H.~Chankowski, S.~Pokorski and J.~Rosiek,
  Phys.\ Lett.\ B {\bf 274} (1992) 191;
  Phys.\ Lett.\ B {\bf 286} (1992) 307.

 \bibitem{Ellis:2000ig}
J.~R.~Ellis and D.~Ross,
  Phys.\ Lett.\ B {\bf 506}, 331 (2001)
  [hep-ph/0012067].


\bibitem{ATLASmu}
G.~Aad \textit{et al.} [ATLAS Collaboration],
Eur. Phys. J. C \textbf{76}, no.1, 6 (2016)
[arXiv:1507.04548 [hep-ex]].
ATLAS Collaboration, \url{https://twiki.cern.ch/twiki/bin/view/AtlasPublic/HiggsPublicResults}.


\bibitem{CMSmu}
S.~Chatrchyan {\it et al.}  [CMS Collaboration],
  JHEP {\bf 1306} (2013) 081
  [arXiv:1303.4571 [hep-ex]];
V.~Khachatryan {\it et al.} [CMS Collaboration],
  Eur.\ Phys.\ J.\ C {\bf 75} (2015) 5,  212
  [arXiv:1412.8662 [hep-ex]].

\bibitem{Ellis:2017djk}
J.~Ellis, J.~L.~Evans, N.~Nagata, D.~V.~Nanopoulos and K.~A.~Olive,
Eur. Phys. J. C \textbf{77}, no.4, 232 (2017)
[arXiv:1702.00379 [hep-ph]].

\bibitem{Ellis:2016qra}
J.~Ellis, K.~Olive and L.~Velasco-Sevilla,
Eur. Phys. J. C \textbf{76}, no.10, 562 (2016)
[arXiv:1605.01398 [hep-ph]].

\bibitem{Reviews}
M.~Raidal, A.~van der Schaaf, I.~Bigi, M.~L.~Mangano, Y.~K.~Semertzidis, S.~Abel, S.~Albino, S.~Antusch, E.~Arganda and B.~Bajc, \textit{et al.}
Eur. Phys. J. C \textbf{57} (2008), 13-182
[arXiv:0801.1826 [hep-ph]];
D.~Croon, T.~E.~Gonzalo, L.~Graf, N.~Ko\v{s}nik and G.~White,
Front. in Phys. \textbf{7} (2019), 76
[arXiv:1903.04977 [hep-ph]].



\bibitem{Altmannshofer:2009ne}
W.~Altmannshofer, A.~J.~Buras, S.~Gori, P.~Paradisi and D.~M.~Straub,
Nucl. Phys. B \textbf{830}, 17-94 (2010)
[arXiv:0909.1333 [hep-ph]].

\bibitem{Gomez:2015ila}
M.~Gomez, S.~Heinemeyer and M.~Rehman,
Eur. Phys. J. C \textbf{75}, no.9, 434 (2015)
[arXiv:1501.02258 [hep-ph]].

\bibitem{Ellis:2010jb}
J.~Ellis, A.~Mustafayev and K.~A.~Olive,
Eur. Phys. J. C \textbf{69}, 219-233 (2010)
[arXiv:1004.5399 [hep-ph]].

\bibitem{Ellis:2016tjc}
J.~Ellis, J.~L.~Evans, A.~Mustafayev, N.~Nagata and K.~A.~Olive,
Eur. Phys. J. C \textbf{76}, no.11, 592 (2016)
[arXiv:1608.05370 [hep-ph]].


\bibitem{Ellis:2019fwf}
J.~Ellis, J.~L.~Evans, N.~Nagata, K.~A.~Olive and L.~Velasco-Sevilla,
[arXiv:1912.04888 [hep-ph]].

\bibitem{Ellis:1979fg}
  J.~R.~Ellis and M.~K.~Gaillard,
  Phys.\ Lett.\  {\bf 88B} (1979) 315.
  
\bibitem{super-GUT}
 L.~Calibbi, Y.~Mambrini and S.~K.~Vempati,
  JHEP {\bf 0709}, 081 (2007)
  [arXiv:0704.3518 [hep-ph]];
  L.~Calibbi, A.~Faccia, A.~Masiero and S.~K.~Vempati,
  Phys.\ Rev.\  D {\bf 74}, 116002 (2006)
  [arXiv:hep-ph/0605139];
  E.~Carquin, J.~Ellis, M.~E.~Gomez, S.~Lola and J.~Rodriguez-Quintero,
  JHEP {\bf 0905} (2009) 026
  [arXiv:0812.4243 [hep-ph]].

\bibitem{emo}
J.~Ellis, A.~Mustafayev and K.~A.~Olive,
  Eur.\ Phys.\ J.\ C {\bf 69}, 201 (2010)
  [arXiv:1003.3677 [hep-ph]];
   J.~Ellis, A.~Mustafayev and K.~A.~Olive,
  Eur.\ Phys.\ J.\ C {\bf 71}, 1689 (2011)
  [arXiv:1103.5140 [hep-ph]].

\bibitem{no-scale1}
E.~Cremmer, S.~Ferrara, C.~Kounnas and D.~V.~Nanopoulos,
  Phys.\ Lett.\ B {\bf 133} (1983) 61.
  
     \bibitem{no-scale2} 
  J.~R.~Ellis, A.~B.~Lahanas, D.~V.~Nanopoulos and K.~Tamvakis,
  Phys.\ Lett.\  {\bf 134B}, 429 (1984).


  \bibitem{LN}
  A.~B.~Lahanas and D.~V.~Nanopoulos,
  Phys.\ Rept.\  {\bf 145} (1987) 1.
  
\bibitem{Siegel:1979wq} 
  W.~Siegel,
  Phys.\ Lett.\  {\bf 84B}, 193 (1979).

\bibitem{Tobe:2003yj}
  K.~Tobe and J.~D.~Wells,
  Phys.\ Lett.\ B {\bf 588}, 99 (2004)
  [hep-ph/0312159].

\bibitem{Hisano:1992jj} 
  J.~Hisano, H.~Murayama and T.~Yanagida,
  Nucl.\ Phys.\ B {\bf 402}, 46 (1993)
  [hep-ph/9207279].
  
\bibitem{Hisano:1992mh} 
  J.~Hisano, H.~Murayama and T.~Yanagida,
  Phys.\ Rev.\ Lett.\  {\bf 69}, 1014 (1992).
  
\bibitem{Hisano:2013cqa} 
  J.~Hisano, T.~Kuwahara and N.~Nagata,
  Phys.\ Lett.\ B {\bf 723}, 324 (2013)
  [arXiv:1304.0343 [hep-ph]].



\bibitem{Hisano:1993zu}
J.~Hisano, H.~Murayama and T.~Goto,
Phys. Rev. D \textbf{49}, 1446-1453 (1994).

\bibitem{Evans:2019oyw}
J.~L.~Evans, N.~Nagata and K.~A.~Olive,
Eur. Phys. J. C \textbf{79}, no.6, 490 (2019)
[arXiv:1902.09084 [hep-ph]].

\bibitem{Borzumati:2009hu}
F.~Borzumati and T.~Yamashita,
Prog. Theor. Phys. \textbf{124}, 761-868 (2010)
[arXiv:0903.2793 [hep-ph]].


  \bibitem{GM}
   G.~F.~Giudice and A.~Masiero,
  Phys.\ Lett.\  B {\bf 206}, 480 (1988).

\bibitem{egno4}
J.~Ellis, M.~A.~G.~Garcia, D.~V.~Nanopoulos and K.~A.~Olive,
JCAP \textbf{10}, 003 (2015)
[arXiv:1503.08867 [hep-ph]].

\bibitem{vcmssm}
  J.~R.~Ellis, K.~A.~Olive, Y.~Santoso and V.~C.~Spanos,
  Phys.\ Lett.\ B {\bf 573} (2003) 162
  [arXiv:hep-ph/0305212],
  and
  Phys.\ Rev.\ D {\bf 70} (2004) 055005
  [arXiv:hep-ph/0405110].

\bibitem{Miura:2016krn}
K.~Abe \textit{et al.} [Super-Kamiokande Collaboration],
Phys. Rev. D \textbf{95} (2017) no.1, 012004
[arXiv:1610.03597 [hep-ex]].

\bibitem{HK}
K.~Abe \textit{et al.} [Hyper-Kamiokande Collaboration],
arXiv:1805.04163 [physics.ins-det].

\bibitem{Goto:1998qg}
T.~Goto and T.~Nihei,
Phys. Rev. D \textbf{59}, 115009 (1999)
[arXiv:hep-ph/9808255 [hep-ph]];
J.~Hisano, D.~Kobayashi, T.~Kuwahara and N.~Nagata,
  JHEP {\bf 1307}, 038 (2013)
  [arXiv:1304.3651 [hep-ph]];
  N.~Nagata and S.~Shirai,
  JHEP {\bf 1403}, 049 (2014)
  [arXiv:1312.7854 [hep-ph]];
  N.~Nagata,
 Ph.D. Thesis,
\url{http://doi.org/10.15083/00006623}.

\bibitem{Sakai:1981pk}
N.~Sakai and T.~Yanagida,
Nucl. Phys. B \textbf{197}, 533 (1982).

\bibitem{Ellis:2015rya}
J.~Ellis, J.~L.~Evans, F.~Luo, N.~Nagata, K.~A.~Olive and P.~Sandick,
Eur. Phys. J. C \textbf{76}, no.1, 8 (2016)
[arXiv:1509.08838 [hep-ph]].

\bibitem{Aoki:2013yxa}
Y.~Aoki, E.~Shintani and A.~Soni,
Phys. Rev. D \textbf{89}, no.1, 014505 (2014)
[arXiv:1304.7424 [hep-lat]].


\bibitem{Aoki:2017puj}
Y.~Aoki, T.~Izubuchi, E.~Shintani and A.~Soni,
Phys. Rev. D \textbf{96}, no.1, 014506 (2017)
[arXiv:1705.01338 [hep-lat]].

\bibitem{TheMEG:2016wtm}
A.~M.~Baldini \textit{et al.} [MEG Collaboration],
Eur. Phys. J. C \textbf{76} (2016) no.8, 434
[arXiv:1605.05081 [hep-ex]].


\bibitem{Baldini:2018nnn}
A.~M.~Baldini \textit{et al.} [MEG II Collaboration],
Eur. Phys. J. C \textbf{78}, no.5, 380 (2018)
[arXiv:1801.04688 [physics.ins-det]].

\bibitem{Bellgardt:1987du}
U.~Bellgardt \textit{et al.} [SINDRUM Collaboration],
Nucl. Phys. B \textbf{299}, 1-6 (1988).

\bibitem{Blondel:2013ia}
A.~Blondel, \textit{et al.} [Mu3e Collaboration], 
arXiv:1301.6113 [physics.ins-det].

\bibitem{Hisano:1995cp}
J.~Hisano, T.~Moroi, K.~Tobe and M.~Yamaguchi,
Phys. Rev. D \textbf{53}, 2442-2459 (1996)
[arXiv:hep-ph/9510309 [hep-ph]].

\bibitem{Arganda:2005ji}
E.~Arganda and M.~J.~Herrero,
Phys. Rev. D \textbf{73}, 055003 (2006)
[arXiv:hep-ph/0510405 [hep-ph]].

\bibitem{Bertl:2006up}
W.~H.~Bertl \textit{et al.} [SINDRUM II Collaboration],
Eur. Phys. J. C \textbf{47}, 337-346 (2006).

\bibitem{Adamov:2018vin}
R.~Abramishvili \textit{et al.} [COMET Collaboration],
PTEP \textbf{2020}, no.3, 033C01 (2020)
[arXiv:1812.09018 [physics.ins-det]].

\bibitem{Abusalma:2018xem}
F.~Abusalma \textit{et al.} [Mu2e Collaboration],
[arXiv:1802.02599 [physics.ins-det]].

\bibitem{Strategy:2019vxc}
R.~K.~Ellis, \textit{et al.},
[arXiv:1910.11775 [hep-ex]].

\bibitem{Czarnecki:1998iz}
A.~Czarnecki, W.~J.~Marciano and K.~Melnikov,
AIP Conf. Proc. \textbf{435}, no.1, 409-418 (1998)
[arXiv:hep-ph/9801218 [hep-ph]]; 
R.~Kitano, M.~Koike and Y.~Okada,
Phys. Rev. D \textbf{66}, 096002 (2002)
[arXiv:hep-ph/0203110 [hep-ph]];
V.~Cirigliano, R.~Kitano, Y.~Okada and P.~Tuzon,
Phys. Rev. D \textbf{80}, 013002 (2009)
[arXiv:0904.0957 [hep-ph]].


\bibitem{Andreev:2018ayy}
V.~Andreev \textit{et al.} [ACME Collaboration],
Nature \textbf{562}, no.7727, 355-360 (2018).

\bibitem{Ibrahim:2007fb}
T.~Ibrahim and P.~Nath,
Rev. Mod. Phys. \textbf{80}, 577-631 (2008)
[arXiv:0705.2008 [hep-ph]].







\bibitem{Ellis:2018jyl}
J.~Ellis, J.~L.~Evans, F.~Luo, K.~A.~Olive and J.~Zheng,
Eur. Phys. J. C \textbf{78}, no.5, 425 (2018)
[arXiv:1801.09855 [hep-ph]].

\bibitem{Bagnaschi:2018igf}
E.~Bagnaschi, H.~Bahl, J.~Ellis, J.~Evans, T.~Hahn, S.~Heinemeyer, W.~Hollik, K.~Olive, S.~Paßehr, H.~Rzehak, I.~Sobolev, G.~Weiglein and J.~Zheng,
Eur. Phys. J. C \textbf{79}, no.2, 149 (2019)
[arXiv:1810.10905 [hep-ph]].

\bibitem{Bagnaschi:2015eha}
E.~Bagnaschi, O.~Buchmueller, R.~Cavanaugh, M.~Citron, A.~De Roeck, M.~Dolan, J.~Ellis, H.~Flächer, S.~Heinemeyer, G.~Isidori, S.~Malik, D.~Martínez Santos, K.~Olive, K.~Sakurai, K.~de Vries and G.~Weiglein,
Eur. Phys. J. C \textbf{75}, 500 (2015)
[arXiv:1508.01173 [hep-ph]].

\bibitem{Bechtle:2015nua}
P.~Bechtle, J.~E.~Camargo-Molina, K.~Desch, H.~K.~Dreiner, M.~Hamer, M.~Krämer, B.~O'Leary, W.~Porod, B.~Sarrazin, T.~Stefaniak, M.~Uhlenbrock and P.~Wienemann,
Eur. Phys. J. C \textbf{76}, no.2, 96 (2016)
[arXiv:1508.05951 [hep-ph]].

\bibitem{Zyla:2020zbs}
P.~A.~Zyla \textit{et al.} [Particle Data Group],
PTEP \textbf{2020}, no.8, 083C01 (2020).

\bibitem{Planck}
   P.~A.~R.~Ade {\it et al.} [Planck Collaboration],
  Astron.\ Astrophys.\  {\bf 594}, A13 (2016)
  [arXiv:1502.01589 [astro-ph.CO]];
N.~Aghanim \textit{et al.} [Planck Collaboration],
Astron. Astrophys. \textbf{641} (2020), A6
[arXiv:1807.06209 [astro-ph.CO]].

\bibitem{Aprile}
See, for example, E.~Aprile \textit{et al.} [XENON],
Phys. Rev. Lett. \textbf{121} (2018) no.11, 111302
[arXiv:1805.12562 [astro-ph.CO]].
\bibitem{nuhm1}
H.~Baer, A.~Mustafayev, S.~Profumo, A.~Belyaev and X.~Tata,
  Phys.\ Rev.\  D {\bf 71} (2005) 095008
  [arXiv:hep-ph/0412059];
            H.~Baer, A.~Mustafayev, S.~Profumo, A.~Belyaev and X.~Tata,
               {\em JHEP} {\bf 0507} (2005) 065,
               hep-ph/0504001;
  J.~R.~Ellis, K.~A.~Olive and P.~Sandick,
  Phys.\ Rev.\  D {\bf 78} (2008) 075012
  [arXiv:0805.2343 [hep-ph]];
 J.~Ellis, F.~Luo, K.~A.~Olive and P.~Sandick,
  Eur.\ Phys.\ J.\ C {\bf 73} (2013) 2403
  [arXiv:1212.4476 [hep-ph]].
  
   \bibitem{nuhm2}
J.~Ellis, K.~Olive and Y.~Santoso,
Phys.\ Lett.\  B~{\bf 539} (2002) 107
[arXiv:hep-ph/0204192];
J.~R.~Ellis, T.~Falk, K.~A.~Olive and Y.~Santoso,
Nucl.\ Phys.\ B {\bf 652} (2003) 259
[arXiv:hep-ph/0210205].
  

  
  
  \bibitem{mc9}
 O.~Buchmueller {\it et al.},
 Eur.\ Phys.\ J.\ C {\bf 74} (2014) no.6,  2922
 [arXiv:1312.5250 [hep-ph]].


\bibitem{mc10}
 O.~Buchmueller {\it et al.},
 Eur.\ Phys.\ J.\ C {\bf 74} (2014) no.12,  3212
 [arXiv:1408.4060 [hep-ph]].

 \bibitem{GAMBIT}
P.~Athron \textit{et al.} [GAMBIT],
Eur. Phys. J. C \textbf{77}, no.12, 824 (2017)
[arXiv:1705.07935 [hep-ph]].


 \bibitem{FeynHiggs}
  H.~Bahl, T.~Hahn, S.~Heinemeyer, W.~Hollik, S.~Passehr, H.~Rzehak and G.~Weiglein,
  arXiv:1811.09073 [hep-ph].

\bibitem{fp}
 J.~L.~Feng, K.~T.~Matchev and T.~Moroi,
  Phys.\ Rev.\ Lett.\  {\bf 84}, 2322 (2000)
  [arXiv:hep-ph/9908309];
  Phys.\ Rev.\  D {\bf 61}, 075005 (2000)
  [arXiv:hep-ph/9909334];
  J.~L.~Feng, K.~T.~Matchev and F.~Wilczek,
  Phys.\ Lett.\  B {\bf 482}, 388 (2000)
  [arXiv:hep-ph/0004043];
  H.~Baer, T.~Krupovnickas, S.~Profumo and P.~Ullio,
  JHEP {\bf 0510} (2005) 020
  [hep-ph/0507282];
   J.~L.~Feng, K.~T.~Matchev and D.~Sanford,
  Phys.\ Rev.\ D {\bf 85}, 075007 (2012)
  [arXiv:1112.3021 [hep-ph]];
  P.~Draper, J.~Feng, P.~Kant, S.~Profumo and D.~Sanford,
  Phys.\ Rev.\ D {\bf 88}, 015025 (2013)
  [arXiv:1304.1159 [hep-ph]].

 \bibitem{susyflavor}
  J.~Rosiek,
  Comput.\ Phys.\ Commun.\  {\bf 188}, 208 (2014)
  [arXiv:1410.0606 [hep-ph]];
  A.~Crivellin, J.~Rosiek, P.~H.~Chankowski, A.~Dedes, S.~Jaeger and P.~Tanedo,
  Comput.\ Phys.\ Commun.\  {\bf 184}, 1004 (2013)
  [arXiv:1203.5023 [hep-ph]];
  J.~Rosiek, P.~Chankowski, A.~Dedes, S.~Jager and P.~Tanedo,
  Comput.\ Phys.\ Commun.\  {\bf 181}, 2180 (2010)
  [arXiv:1003.4260 [hep-ph]].
  
 






































\end{thebibliography}

\end{document}